\documentclass[aps,pra,twocolumn,showpacs,amsmath,amsfonts,amssymb,floatfix]{revtex4-1}
\usepackage{enumerate}
\usepackage{graphicx}
\usepackage{amsmath}
\usepackage{bm}
\usepackage{tensor}
\usepackage{epstopdf}

\begin{document}
\title{Contextual-value approach to the generalized measurement of observables}
\author{J. Dressel}
\author{A. N. Jordan}
\affiliation{Department of Physics and Astronomy, University of Rochester, Rochester, New York 14627, USA}
\date{\today}

\newcommand{\mean}[1]{\big\langle #1 \big\rangle}           
\newcommand{\dmean}[1]{\mean{\widetilde{ #1 }}}           
\newcommand{\cmean}[2]{\tensor[_{#1}]{\dmean{#2}}{}}        
\newcommand{\ccmean}[3]{\tensor[_{#1}]{\dmean{#2}}{_{#3}}}        
\newcommand{\wv}[3]{\tensor*[_{#1}]{\dmean{#2}}{^w_{#3}}}        

\newcommand{\ket}[1]{|#1\rangle}                    
\newcommand{\bra}[1]{\langle #1|}                   
\newcommand{\ipr}[2]{\big\langle #1 , #2 \big\rangle}       
\newcommand{\pipr}[2]{\langle #1 | #2 \rangle}       
\newcommand{\opr}[2]{\ket{#1}\bra{#2}}              
\newcommand{\pprj}[1]{\opr{#1}{#1}}                 

\newcommand{\Tr}[1]{\text{Tr}(#1)}       
\newcommand{\comm}[2]{\big[#1,\,#2\big]}         
\newcommand{\acomm}[2]{\big\{#1,\,#2\big\}}      
\def\R{\mbox{Re}}                                   
\def\I{\mbox{Im}}                                   
\newcommand{\op}[1]{\hat{#1}}                       
\newcommand{\dop}[1]{\check{#1}}                    
\def\prj{\Pi}                                       
\newcommand{\mat}[1]{\mathbf{#1}}                   

\newcommand{\oper}[1]{\mathcal{#1}}                 
\newcommand{\prop}[1]{\textit{#1}}                  

\begin{abstract}
  We present a detailed motivation for and definition of the \emph{contextual values} of an observable, which were introduced by Dressel et al., Phys. Rev. Lett. \textbf{104}, 240401 (2010).  The theory of contextual values is a principled approach to the generalized measurement of observables.  It extends the well-established theory of generalized state measurements by bridging the gap between partial state collapse and the observables that represent physically relevant information about the system.  To emphasize the general utility of the concept, we first construct the full theory of contextual values within an operational formulation of classical probability theory, paying special attention to observable construction, detector coupling, generalized measurement, and measurement disturbance.  We then extend the results to quantum probability theory built as a superstructure on the classical theory, pointing out both the classical correspondences to and the full quantum generalizations of both L\"{u}der's rule and the Aharonov-Bergmann-Lebowitz rule in the process.  As such, our treatment doubles as a self-contained pedagogical introduction to the essential components of the operational formulations for both classical and quantum probability theory.  We find in both cases that the contextual values of a system observable form a generalized spectrum that is associated with the independent outcomes of a partially correlated and generally ambiguous detector; the eigenvalues are a special case when the detector is perfectly correlated and unambiguous.  To illustrate the approach, we apply the technique to both a classical example of marble color detection and a quantum example of polarization detection.  For the quantum example we detail two devices: Fresnel reflection from a glass coverslip, and continuous beam displacement from a calcite crystal.  We also analyze the three-box paradox to demonstrate that no negative probabilities are necessary in its analysis.  Finally, we provide a derivation of the quantum weak value as a limit point of a pre- and postselected conditioned average and provide sufficient conditions for the derivation to hold.  
\end{abstract}


\maketitle
\section{Introduction}
Since the advent of quantum mechanics, practitioners have struggled with an inherent conceptual dualism in its formalism.  On one hand, time evolution of a quantum state is a continuous, deterministic, and reversible process well described by a wave equation.  On the other hand, there is irreducible stochasticity present in the measurement process that leads to discontinuous and generally irreversible state evolution in the form of so-called ``quantum jumps'' or ``state collapse.''  

To cope with the necessary introduction of the stochastic element of the theory while still preserving ties with the deterministic classical mechanics, traditional quantum mechanics \cite{Dirac1930,VonNeumann1932} emphasizes the role of Hermitian observable operators that are analogous to classical observables.  Indeed, we find that observables underlie most of the core concepts in the quantum theory: commutation relations of observables, complete sets of commuting observables, spectral expansions of observables, conjugate pairs of observables, expectation values of observables, uncertainty relations between observables, and time evolution generated by a Hamiltonian observable.  Even the quantum state is introduced as a superposition of observable eigenvectors.  The stochasticity of the theory manifests itself as a single prescription for how to average the omnipresent observables under a deterministically evolving quantum state: the implicit projective quantum jumps corresponding to laboratory measurements are largely hidden by the formalism.

Experimental control of quantum systems has improved since the early days of quantum mechanics, however, so the discontinuous evolution present in the measurement process can now be more carefully investigated.  Modern optical and condensed matter systems, for example, can condition the evolution of a state on the outcomes of weakly coupled measurement devices (e.g. \cite{Korotkov2006,Katz2008,Kim2009}), resulting in \emph{nonprojective} quantum jumps that alter the state more gently, or even resulting in continuous controlled evolution of the state.  Since observables are defined in terms of projective jumps that strongly affect the state, it becomes unclear how to correctly apply a formalism based on observables to such nonprojective measurements.  A refinement of the traditional formalism must be employed to correctly describe the general case.  

To address this need, the theory of \emph{quantum operations}, or generalized measurement, was introduced in the early 1970's by Davies \cite{Davies1970} and Kraus \cite{Kraus1971}, and has been developed over the past forty years to become a comprehensive and mathematically rigorous theory \cite{Kraus1983,Braginski1992,Busch1995,Nielsen2000,Alicki2001,Keyl2002,Breuer2007,Barnett2009,Wiseman2010}.  The formalism of quantum operations has seen the most use in quantum optics, quantum computation, and quantum information communities, where it is indispensable and well-supported by experiment.  However, it has not yet seen wide adoption outside of those communities.  

Unlike the traditional observable formalism, the formalism of quantum operations emphasizes the \emph{states}.  Observables are mentioned infrequently in the quantum operations literature, usually appearing only in the context of projective measurements where they are well-understood.  Some references (e.g. \cite{Keyl2002,Breuer2007,Wiseman2010}) define ``generalized observables'' in terms of the generalized measurements and detector outcome labels, but give no indication about their relationship to traditional observables, if any.  As a result, there is a conceptual gap between the traditional quantum mechanics of observables and the modern treatment of quantum operations that encompasses a much larger class of possible measurements than the traditional observables seemingly allow.

A possible response to this conceptual gap is to declare that traditional observables are meaningless outside the context of projective measurements.  This argument is supported by the fact that any generalized measurement can be understood as a part of a projective measurement being made on a larger joint system that can be associated with a traditional observable in the usual way (i.e. \cite[p. 20]{Keyl2002}).  However, there has been parallel research into the ``weak measurement'' of observables \cite{Aharonov1988,Duck1989,Aharonov1990,Aharonov2005,Aharonov2008,Jozsa2007,DiLorenzo2008,Aharonov2009,Aharonov2010,Tollaksen2007,Geszti2010,Hosoya2010,Shikano2010,Shikano2011,Kagami2011,Wu2011a,Haapasalo2011} that suggests that linking generalized measurements to traditional observables may not be such an outlandish idea.  

Weak measurements were introduced as a consequence of the von Neumann measurement protocol \cite{VonNeumann1932} that uses an interaction Hamiltonian with variable coupling strength to correlate an observable of interest to the generator of translations for a continuous meter observable.  The resulting shift in the meter observable is then used to infer information about the observable of interest in a nonprojective manner.  The technique has been used to great effect in the laboratory \cite{Ritchie1991,Wiseman2002,Wiseman2003,Gisin2004,Pryde2005,Mir2007,Hosten2008,Dixon2009,Lundeen2009,Starling2009,Howell2010,Starling2010,Starling2010b,Goggin2011,Kocsis2011} to measure physical quantities like pulse delays, beam deflections, phase shifts, polarization, and averaged trajectories.  Therefore, we conclude that there must be some meaningful way to reconcile nonprojective measurements with traditional observables more formally.

The primary purpose of the present work is to detail a synthesis between generalized measurements and observables that is powerful enough to encompass projective measurements, weak measurements, and any strength of measurement in between.  The formalism of \emph{contextual values}, which we explicitly introduced in \cite{Dressel2010,Note1} and further developed in \cite{Dressel2011,Dressel2011b,Dressel2012}, forms a bridge between the traditional notion of an observable and the modern theory of quantum operations.  For a concise introduction to the topic in the context of the quantum theory, we recommend reading our letter \cite{Dressel2010}.  

The central idea of the contextual-value formalism is that an observable can be completely measured indirectly using an imperfectly correlated detector by assigning an appropriate set of values to the detector outcomes.  The assigned set of values generally differs from the set of eigenvalues for the observable, and forms a \emph{generalized spectrum} that is associated with the operations of the generalized measurement, rather than the spectral projections for the observable.  Thus, the spectrum that one associates with an observable will depend on the \emph{context} of how the measurement is being performed; such an inability to completely discuss observables without specifying the full measurement context is reminiscent of Bell-Kochen-Specker contextuality \cite{Bell1966,Kochen1967,Mermin1993,Spekkens2005,Leifer2005,Tollaksen2007,Spekkens2008} and motivates the name ``contextual values.''  

The secondary purpose of the present work is to demonstrate that the contextual values formalism for generalized observable measurement is essentially classical in nature.  Hence, it has potential applications outside the usual scope of the quantum theory.  Indeed, we will show that any system that can be described by Bayesian probability theory can benefit from the contextual-value formalism.  

Extending contextual values to the quantum theory from the classical theory clarifies which features of the quantum theory are novel.  The quantum theory can be seen as an extension of a classical probability space to a continuous manifold of incompatible frameworks, where each framework is a copy of the original probability space.  Hence, intrinsically quantum features arise not from the observables defined in any particular framework, but instead from the relative orientations of the incompatible frameworks.  As we shall see, the differences manifest in sequential measurements and conditional measurements due to the probabilistic structure of the incompatible frameworks, rather than the observables or contextual values themselves.

To keep the paper self-contained with these aims in mind, we first develop both the operational approach to measurement and the contextual values formalism completely within the confines of classical probability theory, giving illustrative examples to cement the ideas.  We then port the formalism to the quantum theory and identify the essential differences that arise.  Our analysis therefore doubles as a pedagogical introduction to the operational approaches for both classical and quantum probability theory that should be accessible to a wide audience.

The paper is organized as follows: In Sec. \ref{sec:colorblind}, we provide a simple intuitive example to introduce the concept of contextual values.  In Secs. \ref{sec:csample} through \ref{sec:cdetector}, we develop the classical version of the operational approach to measurement.  In Sec. \ref{sec:ccv}, we introduce the contextual values formalism classically and then give several examples similar to the initial example.  In Secs. \ref{sec:qsample} through \ref{sec:qdetector}, we generalize the classical operations to quantum operations and highlight the key differences with explicit examples.  In Sec. \ref{sec:qcv}, we apply the contextual values formalism to the quantum case and show that it is unchanged.  We also specifically address how to treat weak measurements as a special case of our more general formalism and provide a derivation of the quantum weak value in Sec. \ref{sec:wv}.  Finally, we give our conclusions in Sec. \ref{sec:conclusion}.

\subsection{Example: Colorblind Detector} \label{sec:colorblind}
The idea of the contextual values formalism is deceptively simple.  Its essence can be distilled from the following classical example of an \emph{ambiguous detector}: Suppose we wish to measure a marble that may be colored either red or green.  A person with normal vision can distinguish the colors unambiguously and so would represent an ideal detector for the color state of the marble.  A partially colorblind person, however, may only estimate the color correctly some percentage of the time and so would represent an ambiguous detector of the color state of the marble.  

If the person is only mildly colorblind, then the estimations will be strongly correlated to the actual color of the marble.  The ambiguity would then be perturbative and could be interpreted as \emph{noise} introduced into the measurement.  However, if the person is strongly colorblind, then the estimations may be only mildly correlated to the actual color of the marble.  The ambiguity becomes \emph{nonperturbative}, so the noise dominates the signal in the measurement.

We can design an experimental protocol where an experimenter holds up a marble and the colorblind person gives a thumbs-up if he thinks the marble is green or a thumbs-down if he thinks the marble is red.  Suppose, after testing a large number of known marbles, the experimenter determines that a green marble correlates with a thumbs-up 51\% of the time, while a red marble correlates with a thumbs-down 53\% of the time.  The experimental outcomes of thumbs-up and thumbs-down are thus only weakly correlated with the actual color of the marble.

Having characterized the detector in this manner, the experimenter provides the colorblind person with a very large bag of an unknown distribution of colored marbles.  The colorblind person examines every marble, and for each one records a thumbs-up or a thumbs-down on a sheet of paper, which he then returns to the experimenter.  The experimenter then wishes to reconstruct what the average distribution of marble colors in the bag must be, given only the ambiguous output of his colorblind detector.

For simplicity, the clever experimenter decides to associate the colors with numerical values: $1$ for green (g) and $-1$ for red (r).  In order to compare the ambiguous outputs with the colors, he also assigns them \emph{different} numerical values: $a$ for thumbs-up (u), and $b$ for thumbs-down (d).  He then writes down the following probability constraint equations for obtaining the average marble color, $\mean{\text{color}}$, based on what he has observed,
\begin{align}
  P(u) &= (.51) P(g) + (.49) P(r), \nonumber \\
  P(d) &= (.47) P(g) + (.53) P(r), \nonumber \\
  \mean{\text{color}} &= 1 P(g) - 1 P(r) = a P(u) + b P(d),
  \label{eq:colorblindcv}
\end{align}
which he can rewrite as a matrix equation in the basis of the color probabilities $P(g)$ and $P(r)$,
\begin{align}
  \begin{pmatrix} 1 \\ -1 \end{pmatrix} &= \begin{pmatrix}.51 & .47 \\ .49 & .53\end{pmatrix} \begin{pmatrix}a \\ b\end{pmatrix}.
  \label{eq:colorblindcvmatrix}
\end{align}
After solving this equation, he finds that he must assign the amplified values $a = 25$ and $b = -25$ to the outcomes of thumbs-up and thumbs-down, respectively, in order to compensate for the detector ambiguity.  After doing so, he can confidently calculate the average color of the marbles in the large unknown bag using the identity \eqref{eq:colorblindcv}.  

The classical color observable has eigenvalues of $1$ and $-1$ that correspond to an ideal measurement.  The amplified values of $25$ and $-25$ that must be assigned to the ambiguous detector outcomes are \emph{contextual values} for the same color observable.  The \emph{context} of the measurement is the characterization of the colorblind detector, which accounts for the degree of colorblindness.  The expansion \eqref{eq:colorblindcv} relates the spectrum of the observable to its generalized spectrum of contextual values.  With this identity, both an ideal detector and a colorblind detector can measure the same observable; however, the assigned values must change depending on the context of the detector being used.

\section{Classical probability theory}
To define contextual values more formally, we shall define generalized measurements within the classical theory of probability using the same language as quantum operations.  In particular, rather than representing the observables of classical probability theory in the traditional way as functions, we shall adopt a more calculationally flexible, yet equivalent, \emph{algebraic} representation that closely resembles the operator algebra for quantum observables.

We also briefly comment that the relevant subset of probability theory that is summarized here may slightly differ in emphasis from incarnations that the reader may have encountered previously.  Our treatment acknowledges that probability theory, in its most general incarnation, is a system of formal reasoning about Boolean logic propositions \cite{Cox1946,Cox1961}; specifically, our treatment emphasizes logical inference rather than the traditional frequency analysis of concrete random variable realizations.  However, the ``frequentist'' approach of random variables is not displaced by the logical approach, but is rather subsumed as an important special case pertaining to repeatable experiments with logically independent outcomes.  Due to its clarity and generality, the logical approach has been widely adopted in diverse disciplines under the distinct name ``Bayesian probability theory.''  Several physicists, including (but certainly not limited to) Jaynes \cite{Jaynes2003}, Caves \cite{Caves2002}, Fuchs \cite{Fuchs2010}, Spekkens \cite{Spekkens2007}, Harrigan \cite{Harrigan2010}, Wiseman \cite{Wiseman2010}, and Leifer \cite{Leifer2011}, have also extolled its virtues in recent years.  We follow suit to emphasize the generality of the contextual-value concept.

\subsection{Sample spaces and observables} \label{sec:csample}
\begin{figure}[t]
  \begin{center}
    \includegraphics[width=\columnwidth]{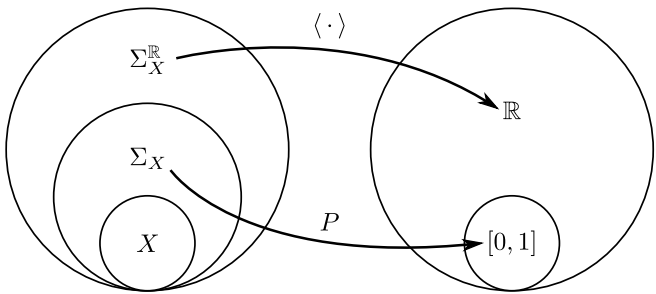}
  \end{center}
  \caption{Diagram of the relationship between the sample space of atomic propositions $X$, the Boolean algebra of propositions $\Sigma_X$, and the algebra of observables $\Sigma_X^\mathbb{R}$.  The probability state $P$ is a measure from $\Sigma_X$ to the interval $[0,1]$.  The expectation functional $\mean{\cdot}$ is a linear extension of $P$ that maps $\Sigma_X^\mathbb{R}$ to the reals $\mathbb{R}$; by construction $\mean{\cdot} = P(\cdot)$ whenever both are defined.}
  \label{fig:venndiagram}
\end{figure}

In what follows, we shall consider the stage on which classical probability theory unfolds---namely its space of observables---to be a commutative algebra over the reals that we denote $\Sigma^\mathbb{R}_X$.  This choice of notation is motivated by the fact that the observable algebra is built from and contains two related spaces, $X$ and $\Sigma_X$, that are conceptually distinct and equally important to the theory.  The three are illustrated in Fig.~\ref{fig:venndiagram} to orient the discussion.  To avoid distracting technical detail, we will briefly describe finite-dimensional versions of these three spaces here, and note straightforward generalizations to the continuous case when needed \cite{Note2}.

\emph{Sample spaces}.---The core of a probability space is a set of mutually exclusive logic propositions, $X$, known as the \emph{sample space} of \emph{atomic propositions}.  In other words, elements of the sample space, such as $g,r\in X$, represent ``yes or no'' questions that cannot be answered ``yes'' simultaneously and cannot be broken into simpler questions.  For example, $g=\text{``Does the marble look green?''}$ and $r=\text{``Does the marble look red?''}$ are valid mutually exclusive atomic propositions.  To be a proper sample space, the propositions should form a complete set, meaning that there must always be exactly one true proposition.  Physically, such propositions typically correspond to mutually independent outcomes of an experiment that probes some system of interest.  Indeed, any accessible physical property must be testable by some experiment, and any experiment can be described by such a collection of yes or no questions.

\emph{Boolean algebra}.---The atomic propositions in $X$ can be extended to more complex propositions by logical combination in order to form the larger space $\Sigma_X$.  Specifically, we can combine them algebraically with a logical \textsc{or} denoted by addition and a logical \textsc{and} denoted by multiplication.  For example, given propositions $x,y,z\in \Sigma_X$, the quantity $xy + yz$ would denote the proposition ``($x$ \textsc{and} $y$) or ($y$ \textsc{and} $z$).''  Importantly, both the sum and the product commute since the corresponding logical operations commute, and the propositions are idempotent so $x^2 = x$ for any $x\in \Sigma_X$.  Furthermore, the product of any two nonequal propositions in $X$ must be trivially false since they are mutually exclusive; we denote the trivially false proposition as $0$ since its product with any proposition is also trivially false.  Similarly, the sum of all propositions in $X$ will be trivially true since one of the atomic propositions must be true by construction; we denote the trivially true proposition as $1_X$ since its product with any proposition $x\in X$ leaves that proposition invariant, $1_X x = x$.  The logical operation of \textsc{not}, or complementation ($x^c$) with respect to $X$, can then be defined as the subtraction from the identity $x^c = 1_X - x$ since $x + x^c = 1_X$ must be true for any proposition $x\in X$ by definition.  The proposition space $\Sigma_X$ contains $X$ and is closed under the operations of \textsc{and}, \textsc{or}, and \textsc{not}; hence, it forms a \emph{Boolean logic algebra} \cite{Note3}.

\emph{Observables}.---Finally, we extend $\Sigma_X$ linearly over the real numbers to obtain the commutative algebra of \emph{observables} $\Sigma_X^\mathbb{R}$.  That is, any linear combination of propositions $F = a x + b y$ with $a,b\in\mathbb{R}$ and $x,y\in\Sigma_X$ is an observable in $\Sigma_X^\mathbb{R}$; similarly any linear combination of observables $H = a' F + b' G$ with $a',b'\in\mathbb{R}$ and $F,G\in\Sigma^\mathbb{R}_X$ is also an observable in $\Sigma_X^\mathbb{R}$.  Countable sums are permitted provided the coefficients converge.  The three spaces $X$, $\Sigma_X$, and $\Sigma_X^\mathbb{R}$ are illustrated in Fig.~\ref{fig:venndiagram}.  

The observables combine logical propositions with numbers that describe the relation of each proposition to some meaningful reference.  For example, one could define a simple observable $A = (1)g + (-1)r$ that assigns a value of $1$ to the proposition asking whether a marble looks green and assigns a value of $-1$ to the proposition asking whether that same marble looks red in order to distinguish the colors by a sign.  Alternatively, one can bestow a physical meaning to the color propositions by defining a wavelength observable instead: $B = (550\text{nm})g + (700\text{nm})r$.  One could even define an observable $C = (\$2)g + (-\$3)r$ that indicates a monetary bet made on the color of the marble, with $\$2$ being awarded for a color of green and $\$3$ being lost for a color of red.  Such numerical labels are always assigned by convention, but indicate physically relevant information about the type of questions being asked by the experimenter that are answerable by the independent propositions.

\emph{Representation}.---The algebra $\Sigma_X$ can be represented as the lattice of projection operators acting on a Hilbert space exactly as in the standard representation of quantum theory \cite{VonNeumann1932,Jauch1968,Alicki2001}.  The elements $\{x\}$ of $X$ correspond to rank-1 projection operators $\{\pprj{x}\}$ onto orthogonal subspaces spanned by orthonormal vectors $\{\ket{x}\}$ in the Hilbert space.  Any sum of $n$ elements of $X$, $x_1 + \dots + x_n$, corresponds to a rank-$n$ projection operator $\pprj{x_1,\dots,x_n}$ onto a subspace spanned by $n$ orthonormal vectors $\{\ket{x_1},\dots,\ket{x_n}\}$ in the Hilbert space.  Hence, we shall casually refer to propositions of the Boolean algebra $\Sigma_X$ as \emph{projections} in what follows.  However, it is important to note that the Boolean algebra $\Sigma_X$ need not be represented in this fashion to be well defined.

Just like the propositions $\Sigma_X$ can be represented as projections on a Hilbert space, the observables $\Sigma_X^\mathbb{R}$ can also be represented as the algebra of Hermitian operators acting on the same Hilbert space.  Hence, we shall casually refer to observables in $\Sigma_X^\mathbb{R}$ as \emph{observable operators} in analogy to the quantum theory.  However, unlike quantum observables, all classical observables commute.  It is important to note that the representation of observables as operators on a Hilbert space in both the classical and the quantum case remains strictly optional for calculational convenience.

\emph{Independent Probability Observables}.---We note that the identity observable $1_X$ can be partitioned into many distinct sets of independent propositions in $\Sigma_X$, such as $\sum_i x_i = 1_X$, which is known as a \emph{closure relation}.  Each partitioning corresponds to a particular detector arrangement that only probes those propositions.  Such a partitioning $\{x_i\}$ has the common mathematical name \emph{projection-valued measure} (PVM) since it forms a measure over the index $i$ and has a representation that consists of orthogonal projections.  However, we shall make an effort to call the propositions $\{x_i\}$  \emph{independent probability observables} to be more physically descriptive.  We will later contrast them with more general probability observables.

General observables can be constructed from independent probability observables by associating a real value $f(x_i)$ to each index $i$ in the sum, $F = \sum_i f(x_i) x_i$.  The product of the observable with any of its constituent probability observables simplifies, $F\, x_i = f(x_i) x_i$; hence, the associated values form the set of \emph{eigenvalues} for the observable.  For a finite observable space $\Sigma_X^\mathbb{R}$, the set of atomic propositions $X$ itself is a maximally refined set of independent probability observables that can construct any observable in the space,
\begin{align}\label{eq:cobs}
  F &= \sum_{x\in X} f(x) x.
\end{align}
In the continuous case the values $f(x)$ form a measurable function that specifies the \emph{spectrum} of the observable; the sum \eqref{eq:cobs} is then commonly written as an integral over the continuous set of propositions $\{\pprj{x}\}$, $F = \int_X f(x) \, d\pprj{x}$.  We use the Hilbert space notation $d\pprj{x}$ in the integral to avoid later confusion with real-valued integrals.

\subsection{States, densities, and collapse}
\emph{Probability measures}.---A \emph{state} $P$ is a \emph{probability measure} over the Boolean algebra $\Sigma_X$, meaning that it is a linear map from $\Sigma_X$ to the interval $[0,1]$ such that $P(1_X) = 1$.  Such a state $P$ assigns a numerical value $P(x)$ to each proposition $x\in \Sigma_X$ that quantifies its degree of \emph{plausibility}; that is, $P(x)$ formally indicates how likely it is that the question $x$ would be answered ``yes'' were it to be answered, with $1$ indicating a certain ``yes'' and $0$ indicating a certain ``no.''  The value $P(x)$ is called the \emph{probability} for the proposition $x$ to be true.  Normalizing $P(1_X) = 1$ ensures that exactly one proposition in the sample space must be true.  For continuous spaces, the state becomes an integral $P(x_0) = \int_{x_0\in \Sigma_X} \, dP(x)$.

\emph{Frequencies}.---Empirically, one can check probabilities by repeatedly asking a proposition in $\Sigma_X$ to identically prepared systems and collecting statistics regarding the answers.  For a particular proposition $x\in\Sigma_X$, the ratio of yes-answers to the number of trials will converge to the probability $P(x)$ as the number of trials becomes infinite.  However, the probability has a well-defined meaning as a plausibility prediction even without actually performing such a repeatable experiment.  Indeed, designing good quality repeatable experiments to check the probabilities assigned by a predictive state is the primary goal of experimental science, and is generally quite difficult to achieve.

\emph{Expectation functionals}.---The linear extension of a state $P$ to the whole observable algebra $\Sigma_X^\mathbb{R}$ is an \emph{expectation functional} that averages the observables, and is traditionally notated with angled brackets $\mean{\cdot}$.  Specifically, for an observable $F = \sum_{x\in X} f(x)\, x$, then,
\begin{align}\label{eq:cexpval}
  \mean{F} &= \sum_{x\in X} f(x) P(x),
\end{align}
is the \emph{expectation value}, or average value, of $F$ under the functional $\mean{\cdot}$ that extends the probability state $P$.  Since $\mean{\cdot}$ is linear, it passes through the sum and the constant factors of $f(x)$ to apply directly to the propositions $x$.  The restriction of $\mean{\cdot}$ to $\Sigma_X$ is $P$, so $\mean{x} = P(x)$ as written in \eqref{eq:cexpval}.  That is, the expectation value $\mean{x}$ of a pure proposition $x$ is the probability of that proposition.  The probability state $P$ and its linear extension $\mean{\cdot}$ are illustrated in Fig.~\ref{fig:venndiagram}.  For continuous spaces the sum \eqref{eq:cexpval} becomes an integral of the measurable function $f(x)$, $\mean{F} = \mean{\int_X f(x) \, d\pprj{x}} = \int_X f(x) \, dP(x)$.  

\emph{Moments}.---The $n$\textsuperscript{th} statistical moment of $F$ is $\mean{F^n} = \sum_{x\in X} f^n(x) P(x)$ and empirically corresponds to measuring the observable $F$ $n$ times in a row per trial on identical systems and averaging the repeated results.  Hence, the moments quantify the fluctuations of the observable measurements that stem from uncertainty in the state.  For continuous spaces, the higher moments also become integrals $\mean{F^n} = \int_X f^n(x) \, dP(x)$.

\emph{Densities}.---States can often be represented as \emph{densities} with respect to some \emph{reference measure} $\mu$ from $\Sigma_X$ to $\mathbb{R}^+$, which can be convenient for calculational purposes.  Just as the state $P$ can be linearly extended to an expectation functional $\mean{\cdot}$, any reference measure $\mu$ can be linearly extended to a functional $\mean{\cdot}_\mu$.  For continuous spaces, such a reference functional takes the form of an integral $\mean{F}_\mu = \int_X f(x) d\mu(x)$.  The representation of a state as a density follows from changing the integration measure for the state to the reference measure $\mean{F} = \int_X f(x) \, dP(x) = \int_X f(x) (dP/d\mu)(x) \, d\mu(x)$.  The Jacobian conversion factor $dP/d\mu$ from the integral over $dP(x)$ to the integral over a different measure $d\mu(x)$ is the \emph{probability density} for $P$ with respect to $\mu$, if it exists \cite{Note4}.  We can then define a \emph{state density observable} $P_\mu = \int_X (dP/d\mu)(x) \, d\pprj{x}$ that relates the expectation functional $\mean{\cdot}$ to the reference functional $\mean{\cdot}_\mu$ directly according to the relation $\mean{P_\mu\, F}_\mu = \mean{F}$.  

For continuous spaces, the standard integral is most frequently used as a reference.  Hence, the probability density with respect to the standard integral is given the simple notation $p(x)$ such that $\mean{F} = \int_X f(x) p(x)\, dx$.  Importantly, the probability for $x$ is not the density $p(x)=(dP/dx)(x)$, but is the (generally infinitesimal) integral of the density over a single point $P(x) = \int_{x\in X} p(x)\, dx$ \cite{Robinson1966,Nelson1987}, commonly notated $p(x) dx$. 

In discrete spaces we apply the same idea by defining a state density observable directly in terms of measure ratios,
\begin{align}\label{eq:cdensity}
  P_\mu = \sum_{x\in X} \frac{P(x)}{\mu(x)}\, x.
\end{align}
Then by definition and linearity, $\mean{P_\mu F}_\mu = \sum_{x\in X} [P(x)/\mu(x)] f(x) \mu(x) = \sum_{x\in X} f(x) P(x) = \mean{F}$, as required.  Evidently, the measure $\mu$ must be nonzero for all propositions $x$ for which $P$ is nonzero in order for such a state density to be well defined.  This definition as a ratio of functionals will correctly reproduce the Jacobian derivative in the continuous case using a limiting prescription.

\emph{Trace}.---An important reference measure which is nonzero for any nonzero proposition is the counting measure, or \emph{trace} $\text{Tr}$, which evaluates to the rank of any proposition in $\Sigma_X$; for example, given $x,y,z\in X$ then $(x+y+z)\in \Sigma_X$ is a rank-3 proposition and $\text{Tr}(x + y + z) = \text{Tr}(x) + \text{Tr}(y) + \text{Tr}(z) = 1 + 1 + 1 = 3$.  Since the trace evaluates to unity on any atomic proposition, \emph{any} state has a trace-density defined by equation \eqref{eq:cdensity} that is traditionally notated as $\rho$.  
\begin{align}\label{eq:rho}
  \rho &= \sum_{x\in X} P(x)\, x.
\end{align}
The trace-density is the only state density that is always defined and exactly determined by the probabilities of the atomic propositions $P(x)$.  Because of this, the trace-representation of a state can be naturally interpreted as an inner product,
\begin{align}\label{eq:ipr}
  \ipr{\rho}{F} &= \text{Tr}(\rho\, F) = \mean{F},
\end{align}
between the trace-density and the observable, known as the Hilbert-Schmidt inner product.  The trace will become particularly important when we generalize to quantum mechanics, which is why we mention it here.  Indeed, the trace-density $\rho$ will be equivalent to the quantum mechanical density operator when extended to the noncommutative case.  For continuous spaces the integral is traditionally preferred to the trace as a reference because the trace can frequently diverge.  

\emph{State collapse}.---If a question on the probability space is answered by some experiment, then the state indicating the plausibilities for future answers must be updated to reflect the acquired answer.  The update process is known as \emph{Bayesian state conditioning}, or \emph{state collapse}.  Specifically, if a proposition $y\in \Sigma_X$ is verified to be true, then the experimenter updates the expectation functional to the conditioned functional,
\begin{align}\label{eq:ccollapse}
  \mean{F}_y &= \frac{\mean{y\, F}}{P(y)},
\end{align}
that reflects the new information.  For a proposition $x\in \Sigma_X$, the conditional probability $\mean{x}_y = P(yx)/P(y)$ has the traditional notation $P(x | y)$ and is read as ``the probability of $x$ given $y$.''  

From \eqref{eq:ccollapse}, any state density corresponding to $P$ will be similarly updated to a new density via a product, 
\begin{align}\label{eq:ccollapsedensity}
  P_{\mu|y} = \frac{P_\mu \,y}{P(y)}.
\end{align}
Notably, conditioning the trace-density $\rho$ on an atomic proposition $y\in X$ will collapse the density to become the proposition itself, $\rho_y = \rho\, y / P(y) = y$.

Note that the proposition $y$ serves a dual role in the conditioning procedure.  First, it is used to compute the normalization probability $P(y)$.  Second, it directly updates the state via a product action.  The product indicates that future questions will be logically linked to the answered question with the \emph{and} operation; that is, the knowledge about the system has been refined by the answered question.  The process of answering a question about the system and then conditioning the state on the new information is called a \emph{measurement}; moreover, since the proposition $y$ is a projection acting on the density, this kind of measurement is called a \emph{projective measurement}.

\emph{Bayes' rule}.---If we pick another proposition $z\in\Sigma_X$ as the observable in \eqref{eq:ccollapse} we can derive \emph{Bayes' rule} as a necessary consequence by interchanging $y$ and $z$ and then equating the joint probabilities $P(yz)$, 
\begin{align}\label{eq:cbayes}
  P(z|y) = P(y|z) \frac{P(z)}{P(y)}.
\end{align}
Bayes' rule relates conditioned expectation functionals to one another and so is a powerful logical inference tool that drives much of the modern emphasis on the logical approach to probability theory.  

\emph{Disturbance}.---Conditioning, however, is not the only way that one can alter a state.  One can also \emph{disturb} a state without learning any information about it, which creates a transition to an updated expectation functional that we denote with a tilde $\dmean{\,\cdot\,}$ according to,
\begin{subequations}\label{eq:cdisturb}
\begin{align}
  \dmean{F} &= \mean{\mathcal{D}(F)}, \\
  \mathcal{D}(F) &= \sum_{x\in X} \mean{F}_{D_x} \, x, \\
  \mean{F}_{D_x} &= \sum_{x'\in X} f(x') D_x(x').
\end{align}
\end{subequations}
Here the disturbance $\mathcal{D}$ is a map from $\Sigma_X^\mathbb{R}$ to $\Sigma_X^\mathbb{R}$ that is governed by a collection of states $\{D_x\}$ that specify \emph{transition probabilities} $D_x(x')$ from old propositions $x$ to new propositions $x'$.  To be normalized, the transition states must satisfy $D_x(1_X) = 1$, so that $\mean{1_X}_{D_x} = 1_X$ and therefore $\mathcal{D}(1_X) = 1_X$.  Updating the state according to \eqref{eq:cdisturb} is also known as \emph{Bayesian belief propagation} \cite{Leifer2011} and is more commonly written in the fully expanded form $\dmean{F} = \sum_{x\in X} P(x) \sum_{x'\in X} D_x(x') f(x')$.

\emph{Time evolution}.---As an important special case, the time evolution of a Markovian stochastic process is a form of disturbance $\mathcal{D}_t$, known as a \emph{propagator}, that is parametrized by a time interval $t$.  No information is learned as the system evolves, so the knowledge about the system as represented by the expectation functional can only propagate according to the laws governing the time evolution.  For a Hamiltonian system, the time evolution is of Liouville form; that is, if we define a time-evolving observable as $F(t) = \mathcal{D}_t(F)$ then we have $dF(t)/dt = \{F(t),H\}_p$, where $\{\cdot,\cdot\}_p$ is defined point wise as the Poisson bracket.  The differential equation implicitly specifies the form of the disturbance $\mathcal{D}_t$.

\emph{Correlation functions}.---Correlations between observables at different times can be obtained by inserting a time-evolution disturbance between the observable measurements,
\begin{align}\label{eq:ccorrfun}
  \mean{F(0)G(t)} &= \mean{F\mathcal{D}_t(G)}, \\
  &= \sum_{x\in X} P(x) f(x) \sum_{x'\in X} D_{x,t}(x') g(x'). \nonumber
\end{align}
Operationally this corresponds to measuring the observable $F$, waiting an interval of time $t$, then measuring the observable $G$.  Similarly, $n$-time correlations can be defined with $n-1$ time-evolution disturbances between the observable measurements $\mean{F_1\mathcal{D}_{t_1}(F_2\cdots \mathcal{D}_{t_{n-1}}(F_n)\cdots)}$.  Computing the correlation of an observable with itself at the same time will produce a higher moment $\mean{F^n}$.

\emph{Invasive measurement}.---A system may also be disturbed during the physical process that implements conditioning, which will alter the state above and beyond the pure conditioning expression \eqref{eq:ccollapse}.  With such an \emph{invasive measurement}, one conditions a state after a disturbance induced by the measurement process has occurred; hence, one obtains a new state,
\begin{align}\label{eq:cmdisturbpost}
  \dmean{F}_y &= \frac{\mean{\mathcal{D}(y\, F)}}{\mean{\mathcal{D}(y)}}, \\
  &= \frac{\sum_{x\in X} P(x) \sum_{x'\in X} D_x(y\, x') f(x')}{\sum_{x\in X} P(x)D_x(y)}, \nonumber
\end{align}
which is a composition of the measurement disturbance \eqref{eq:cdisturb} followed by the pure conditioning \eqref{eq:ccollapse}.  

As we shall see later in Sec. \ref{sec:qcollapse}, the quantum projection postulate (L\"{u}der's Rule) can be understood as an invasive measurement similar to \eqref{eq:cmdisturbpost}, but not as pure conditioning similar to \eqref{eq:ccollapse}.  This observation has also been recently emphasized by \citet{Leifer2011}, who show that a careful extension of \eqref{eq:ccollapse} to the noncommutative quantum setting does not reproduce the projection postulate.  Hence, better understanding classical invasive measurement should provide considerable insight into the quantum measurement process.  However, to properly understand the implications of invasive measurements on the measurement of observables, we must consider the measurement process in more detail.

\subsection{Detectors and probability observables} \label{sec:cdetector}
For a single ideal experiment that answers questions of interest with perfectly correlated independent outcomes, knowing the spectrum of an observable for that experiment is completely sufficient.  However, in many (if not most) cases the independent propositions corresponding to the experimental outcomes are only \emph{imperfectly correlated} with the questions of interest about the system.  Since in such a case one may not have direct access to the questions of interest, one also may not have direct access to the observables of interest.  One must instead \emph{infer} information about the observables of interest \emph{indirectly} from the correlated outcomes of the detector to which one does have access.

\emph{Joint sample space}.---To handle this case formally, we first enlarge the sample space to include both the sample space of interest, which we call the \emph{system}, $X$ and the accessible sample space, which we call the \emph{detector}, $Y$.  Questions about the system and the detector can be asked independently, so every question for the system can be paired with any question from the detector; therefore, the resulting joint sample space must be a product space, $XY = \{x y\,|\, x \in X, y\in Y\}$, where the products of propositions from different sample spaces commute.  The Boolean algebra $\Sigma_{XY}$ and observable algebra $\Sigma_{XY}^\mathbb{R}$ are constructed in the usual way from the joint sample space, and contain the algebras $\Sigma_X$, $\Sigma_Y$, $\Sigma_X^\mathbb{R}$, and $\Sigma_Y^\mathbb{R}$ as subalgebras.  When represented as operators on a Hilbert space, the corresponding joint representation exists within the tensor product of the system and detector space representations.

\emph{Product states}.---If the probabilities of the system propositions are uncorrelated with the probabilities of the detector propositions under a joint state $P$ on the joint sample space, then the joint state can be written as a \emph{composition} of independent states that are restricted to the sample spaces of the system and detector, $P = P_X\circ P_Y$.  Just as the state $P$ has a linear extension to $\mean{\cdot}$, its restrictions $P_X$ and $P_Y$ have linear extensions $\mean{\cdot}_X$ and $\mean{\cdot}_Y$, respectively.  Thus, for any joint observable $F$ an uncorrelated expectation has the form $\mean{F} = \mean{\mean{F}_Y}_X = \mean{\mean{F}_X}_Y$.  Such an uncorrelated joint state is known as a \emph{product state}.  The name stems from the fact that for a simple product $F_X F_Y$ of system and detector observables the corresponding joint expectation decouples into a product of system and detector expectations separately, $\mean{F_X F_Y} = \mean{F_X}_X \mean{F_Y}_Y$.  

Similarly, general measures on the joint sample space can be product measures.  A particularly useful example is the trace $\text{Tr} = \text{Tr}_X \circ \text{Tr}_Y$ on $XY$, which is composed of the \emph{partial traces}, $\text{Tr}_X$ and $\text{Tr}_Y$.  The trace serves as a convenient reference measure since it is a product measure for which any joint state has a corresponding density.  On continuous spaces the standard integral is also a product measure, $\mean{F} = \int_X [\int_Y f(x,y) p(x,y)\, dy]\,dx = \int_Y [\int_X f(x,y) p(x,y)\, dx]\, dy$, which tends to have noninfinitesimal densities.

\emph{Correlated states}.---In addition to product states, the joint space admits a much larger class of \emph{correlated} states where the detector and system questions are dependent on one another.  With such a correlated state a measurement on the detector cannot be decoupled in general from a measurement on the system.  Information gathered from a measurement on a detector under a correlated state will also indirectly provide information about the system, thus motivating the term ``detector.''

\emph{Reduced states}.---For a pure system observable $F_X$ or a pure detector observable $F_Y$, the average under a joint state will be equivalent to the average under a state restricted to either the system or the detector space, known as a \emph{reduced state}, or a \emph{marginalized state}.  We can define such a reduced state by using the joint state density under any reference \emph{product} measure $\mu = \mu_X \circ \mu_Y$, such as the trace $\text{Tr}$.  It then follows that,
\begin{subequations}
\begin{align}
  \mean{F_X} &= \mean{\mean{P_\mu}_{\mu_Y} F_X}_{\mu_X} = \mean{P_{\mu_X} F_X}_{\mu_X}, \\
  \mean{F_Y} &= \mean{\mean{P_\mu}_{\mu_X} F_Y}_{\mu_Y} = \mean{P_{\mu_Y} F_Y}_{\mu_Y}. 
\end{align}
\end{subequations}
The quantities $P_{\mu_X} = \mean{P_\mu}_{\mu_Y}$ and $P_{\mu_Y} = \mean{P_\mu}_{\mu_X}$ are the \emph{reduced state densities} that define the reduced states $P_X$ and $P_Y$ with expectation functionals, 
\begin{subequations}
\begin{align}
  \mean{F_X}_X &= \mean{P_{\mu_X} F_X}_{\mu_X}, \\
  \mean{F_Y}_Y &= \mean{P_{\mu_Y} F_Y}_{\mu_Y}.
\end{align}
\end{subequations}
By definition, $\mean{F_X} = \mean{F_X}_X$ and $\mean{F_Y} = \mean{F_Y}_Y$.  However, in general $\mean{F} \neq \mean{\mean{F}_Y}_X$, $\mean{F} \neq \mean{\mean{F}_X}_Y$, and $\mean{\mean{F}_Y}_X \neq \mean{\mean{F}_X}_Y$ unless $P$ is a product state.  The resulting reduced expectations $\mean{\cdot}_X$ and $\mean{\cdot}_Y$ are independent of the choice of reference product functional $\mu$.

\emph{Probability observables}.---Any correlation between the system and detector in the joint state allows us to directly relate propositions on the detector to \emph{observables} on the system.  We can compute the relationship directly by using a closure relation and rearranging the conditioning procedure \eqref{eq:ccollapse} to find,
\begin{align}
  \label{eq:cpovmprob}
  P(y) =& \sum_{x\in X} P(x)P(y|x) = \mean{\sum_{x\in X} P(y|x)\, x} = \mean{E_y}_X, \\
  \label{eq:cpovm}
  E_y =& \sum_{x\in X} P(y|x)\, x.
\end{align}
The resulting set of system observables $\{E_y\}$ exactly correspond to the detector outcomes $\{y\}$.  Analogously to a set of independent probability observables, they form a partition of the system identity, but are indexed by detector propositions rather than by system propositions, $\sum_{y\in Y} E_y = 1_X$.  Such a set $\{E_y\}$ has the common mathematical name \emph{positive operator-valued measure} (POVM) \cite{Nielsen2000}, since it forms a measure over the detector sample space $Y$ consisting of positive operators.  However, we shall make an effort to refer to them as general \emph{probability observables} to emphasize their physical significance.  As long as the detector outcomes are not mutually exclusive with the system, the probability observables \eqref{eq:cpovm} will be a faithful representation of the reduced state of the detector in the observable space of the system.  

\emph{Process tomography}.---The probability observables are completely specified by the \emph{conditional likelihoods} $P(y|x)$ for a detector proposition $y$ to be true given that a system proposition $x$ is true.  Such conditional likelihoods are more commonly known as \emph{response functions} for the detector and can be determined via independent \emph{detector characterization} using known reduced system states; such characterization is also known as \emph{detector tomography}, or \emph{process tomography}.  Any good detector will then maintain its characterization with any \emph{unknown} reduced system state.  That is, a noninvasive coupling of such a good detector to an unknown system produces a correlated joint state according to $P(xy) = P_X(x)P(y|x)$, where $P_X$ is the unknown reduced system state prior to the interaction with the detector.  

\emph{Generalized state collapse}.---In addition to allowing the computation of detector probabilities, $P(y) = \mean{E_y}_X$, probability observables also have the dual role of updating the reduced system state following a measurement on the detector.  To see this, we apply the general rule for state collapse \eqref{eq:ccollapse} for a detector proposition $y$ on the joint state to find,
\begin{align}
  \label{eq:cgcollapse}
  \mean{F_X}_y &= \frac{\mean{y\, F_X}}{P(y)} = \sum_{x\in X} f_X(x) P(y|x)\frac{P_X(x)}{P(y)}, \\
  &= \frac{\mean{E_y F_X}_X}{\mean{E_y}_X}, \nonumber
\end{align}
which can be seen as a generalization of the Bayesian conditioning rule \eqref{eq:ccollapse} to account for the effect of an imperfectly correlated detector, and can also be understood as a form of \emph{Jeffrey's conditioning} \cite{Jeffrey1965}.  For this reason, probability observables are commonly called \emph{effects} of the \emph{generalized measurement}.  A reduced state density $P_{\mu_X}$ for the system updates as $P_{\mu_X|y} = P_{\mu_X}\, E_y / \mean{E_y}_X$.  Such a \emph{generalized measurement} is nonprojective, so is not constrained to the disjoint questions on the sample space of the system.  As a result, it answers questions on the system space \emph{ambiguously} or \emph{noisily}.

\emph{Weak measurement}.---The extreme case of such an ambiguous measurement is a \emph{weak measurement}, which is a measurement that does not (appreciably) collapse the system state.  Such a measurement is inherently ambiguous to the extent that only a minuscule amount of information is learned about the system with each detection.  Formally, the probability observables for a weak measurement are all nearly proportional to the identity on the system space.  Typically, an experimenter has access to some control parameter $\epsilon$ (such as the correlation strength) that can alter the weakness of the measurement such that,
\begin{align}\label{eq:cweak}
  \forall y, \lim_{\epsilon\to 0} E_y(\epsilon) = P_Y(y) 1_X,
\end{align}
where $P_Y(y) \in (0,1)$ is the nonzero probability of obtaining the detector outcome $y$ in the absence of any interaction with the system.  Then for small values of $\epsilon$ the measurement leaves the system state nearly unperturbed, $P_{\mu_X|y} = P_{\mu_X}\, E_y(\epsilon) / \mean{E_y(\epsilon)}_X \approx P_{\mu_X}$.  The limit as such a control parameter $\epsilon\to 0$ is known as the \emph{weak measurement limit} and is a formal idealization not strictly achievable in an experiment.

\emph{Strong measurement}.---The opposite extreme case is a \emph{strong measurement} or projective measurement, which is a measurement for which all outcomes are independent, as in \eqref{eq:cobs}.  In other words, the probability observables are independent for a strong measurement.  The projective collapse rule \eqref{eq:ccollapse} can therefore be seen as a special case of the general collapse rule \eqref{eq:cgcollapse} from this point of view.

\emph{Measurement sequences}.---A further benefit of the probability observable representation of a detector is that it becomes straightforward to discuss sequences of generalized measurements performed on the same system.  For example, consider two detectors that successively couple to a system and have the outcomes $y$ and $z$ measured, respectively.  To describe the full joint state of the system and both detectors requires a considerably enlarged sample space.  However, if the detectors are characterized by two sets of probability observables $\{E_y\}$ and $\{E'_z\}$ we can immediately write down the probability of both outcomes to occur as well as the resulting final collapsed system state without using the enlarged sample space,
\begin{subequations}
\begin{align}\label{eq:cgseqcollapse}
  P(yz) &= \mean{E'_z E_y}_X, \\
  \mean{F_X}_{yz} &= \frac{\mean{E'_z E_y F_X}_X}{\mean{E'_z E_y}_X}.
\end{align}
\end{subequations}
Similarly, a conditioned density takes the form $P_{\mu_X|yz} = P_{\mu_X}\, E'_z E_y / \mean{E'_z E_y}_X$.  The detectors have been \emph{abstracted} away to leave only their effect upon the system of interest. 

\emph{Generalized invasive measurement}.---The preceding discussion holds provided that the detector can be noninvasively coupled to a reduced system state $P_X$ to produce a joint state $P(xy) = P_X(x)P(y|x)$.  However, more generally the process of coupling a reduced detector state $P_Y$ to the reduced system state $P_X$ will \emph{disturb} both states as discussed for \eqref{eq:cdisturb}.  The disturbance produces a joint state from the original product state of the system and detector according to,
\begin{align}\label{eq:cdisturbcouple}
  \dmean{xy} &= \mean{\mean{\mathcal{D}(xy)}_Y}_X, \\
  \mathcal{D}(xy) &= \sum_{x'\in X}\sum_{y'\in Y} D_{x',y'}(xy)\, x'y',
\end{align}
where $D_{x',y'}$ are states specifying the joint transition probabilities for the disturbance.  The noninvasive coupling $P(xy) = P_X(x)P(y|x)$ is a special case of this where the reduced system state is unchanged by the coupling.

As a result, we must slightly modify the derivation of the probability observables \eqref{eq:cpovmprob} to properly include the disturbance,
\begin{subequations}\label{eq:cpovmdisturb}
\begin{align}
  \dmean{y} &= \mean{\mean{\mathcal{D}(y)}_Y}_X = \mean{\tilde{E}_y}_X, \\
  \tilde{E}_y &= \mean{\mathcal{D}(y)}_Y, \\
  &= \sum_{x\in X}\sum_{y'\in Y} P_Y(y') D_{x,y'}(y) \, x. \nonumber
\end{align}
\end{subequations}
The modified probability observable $\tilde{E}_y$ includes both the initial detector state $P_Y$ and the disturbance from the measurement.  Detector tomography will therefore find the effective characterization probabilities $\tilde{P}(y|x) = \sum_{y'\in Y} D_{x,y'}(y) P_Y(y')$.

The generalized collapse rule similarly must be modified to include the disturbance,
\begin{align}
  \label{eq:cgcollapsedisturb}
  \dmean{F_X}_y &= \frac{\mean{\mean{\mathcal{D}(y\, F_X)}_Y}_X}{\mean{\mean{\mathcal{D}(y)}_Y}_X} = \frac{\mean{\mathcal{E}_y(F_X)}_X}{\mean{\tilde{E}_y}_X}, \\
  \label{eq:cgoperations}
  \mathcal{E}_y(F_X) &= \mean{\mathcal{D}(y\, F_X)}_Y, \\
  &= \sum_{x'\in X} x' \sum_{y'\in Y} P_Y(y') \sum_{x\in X} D_{x',y'}(y\, x) f(x). \nonumber 
\end{align}
Surprisingly, we can no longer write the conditioning in terms of just the probability observables $\tilde{E}_y$; instead we must use an \emph{operation} $\mathcal{E}_y$ that takes into account both the coupling of the detector and the disturbance of the measurement in an active way.  The measurement operation is related to the effective probability observable according to, $\mathcal{E}_y(1_X) = \tilde{E}_y$.

The change from observables to operations when the disturbance is included becomes particularly important for a sequence of invasive measurements.  Consider an initial system state $P_X$ that is first coupled to a detector state $P_Y$ via a disturbance $\mathcal{D}_1$, then conditioned on the detector proposition $y$, then coupled to a second detector state $P_Z$ via a disturbance $\mathcal{D}_2$, and finally conditioned on the detector proposition $z$.  The joint probability for obtaining the ordered sequence $(y,z)$ can be written as
\begin{align}\label{eq:cgseqcollapsedisturb}
  \mean{\mean{\mathcal{D}_1(y\, \mean{\mathcal{D}_2(z)}_Z)}_Y}_X &= \mean{\mathcal{E}_y(\tilde{E}'_z)}_X.
\end{align}
The effective probability observable $\mathcal{E}_y(\mathcal{E}'_z(1_X)) = \mathcal{E}_y(\tilde{E}'_z)$ for the ordered measurement sequence $(y,z)$ is no longer a simple product of the probability observables $\tilde{E}_y$ and $\tilde{E}'_z$ as in \eqref{eq:cgseqcollapse}, but is instead an ordered \emph{composition of operations}.

The ordering of operations also leads to a new form of \emph{postselected} conditioning.  Specifically, if we condition only on the second measurement of $z$ in an invasive sequence $(y,z)$, we obtain,
\begin{align}\label{eq:cgpostselection}
  \cmean{z}{y} &= \frac{\mean{\mathcal{E}_y(\tilde{E}'_z)}_X}{\sum_{y'\in Y} \mean{\mathcal{E}_{y'}(\tilde{E}'_z)}_X} = \frac{\mean{\mathcal{E}_y(\tilde{E}'_z)}_X}{\mean{\mathcal{E}(\tilde{E}'_z)}_X}, \\
  \mathcal{E}(\tilde{E}'_z) &= \sum_{y'\in Y} \mathcal{E}_{y'}(\tilde{E}'_z) = \mean{\mathcal{D}(\tilde{E}'_z)}_Y.
\end{align}
The different position of the subscript serves to distinguish the postselected probability $\cmean{z}{y}$ from the preselected probability $\dmean{y}_z = \mean{\mathcal{E}'_z(\tilde{E}_y)}_X / \mean{\tilde{E}'_z}_X$ corresponding to the reverse measurement ordering of $(z,y)$.  The operation $\mathcal{E}$ appearing in the denominator is called a \emph{nonselective measurement} since it includes the disturbance induced by the measurement coupling, but does not condition on any particular detector outcome.  When the disturbance to the reduced system state vanishes, the conditioning becomes order-independent and both types of conditional probability reduce to $P(y|z) = \mean{E_y E'_z}_X / \mean{E'_z}_X$.   

The two forms of conditioning for invasive measurements in turn lead to a modified form of Bayes' rule that relates the preselected conditioning of a sequence to the postselected conditioning of the same sequence,
\begin{align}\label{eq:cgbayesdisturb}
  \cmean{z}{y} &= \dmean{z}_y \frac{\mean{E_y}_X}{\mean{\mathcal{E}(\tilde{E}'_z)}_X}.
\end{align}
When the disturbance to the reduced system state vanishes, the nonselective measurement $\mathcal{E}$ reduces to the identity operation, $\cmean{z}{y}$ reduces to $P(y|z)$, $\dmean{z}_y$ reduces to $P(z|y)$, and we correctly recover the noninvasive Bayes' rule \eqref{eq:cbayes}. 

\subsection{Contextual values} \label{sec:ccv}
\emph{Observable correspondence}.---With the preliminaries about generalized state conditioning out of the way, we are now in a position to discuss the measurement of observables in more detail.  First we observe an important corollary of the observable representation of the detector probabilities $P(y) = \mean{E_y}_X$ from \eqref{eq:cpovmprob}:  \emph{detector} observables can be mapped into equivalent \emph{system} observables,
\begin{align}
  \mean{F_Y} &= \sum_{y\in Y} f_Y(y) P(y) = \mean{F_X}_X, \\
  \label{eq:ccgrain}
  F_X &= \sum_{y\in Y} f_Y(y) E_y. 
\end{align}
Note that the eigenvalues $f_X(x) = \sum_{y\in Y} f_Y(y) P(y|x)$ of the equivalent system observable $F_X$ are not the same as the eigenvalues $f_Y(y)$ of the original detector observable $F_Y$, but are instead their average under the detector response.  If the system propositions were accessible then the system observable $F_X$ would allow nontrivial inference about the detector observable $F_Y$, provided that the probability observables were nonzero for all $y$ in the support of $F_Y$.  

\emph{Contextual values}.---A more useful corollary of the expansion \eqref{eq:ccgrain} is that any \emph{system} observable that can be expressed as a combination of probability observables may be equivalently expressed as a \emph{detector} observable,
\begin{align}\label{eq:cpovmexpand}
  F_X &= \sum_{y\in Y} f_Y(y)\, E_y \implies F_Y = \sum_{y\in Y} f_Y(y)\, y,
\end{align}
which is the classical form of our main result.  Using this equivalence, \emph{we can indirectly measure such system observables using only the detector}.  We dub the eigenvalues of the detector observable $f_Y(y)$ the \textbf{contextual values} (CVs) of the system observable $F_X$ under the \emph{context of the specific detector} characterized by a specific set of probability observables $\{E_y\}$.  The CVs form a \emph{generalized spectrum} for the observable since they are associated with general probability observables for a generalized measurement and not independent probability observables for a projective measurement; the eigenvalues are a special case when the probability observables are the spectral projections of the observable being measured.  

With this point of view, we can understand an observable as an \emph{equivalence class} of possible measurement strategies for the same average information.  That is, using appropriate pairings of probability observables and CVs, one can measure the same observable average in many different ways, $\mean{F_X} = \sum_{x\in X} f_X(x) P(x) = \sum_{y\in Y} f_Y(y) \mean{E_y}_X$.  Each such expansion corresponds to a different experimental setup.

\emph{Moments}.---Similarly, the $n$\textsuperscript{th} statistical moment of an observable can be measured in many different, yet equivalent, ways.  For instance, the $n$\textsuperscript{th} moment of an observable $F_X$ can be found from the expansion \eqref{eq:cpovmexpand} as,
\begin{align}\label{eq:cnthmom}
  \mean{(F_X)^n} &= \mean{ (\sum_{y\in Y} f_Y(y)E_y)^n }_X, \\
  &= \sum_{y_1,\ldots,y_n\in Y} f_Y(y_1)\cdots f_Y(y_n)\mean{E_{y_1}\cdots E_{y_n}}_X. \nonumber
\end{align}
By examining the general collapse rule for measurement sequences \eqref{eq:cgseqcollapse} we observe that the quantity $\mean{E_{y_1} \cdots E_{y_n}}_X$ must be the joint probability for a sequence $(y_1,\cdots,y_n)$ of $n$ \emph{noninvasive} measurements that couple the same detector to the system $n$ times in succession.  Furthermore, the average in \eqref{eq:cnthmom} is explicitly different from the $n$\textsuperscript{th} statistical moment of the raw detector results, $\mean{(F_Y)^n} = \sum_{y\in Y} (f_Y(y))^n P(y)$.

We conclude that, \emph{for imperfectly correlated noninvasive detectors, one must perform measurement sequences to obtain the correct statistical moments of an observable using a particular set of CVs}.  Only for unambiguous measurements with independent probability observables do such measurement sequences reduce to simple powers of the eigenvalues being averaged with single measurement probabilities.  If a single measurement by the detector is done per trial, then only the statistical moments of the \emph{detector} observable $F_Y$ can be inferred from that set of CVs, as opposed to the true statistical moments of the inferred system observable $F_X$.

We can, however, change the CVs to define new observables that correspond to powers of the original observable, such as $G_X = (F_X)^n = \sum_{y\in Y} g_Y(y) E_y$.  These new observables can then be measured indirectly using the same experimental setup without the need for measurement sequences.  The CVs $g_Y(y)$ for the $n$\textsuperscript{th} power of $F_X$ will not be simple powers of the CVs $f_Y(y)$ for $F_X$ unless the measurement is unambiguous.

\emph{Invasive measurements}.---If the measurement is invasive, then the disturbance forces us to associate the CVs with the measurement \emph{operations} $\{\mathcal{E}_y\}$ and not solely with their associated probability operators $\{\tilde{E}_y\}$ in order to properly handle measurement sequences as in \eqref{eq:cgoperations}.  Specifically, we must define the \emph{observable operation},
\begin{align}\label{eq:coperexpand}
  \mathcal{F}_X &= \sum_{y\in Y} f_Y(y) \mathcal{E}_y,
\end{align}
which produces the identity $\mathcal{F}_X(1_X) = \sum_{y\in Y} f_Y(y) \tilde{E}_y = F_X$ similar to \eqref{eq:cpovmexpand}.  

Correlated sequences of invasive observable measurements can be obtained by composing the observable operations,
\begin{align}\label{eq:coperseq}
  \mean{ (\mathcal{F}_X)^n(1_X) }_X &= \sum_{y_1,\ldots ,y_n}f_Y(y_1)\cdots f_Y(y_n) \times \nonumber \\
  &\qquad \mean{\mathcal{E}_{y_1}(\mathcal{E}_{y_2}(\cdots (\tilde{E}_{y_n})\cdots))}_X.
\end{align}
Such an $n$-measurement sequence reduces to the $n$\textsuperscript{th} moment \eqref{eq:cnthmom} when the disturbance vanishes.  

If time evolution disturbance $\mathcal{D}_t$ is inserted between different invasive observable measurements, then we obtain an invasive \emph{correlation function} instead,
\begin{align}\label{eq:copercorrfun}
  \dmean{F_X(0)G_X(t)} &= \mean{\mathcal{F}_X(\mathcal{D}_t(\mathcal{G}_X(1_X)))}_X.
\end{align}
When the observable measurements become noninvasive, then this correctly reduces to the noninvasive correlation function \eqref{eq:ccorrfun}.  Similarly, $n$-time invasive correlations can be defined with $n-1$ time-evolution disturbances between the invasive observable measurements $\mean{\mathcal{F}_1(\mathcal{D}_{t_1}(\mathcal{F}_2(\cdots \mathcal{D}_{t_{n-1}}(\mathcal{F}_n(1_X))\cdots)))}$.  

\emph{Conditioned averages}.---In addition to statistical moments of the observable, we can also use the CVs to construct principled \emph{conditioned averages} of the observable.  Recall that in the general case of an invasive measurement sequence we can condition the observable measurement in two distinct ways.  If we condition on an outcome $z$ before the measurement of $F_X$ we obtain the \emph{preselected conditioned average} $\dmean{F_X}_z$ defined in \eqref{eq:cgcollapsedisturb}.  On the other hand, if the invasive conditioning measurement of $z$ happens after the invasive observable measurement then we must use the postselected conditional probabilities \eqref{eq:cgpostselection} to construct a \emph{postselected conditioned average},
\begin{align}\label{eq:ccondav}
  \cmean{z}{F_X} &= \sum_{y\in Y} f_Y(y)\, \cmean{z}{y}, \\
  &= \frac{\sum_{y\in Y} f_Y(y) \mean{\mathcal{E}_y(\tilde{E}'_z)}_X}{\sum_{y\in Y} \mean{\mathcal{E}_y(\tilde{E}'_z)}_X} = \frac{\mean{\mathcal{F}_X(\tilde{E}'_z)}_X}{\mean{\mathcal{E}(\tilde{E}'_z)}}. \nonumber
\end{align}
The observable operation $\mathcal{F}_X$ and the nonselective measurement $\mathcal{E}$ encode the relevant details from the first measurement.  When the disturbance to the reduced system state vanishes, both the preselected and the postselected conditioned averages simplify to the pure conditioned average $\mean{F_X}_z$ defined in \eqref{eq:cgcollapse} that depends only on the system observable $F_X$.

While the pure conditioned average $\mean{F_X}_z$ is independent of the order of conditioning and is always constrained to the eigenvalue range of the observable, the postselected invasive conditioned average $\cmean{z}{F_X}$ can, perhaps surprisingly, stray outside the eigenvalue range with ambiguous measurements.  The combination of the amplified CVs and the disturbance can lead to a postselected average that lies anywhere in the full CV range, rather than just the eigenvalue range.  We will see an example of this in Sec. \ref{sec:cmarbledisturb}.

\emph{Inversion}.---So far we have treated the CVs in the expansion \eqref{eq:cpovmexpand} as known quantities.  However, for a realistic detector situation, the CVs will need to be experimentally determined from the characterization of the detector and the observable that one wishes to measure.  The reduced system state $P_X$ will generally not be known \emph{a priori}, since the point of a detector is to learn information about the system in the \emph{absence} of such prior knowledge.  We can still solve for the CVs without knowledge of the system state, however, since the probability observables are only specified by the conditional likelihoods $P(y|x)$ that can be obtained independently from detector tomography.  

To solve for the CVs when the system state is presumed unknown, we rewrite \eqref{eq:cpovmexpand} in the form,
\begin{align}
  \label{eq:cpovmmap}
  F_X &= \sum_{x\in X} x\, \sum_{y\in Y}P(y|x)f_Y(y), \\
  &= \sum_{x\in X} x\, \mean{F_Y}_x = \mathcal{S}(F_Y), \nonumber
\end{align}
where $\mathcal{S} = \sum_x x\, \mean{\cdot}_x$ is the map that converts observables in the detector space to observables in the system space $\mathcal{S} : \Sigma_Y^\mathbb{R} \to \Sigma_X^\mathbb{R}$.  Our goal is to invert this map and solve for the required spectrum of $F_Y$ given a desired system observable $F_X$.  However, the inverse of such a map is not generally unique; for it to be uniquely invertible it must be one-to-one between system and detector spaces of equal size.  If the detector space is smaller than the system, then no exact inverse solutions are possible; it may be possible, however, to find course-grained solutions that lose some information.  Perhaps more alarmingly, if the detector space is larger than the system, then it is possible to have an infinite set of exact solutions.

When disturbance is taken into account as in \eqref{eq:cpovmdisturb}, the equality \eqref{eq:cpovmmap} becomes,
\begin{align}\label{eq:cpovmmapdisturb}
  F_X &= \mean{\mathcal{D}(F_Y)}_Y = \mathcal{S}(F_Y),
\end{align}
so the composition of the disturbance $\mathcal{D}$ and the detector expectation $\mean{\cdot}_Y$ produces the map $\mathcal{S}$ that must be inverted.  Equation \eqref{eq:cpovmmap} is a special case when the reduced system state is unchanged by the coupling disturbance.

\emph{Pseudoinversion}.---The entire set of possible solutions to \eqref{eq:cpovmmapdisturb} may be completely specified using the \emph{Moore-Penrose pseudoinverse} of the map $\mathcal{S}$, which we denote as $\mathcal{S}^+$.  The pseudoinverse is the inverse of the restriction of $\mathcal{S}$ to the space $\Sigma_Y^\mathbb{R} \setminus \{F \in \Sigma_Y^\mathbb{R}\, |\, \mathcal{S}(F) = 0 \}$; that is, the null space of $\mathcal{S}$ is removed from the detector space before constructing the inverse.  We will show a practical method for computing the pseudoinverse using the singular value decomposition in the examples to follow.  

Using the pseudoinverse, all possible solutions of \eqref{eq:cpovmmapdisturb} can be written compactly as,
\begin{align}\label{eq:pseudoinversion}
  F_Y = \mathcal{S}^+(F_X) + (\mathcal{I} - \mathcal{S}^+\mathcal{S})(G),
\end{align}
where $\mathcal{I}$ is the identity map and $G \in \Sigma_Y^\mathbb{R}$ is an arbitrary detector observable.  The solutions specified by the pseudoinverse in this manner contain exact inverses and course-grainings as special cases.  

\emph{Detector variance}.---Since $(\mathcal{I} - \mathcal{S}^+\mathcal{S})$ is a projection operation to the null space of $\mathcal{S}$, the second term of \eqref{eq:pseudoinversion} lives in the null space of $\mathcal{S}$ and is orthogonal to the first term.  Therefore, the norm squared of $F_Y$ has the form,
\begin{align}\label{eq:pseudonorm}
  ||F_Y||^2 &= \sum_y (f_Y(y))^2, \\
  &= ||\mathcal{S}^+(F_X)||^2 + ||(\mathcal{I}-\mathcal{S}^+\mathcal{S})(G)||^2, \nonumber
\end{align}
making the $G=0$ solution have the smallest norm.  

The norm $||F_Y||$ of the CV solution is relevant because the second moment of the detector observable $F_Y$ is simply bounded by the norm squared $\mean{ (F_Y)^2} = \sum_y P(y) (f_Y(y))^2 \leq ||F_Y||^2$.  The second moment is similarly an upper bound for the variance of the detector observable $\text{Var}(F_Y) = \mean{ (F_Y)^2} - (\mean{F_Y})^2 \leq \mean{(F_Y)^2}$.  Therefore, the norm squared is a reasonable upper bound for the detector variance that one can make without prior knowledge of the state.  

\emph{Mean-squared error}.---The variance of $F_Y$ governs the mean-squared error of any estimation of its average with a finite sample, such as an empirically measured sample in a laboratory.  Specifically, one measures a sequence of detector outcomes of length $n$, $(y_1, y_2, \dots, y_n)$, and uses this finite sequence to estimate the average of $F_Y$ via the \emph{unbiased estimator},
\begin{align}\label{eq:meanestimator}
  \overline{F_Y} &= \frac{1}{n}\sum_i^n f_Y(y_i),
\end{align}
that converges to the true mean value $\mean{F_Y}_Y = \mean{F_X}$ as $n\to\infty$.  The mean squared error of this estimator $\text{MSE}(\overline{F_Y})$ from the true mean is the variance over the number of trials in the sequence $\text{Var}(F_Y) / n$.  Hence, the maximum mean squared error for a finite sequence of length $n$ must be bounded by the norm squared of the CVs divided by length of the sequence,
\begin{align}\label{eq:mse}
  \text{MSE}(\overline{F_Y}) = \frac{\text{Var}(F_Y)}{n} \leq \frac{||F_Y||^2}{n}.
\end{align}
That is, the norm bounds the number of trials necessary to obtain an experimental estimation of observable averages to a desired precision using the imperfect detector.  
  
\emph{Pseudoinverse prescription}.---Choosing the arbitrary observable to be $G=0$ therefore not only picks the solution $F_Y = \mathcal{S}^+ (F_X)$ that is uniquely related to $F_X$ by discarding the irrelevant null space of $\mathcal{S}$, but also picks the solution with the smallest norm, which places a reasonable upper bound on the statistical error.  Without prior knowledge of the system state, the pseudoinverse solution does a reasonable job at obtaining an optimal fit to the relation \eqref{eq:cpovmmapdisturb}.  Moreover, when \eqref{eq:cpovmmapdisturb} is not satisfied by the direct pseudoinverse then an exact solution is impossible, but the pseudoinverse still gives the ``best fit'' coursegraining of an exact solution in the least-squares sense.  As such, we consider the direct pseudoinverse of $F_X$ to be the preferred solution in the absence of other motivating factors stemming from prior knowledge of the state being measured.  

\subsubsection{Example: Ambiguous marble detector}\label{sec:cmarble}
As an illustrative example similar to the one given in the introduction, suppose that one wishes to know whether the color of a marble is green or red, but one is unable to examine the marble directly.  Instead, one only has a machine that can display a blue light or a yellow light after it examines the marble color.  In such a case, the marble colors are the propositions of interest, but the machine lights are the only accessible propositions.  The lights may be correlated imperfectly with the marble color; for instance, if a blue light is displayed one may learn something about the possible marble color, but it may still be partially \emph{ambiguous} whether the marble is actually green or actually red.  

The relevant Boolean algebra for the system is $\Sigma_X = \{0, g, r, 1_X\}$, where $g$ is the proposition for the color green, $r$ is the proposition for the color red, and $1_X = g + r$ is the logical \emph{or} of the two possible color propositions.  We consider the task of measuring a simple color observable $F_X = (+1)g + (-1)r$ that distinguishes the colors with a sign using an imperfectly correlated detector.

The relevant Boolean algebra for the detector is $\Sigma_Y = \{0, b, y, 1_Y\}$, where $b$ is the proposition for the blue light, $y$ is the proposition for the yellow light, and $1_Y = b + y$.  In order to measure the marble observable $F_X$ using only the detector, the experimenter must determine the proper form of the corresponding detector observable $F_Y$.

First, the experimenter characterizes the detector by sending in known samples and observing the outputs of the detector.  After many characterization trials, the experimenter determines to some acceptable precision the four conditional probabilities,
\begin{subequations}\label{eq:marblecharacter}
\begin{align}
  P(b|g) &= 0.6, & P(y|g) &= 0.4, \\
  P(b|r) &= 0.2, & P(y|r) &= 0.8,
\end{align}
\end{subequations}
for the detector outcomes $b$ and $y$ given specific marble preparations $g$ and $r$.  These characterization probabilities completely determine the detector response in the form of its \emph{probability observables} \eqref{eq:cpovm},
\begin{subequations}\label{eq:marblepovm}
\begin{align}
  E_b &= P(b|g) g + P(b|r) r, \\
  E_y &= P(y|g) g + P(y|r) r.
\end{align}
\end{subequations}
By construction, $E_b + E_y = g + r = 1_X$.

Second, the experimenter expands the system observable $F_X$ using the detector probability observables \eqref{eq:marblepovm} and unknown \emph{contextual values} (CVs) $f_Y(b)$ and $f_Y(y)$ \eqref{eq:cpovmexpand},
\begin{align}
  F_X &= (+1)g + (-1)r = f_Y(b) E_b + f_Y(y) E_y. 
\end{align}
After expressing this relation as the equivalent matrix equation,
\begin{align}\label{eq:marblematrix}
  \begin{pmatrix}+1 \\ -1\end{pmatrix} &= \begin{pmatrix}P(b|g) & P(y|g) \\ P(b|r) & P(y|r)\end{pmatrix} \begin{pmatrix}f_Y(b) \\ f_Y(y)\end{pmatrix},
\end{align}
it can be directly inverted to find the CVs \eqref{eq:pseudoinversion},
\begin{align}\label{eq:marblecv}
  f_Y(b) &= 3, & f_Y(y) &= -2.
\end{align}
Therefore,
\begin{align}
  F_X &= (+1)g + (-1)r = (3) E_b + (-2) E_y,
\end{align}
so $F_X$ can be inferred from a measurement of the equivalent detector observable $F_Y = (3)b + (-2)y$.

Notably, the CVs \eqref{eq:marblecv} are amplified from the eigenvalues of $\pm 1$ due to the \emph{ambiguity} of the detector.  The amplification compensates for the ambiguity so that the correct \emph{average} can be obtained after measuring an ensemble of many unknown marbles described by the initial marble state $P_X$.  The amplification also leads to a larger upper bound for the variance \eqref{eq:pseudonorm} of the detector,
\begin{align}
  ||F_Y||^2 &= 13.
\end{align}
Hence, we can expect the imperfect detector to display a root-mean-squared (RMS) error \eqref{eq:mse} in the reported average color that is no larger than $\sqrt{13/n} \approx 3.6/\sqrt{n}$ after $n$ repeated measurements.  For contrast, a perfect detector would display an RMS error no larger than $\sqrt{2/n} \approx 1.4/\sqrt{n}$ after $n$ repeated measurements.

\subsubsection{Example: Invasive ambiguous detector} \label{sec:cmarbledisturb}
The detector apparatus in the last example could be generally \emph{invasive}.  In such a case, the characterization probabilities \eqref{eq:marblecharacter} composing the probability observables \eqref{eq:marblepovm} would be a combination of the initial state of the detector lights $P_Y$ and a \emph{disturbance} $\mathcal{D}$ from the measurement coupling according to \eqref{eq:cpovmdisturb},
\begin{subequations}
\begin{align}
  \tilde{P}(b|g) &= P_Y(b)(D_{g,b}(g\, b) + D_{g,b}(r\, b)) \\
  &\quad + P_Y(y)(D_{g,y}(g\, b) + D_{g,y}(r\, b)), \nonumber \\
  \tilde{P}(y|g) &= P_Y(b)(D_{g,b}(g\, y) + D_{g,b}(r\, y)) \\
  &\quad + P_Y(y)(D_{g,y}(g\, y) + D_{g,y}(r\, y)), \nonumber \\
  \tilde{P}(b|r) &= P_Y(b)(D_{r,b}(g\, b) + D_{r,b}(r\, b)) \\
  &\quad + P_Y(y)(D_{r,y}(g\, b) + D_{r,y}(r\, b)), \nonumber \\
  \tilde{P}(y|r) &= P_Y(b)(D_{r,b}(g\, y) + D_{r,b}(r\, y)) \\
  &\quad + P_Y(y)(D_{r,y}(g\, y) + D_{r,y}(r\, y)), \nonumber
\end{align}
\end{subequations}
where we have used the marginalization identity $D_{c,d}(b) = D_{c,d}(g\, b) + D_{c,d}(r\, b)$ for $c\in\{g,r\}$ and $d\in\{b,y\}$.  For a noninvasive detector, the transition probabilities that involve marbles changing color must be zero $D_{g,b}(r\, b) = D_{g,b}(r\,y) = D_{g,y}(r\,y) = D_{g,y}(r\,b) = D_{r,b}(g\,b) = D_{r,b}(g\,y) = D_{r,y}(g\,b) = D_{r,y}(g\,y) = 0$.  However, they need not be zero for a general invasive detector.

As an example, suppose that the initial detector state is unbiased, $P_Y(b) = P_Y(y) = 1/2$, and that the detector has a 10\% chance of flipping the color of a given marble.   The following possible values for the sixteen transition probabilities would then lead to the same effective characterization probabilities \eqref{eq:marblecharacter} as before,
\begin{subequations}\label{eq:marbledisturb}
\begin{align}
D_{g,b}(g\,b) &= 0.5 & D_{g,y}(g\,b) &= 0.5, \\
D_{g,b}(g\,y) &= 0.3 & D_{g,y}(g\,y) &= 0.3, \\
D_{r,b}(r\,b) &= 0.1 & D_{r,y}(r\,b) &= 0.1, \\
D_{r,b}(r\,y) &= 0.7 & D_{r,y}(r\,y) &= 0.7, \\
D_{g,b}(r\,b) &= 0.1 & D_{g,y}(r\,b) &= 0.1, \\
D_{g,b}(r\,y) &= 0.1 & D_{g,y}(r\,y) &= 0.1, \\
D_{r,b}(g\,b) &= 0.1 & D_{r,y}(g\,b) &= 0.1, \\
D_{r,b}(g\,y) &= 0.1 & D_{r,y}(g\,y) &= 0.1.
\end{align}
\end{subequations}
Since the effective characterization probabilities are the same, the probability observables are the same as \eqref{eq:marblepovm}, leading to the same CVs as \eqref{eq:marblecv} to measure the observable $F_X = (+1)g + (-1)r$.

The disturbance of the reduced marble state will become apparent only when making a second measurement after the first one.  Suppose we make a second measurement of the marble colors $g$ and $r$ directly.  The probability of obtaining a detector outcome $d\in\{b,y\}$ and then observing a specific marble color $c\in\{g,r\}$ will then be $P_X(g)(P_Y(b)D_{g,b}(c\, d) + P_Y(y)D_{g,y}(c\, d)) + P_X(r)(P_Y(b)D_{r,b}(c\, d) + P_Y(y)D_{r,y}(c\, d))$.  If we define an \emph{operation} as in \eqref{eq:cgoperations} to be,
\begin{align}
  \mathcal{E}_d(c) &= \mean{\mathcal{D}(c\, d)}_Y, \\
  &= g\, (P_Y(b)D_{g,b}(c\, d) + P_Y(y)D_{g,y}(c\, d)) \nonumber \\
  &\quad + r\, (P_Y(b)D_{r,b}(c\, d) + P_Y(y)D_{r,y}(c\, d)), \nonumber
\end{align}
then we can express the probability for the sequence compactly as $\mean{\mathcal{E}_d(c)}_X$.

Averaging the outcomes for the detector lights using the CVs \eqref{eq:marblecv} and then conditioning on a particular marble color $c$ in the second measurement produces a \emph{postselected conditioned average} of the marble colors \eqref{eq:ccondav} as reported by the invasive ambiguous detector,
\begin{align}\label{eq:marblecondav}
  \cmean{c}{F_X} &= \frac{f_Y(b)\, \mean{\mathcal{E}_b(c)}_X + f_Y(y)\, \mean{\mathcal{E}_y(c)}_X}{\mean{\mathcal{E}_b(c)}_X + \mean{\mathcal{E}_y(c)}_X}. 
\end{align} 

If we also preselect the marbles to be a particular color, we can compute the pre- and postselected conditioned averages of the marble colors as reported by the invasive ambiguous detector from \eqref{eq:marblecv}, \eqref{eq:marbledisturb}, and \eqref{eq:marblecondav},
\begin{subequations}\label{eq:marblecondavs}
\begin{align}
  \ccmean{g}{F_X}{g} &= 1.125, \\
  \ccmean{r}{F_X}{g} &= 0.5, \\
  \ccmean{g}{F_X}{r} &= 0.5, \\
  \ccmean{r}{F_X}{r} &= -1.375. 
\end{align}
\end{subequations}
Due to a combination of the invasiveness and the ambiguity of the measurement, the postselected conditioned averages can stray outside the eigenvalue range $[-1,1]$ for the observable $F_X$.  However, they remain within the CV range $[-2,3]$.  When the measurement is noninvasive, then the pre- and postselected conditioned averages in \eqref{eq:marblecondavs} that remain well-defined reduce to the pure conditioned averages $\mean{F_X}_g = 1$ and $\mean{F_X}_r = -1$.

\subsubsection{Example: Redundant ambiguous detector}
Consider a similar marble detection setup to the previous examples, but where the detector apparatus has \emph{three} independent outcome lights: blue, yellow, and purple.  Hence, the detector Boolean algebra is $\Sigma_Y = \{0, b, y, p, b+y, b+p, y+p, 1_Y\}$, where $p$ is the new proposition for the purple light, and $1_Y = b + y + p$.  After characterizing the detector the experimenter finds the conditional probabilities,
\begin{subequations}
\begin{align}
  P(b|g) &= 0.5, & P(y|g) &= 0.3, & P(p|g) &= 0.2, \\
  P(b|r) &= 0.1, & P(y|r) &= 0.7, & P(p|r) &= 0.2,
\end{align}
\end{subequations}
that define the probability observables,
\begin{subequations}
\begin{align}
  E_b &= P(b|g)g + P(b|r)r, \\
  E_y &= P(y|g)g + P(y|r)r, \\
  E_p &= P(p|g)g + P(p|r)r.
\end{align}
\end{subequations}
By construction, $E_b + E_y + E_p = 1_X$.  Furthermore, $E_p = (0.2)1_X$, so the purple outcome cannot distinguish whether the marble is green or red and can be imagined as a generic detector malfunction outcome.

The experimenter now has a choice for how to assign CVs to a detector observable $F_Y$ in order to infer the marble observable $F_X = (+1)g + (-1)r$.  A simple choice is to ignore the redundant (and nondistinguishing) purple outcome by zeroing out its CV $f_Y(p) = 0$, and then invert the remaining relationship analogously to \eqref{eq:marblematrix} to find $f_Y(b) = 3.125$ and $f_Y(y) = -1.875$.  The variance bound for this simple choice is $||F_Y||^2 = 13.2813$, leading to a root-mean-squared error no larger than $\sqrt{13.2813/n} \approx 3.6/\sqrt{n}$ after $n$ repeated measurements.

However, a better choice is to find the preferred values for all three outcomes using the pseudoinverse \eqref{eq:pseudoinversion} of the map between $F_Y$ and $F_X$.  To do this, we write a matrix equation similar to \eqref{eq:marblematrix} that uses all three outcomes,
\begin{subequations}\label{eq:marblematrixredundant}
\begin{align}
  \begin{pmatrix}+1 \\ -1\end{pmatrix} &= \mathcal{S}\begin{pmatrix}f_Y(b) \\ f_Y(y)\end{pmatrix}, \\
    \mathcal{S} &= \begin{pmatrix}P(b|g) & P(y|g) & P(p|g) \\ P(b|r) & P(y|r) & P(p|r)\end{pmatrix}. 
\end{align}
\end{subequations}
The pseudoinverse $\mathcal{S}^+$ can be constructed by using the singular value decomposition, $\mathcal{S} = \mathcal{U}\Sigma\mathcal{V}^T$, where $\mathcal{U}$ is an orthogonal matrix composed of the normalized eigenvectors of $\mathcal{S}\mathcal{S}^T$, $\mathcal{V}$ is an orthogonal matrix composed of the normalized eigenvectors of $\mathcal{S}^T\mathcal{S}$, and $\Sigma$ is a diagonal matrix composed of the singular values of $\mathcal{S}$ (which are the square roots of the eigenvalues of $\mathcal{S}\mathcal{S}^T$ and $\mathcal{S}^T\mathcal{S}$).  After computing the singular value decomposition, the pseudoinverse can be constructed as $\mathcal{S}^+ = \mathcal{V}\Sigma^+\mathcal{U}^T$, where $\Sigma^+$ is the diagonal matrix constructed by inverting all nonzero elements of $\Sigma^T$.  Performing this inversion we find the following preferred CV,
\begin{subequations}
\begin{align}
  \mathcal{S}^+ &= \frac{5}{36}\begin{pmatrix} 15 & -7 \\ -3 & 11 \\ 3 & 1 \end{pmatrix}, \\
    \begin{pmatrix}f_Y(b) \\ f_Y(y) \\ f_Y(p)\end{pmatrix} &= \mathcal{S}^+ \begin{pmatrix}+1\\ -1\end{pmatrix} = \frac{5}{18}\begin{pmatrix}11 \\ -7 \\ 1\end{pmatrix} = \begin{pmatrix}3.0\bar{5} \\ -1.9\bar{4} \\ 0.2\bar{7}\end{pmatrix}.
\end{align}
\end{subequations}
This preferred solution has the smallest variance bound of $||F_Y||^2 = 13.1944$.

We find (perhaps counterintuitively) that even though the purple outcome itself cannot distinguish the marble color, the fact that one obtains a purple outcome at all still provides some useful information to the experimenter due to the asymmetry of the blue and yellow outcomes.  Indeed, if for the red marble we instead found the symmetric detector response $P(b|r) = 0.3$, $P(y|r) = 0.5$, and $P(p|r) = 0.2$, the pseudoinverse would produce the preferred CVs $f_Y(b) = 5$, $f_Y(y) = -5$, and $f_Y(p) = 0$, indicating that the purple outcome was truly noninformative. 

A less principled approach to solving \eqref{eq:marblematrixredundant} would be for the experimenter to assign a completely arbitrary value to one outcome, like $f_Y(b) = B$.  The CV relation still produces a matrix equation, 
\begin{align}
  \begin{pmatrix}+1 - B P(b|g) \\ -1 - B P(b|r)\end{pmatrix} &= \begin{pmatrix}P(y|g) & P(p|g) \\ P(y|r) & P(p|r)\end{pmatrix} \begin{pmatrix}f_Y(y) \\ f_Y(p)\end{pmatrix},
\end{align}
that can be solved to find,
\begin{align}
  f_Y(y) &= B - 5, & f_Y(p) &= 12.5 - 4 B.
\end{align}
The bound for the variance of this solution is $||F_Y||^2 = 18 B^2 - 110 B + 181.25 \geq 13.1944$; the value of $B$ that minimizes the bound is $B = 3.0\bar{5}$, which recovers the pseudoinverse solution.  

Although picking an arbitrary solution gives mathematically equivalent results, the experimenter will only increase the norm of the solution without any physical motivation.  As such, the higher moments of the detector observable $F_Y$ can be correspondingly larger, and more trials may be necessary for the estimated average of the system observable $F_X$ to reach the desired precision.  

\subsubsection{Example: Continuous Detector}\label{sec:ccontinuousexample}
Consider the extreme example of a marble color detector that has a continuum of outcomes, such as the position of impact of a marble on a continuous screen.  In such a case, the detector sample space $Y$ is indexed by a real parameter $y\in \mathbb{R}$, and the relevant Boolean algebra $\Sigma_Y$ can be chosen to be the set of all Borel subsets of the real line \cite{Rudin1987,Szekeres2004}.  

After characterizing the detector, the experimenter finds that the detector displaces its initial probability distribution $dP_Y(y) = p_Y(y)\,dy$ by an amount $z$ from the zero-point according to which marble-color is sent into the detector,
\begin{align}
  dP(y|g) &= dP_Y(y - z), & dP(y|r) &= dP_Y(y + z).
\end{align}
These probabilities define the probability observables,
\begin{align}
  dE(y) &= g\, dP(y|g) + r\, dP(y|r),
\end{align}
such that $\int_\mathbb{R} dE(y) = 1_X$.

To infer information about the marble observable $F_X$ using this detector, the experimenter must assign a continuum of CVs $f_Y(y)$ such that,
\begin{align}\label{eq:marblecontobs}
  F_X &= (+1)g + (-1)r = \int_\mathbb{R} f_Y(y) \, dE(y),
\end{align}
or in matrix form,
\begin{align}
  \begin{pmatrix}+1\\-1\end{pmatrix} &= \mathcal{S}[f_Y] = \begin{pmatrix}\int_\mathbb{R} f_Y(y) \, dP_Y(y-z) \\ \int_\mathbb{R} f_Y(y) \, dP_Y(y+z)\end{pmatrix}.
\end{align}
Since $f_Y$ is a function, $\mathcal{S}$ is a vector-valued functional, which is why we adopt the square-bracket notation.

In this case, the detector outcomes are overwhelmingly redundant.  However, we can pick the least norm solution using the pseudoinverse of the map $\mathcal{S}$ as before.  To do so, we first calculate $\mathcal{S}\mathcal{S}^T$,
\begin{subequations}
\begin{align}
  \mathcal{S}^T &= \begin{pmatrix}p_Y(y-z) & p_Y(y+z)\end{pmatrix}, \\
  \mathcal{S}\mathcal{S}^T &= \begin{pmatrix} a & b(z) \\ b(z) & a \end{pmatrix}, 
\end{align}
\end{subequations}
where,
\begin{subequations}\label{eq:marblecont}
\begin{align}
    a &= \int_\mathbb{R} p_Y(y) \, dP_Y(y) = \int_\mathbb{R} p^2_Y(y)\, dy, \\
    b(z) &= \int_\mathbb{R} p_Y(y+z)p_Y(y-z)\, dy, 
\end{align}
\end{subequations}
and we find its eigenvalues of $a + b(z)$ with corresponding normalized eigenvector $(1,1)/\sqrt{2}$ and $a - b(z)$ with corresponding normalized eigenvector $(-1,1)/\sqrt{2}$.  We can then construct the orthogonal matrix $\mathcal{U}$ composed of the normalized eigenvectors of $\mathcal{S}\mathcal{S}^T$ and the diagonal matrix $\Sigma$ composed of the square roots of the eigenvalues of $\mathcal{S}\mathcal{S}^T$,
\begin{align}
  \mathcal{U} &= \frac{1}{\sqrt{2}}\begin{pmatrix}1 & -1 \\ 1 & 1\end{pmatrix}, \\
  \Sigma &= \begin{pmatrix}\sqrt{a + b(z)} & 0 \\ 0 & \sqrt{a - b(z)}\end{pmatrix}.
\end{align}

Next we calculate the relevant eigenfunctions of $\mathcal{S}^T\mathcal{S}$ that correspond to the same nonzero eigenvalues $a \pm b(z)$ of $\mathcal{S}\mathcal{S}^T$; the remaining eigenfunctions belong to the nullspace of $\mathcal{S}$ and do not contribute.  Specifically, we have,
\begin{align}
  \mathcal{S}^T\mathcal{S}[h](y) =& p_Y(y-z)\int_\mathbb{R} h(y)\, dP_Y(y-z) \\
  &+ p_Y(y+z)\int_\mathbb{R} h(y) dP_Y(y+z), \nonumber
\end{align}
where $h$ is an arbitrary function.  Then the equations,
\begin{subequations}
\begin{align}
  \mathcal{S}^T\mathcal{S}[v_+](y) =& (a + b(z))\, v_+(y), \\
  \mathcal{S}^T\mathcal{S}[v_-](y) =& (a - b(z))\, v_-(y), 
\end{align}
\end{subequations}
define the normalized eigenfunctions,
\begin{subequations}
\begin{align}
  v_+(y) &= \frac{p_Y(y-z) + p_Y(y+z)}{\sqrt{2(a + b(z))}}, \\
  v_-(y) &= - \frac{p_Y(y-z) - p_Y(y+z)}{\sqrt{2(a - b(z))}},
\end{align}
\end{subequations}
which allows us to construct the relevant part of the orthogonal map $\mathcal{V}^T$,
\begin{align}
  \mathcal{V}^T[h] &= \begin{pmatrix}\int v_+(y)h(y)\, dy & \int v_-(y)h(y)\, dy \end{pmatrix},
\end{align}
completing the nonzero part of the singular value decomposition of $\mathcal{S} = \mathcal{U}\Sigma\mathcal{V}^T$.

Finally, we construct the pseudoinverse,
\begin{align}
  \mathcal{S}^+ &= \mathcal{V}\Sigma^+\mathcal{U}^T, \\
  &= \begin{pmatrix}\frac{v_+(y)}{\sqrt{2(a + b(z))}} - \frac{v_-(y)}{\sqrt{2(a - b(z))}} & \frac{v_+(y)}{\sqrt{2(a + b(z))}} + \frac{v_-(y)}{\sqrt{2(a - b(z))}}\end{pmatrix}, \nonumber
\end{align}
and solve for the CV,
\begin{align}\label{eq:marblecontcv}
  f_Y(y) &= \mathcal{S}^+ \begin{pmatrix}+1\\-1\end{pmatrix} = \frac{p_Y(y - z) - p_Y(y + z)}{a - b(z)},
\end{align}
where $a$ and $b(z)$ are as defined in \eqref{eq:marblecont}. 

The pseudoinverse solution \eqref{eq:marblecontcv} contains only the physically relevant detector state density $p_Y$ and provides direct physical intuition about the detection process.  Namely, everything in the shifted distribution corresponding to the green marble $p_Y(y - z)$ is associated with the eigenvalue $+1$, while everything in the shifted distribution corresponding to the red marble $p_Y(y + z)$ is associated with the eigenvalue $-1$.  The overall amplification factor $a - b(z)$ indicates the discrepancy between the overlap of the shifted distributions and the distribution autocorrelation.  The more the shifted distributions overlap, the more ambiguous the measurement will be, so the amplification factor makes the CVs larger to compensate.  If the shifted distributions do not overlap, then $b(z)\to 0$ and the only amplification comes from the autocorrelation $a$ that indicates the ambiguity of the intrinsic profile of the detector state.  Moreover, the support of the CVs is equal to the support of both shifted detector distributions, which is physically satisfying.

The bound for the detector variance using the pseudoinverse solution is $||f_Y||^2 = 2 / [a - b(z)]$, which depends solely on the amplification factor in the denominator.  If the measurement is strong, such that $a - b(z) = 1$, then the variance bound reduces to the ideal variance bound of $2$, as expected, leading to a maximum RMS error of $\sqrt{2/n}$.  Any additional ambiguity of the measurement stemming from distribution overlap or distributed autocorrelation amplifies the maximum RMS error by a factor of $\sqrt{1/[a-b(z)]}$.

Contrast these preferred values with the generic linear solution $f_Y(y) = y/z$, which also satisfies \eqref{eq:marblecontobs} when $p_Y$ is symmetric about its mean \cite{Aharonov1988,DiLorenzo2008,Dressel2010}.  While the generic solution could be argued to be simpler in form, it provides no information about the detector and provides no physical insight into the meaning or origin of the values themselves.  It has nonzero support in areas where the detector has zero support and even gets progressively larger in regions that will not contribute to the average.  Moreover, the bound for the detector variance diverges, indicating that the RMS error can in principle be unbounded.  Hence, despite the mathematical equivalence, the linear solution is \emph{physically} inferior as a solution when compared to the pseudoinverse \eqref{eq:marblecontcv}.

\section{Quantum probability theory}
To transition from the classical theory of probability to the quantum theory we shall take a somewhat unconventional approach that leverages what we have already derived in the classical theory.  Specifically, we shall construct the quantum theory as a superstructure over the existing classical theory, rather than developing it as an independent logical system \cite{VonNeumann1932,Jauch1968} or as a restriction of a larger classical theory \cite{Spekkens2007,Fuchs2010}.  This approach serves to illustrate the myriad similarities between the quantum and classical theories, while also highlighting their key differences.  We shall see that the contextual-value formalism is essentially unchanged, despite the modifications that must be made to the operational theory of measurement.

\subsection{Sample spaces and observables} \label{sec:qsample}
\emph{Quantum sample space}.---The quantum theory of probability forms a superstructure on the classical theory of probability in the following sense: given a classical sample space $X$, the corresponding quantum sample space can be obtained as the orbit of $X$ under the action of the special unitary group of rotations.  That is, the entire classical sample space $X$ can be rotated to a different classical sample space $X' = \mathcal{U}(X)$ with some special unitary rotation $\mathcal{U}$.  We call each classical sample space generated in this fashion a \emph{framework} to be consistent with other recent work \cite{Griffiths2011}.  The collection of all such continuously connected classical sample spaces is the quantum sample space, which we will notate as $\mathcal{Q}(X)$ to emphasize that it can be generated from $X$.  

\emph{Representation}.---If the sample space $X$ is represented as a set of orthogonal rank-1 projections $\{\pprj{x}\}$ on a Hilbert space, the rotated sample space $X'=\mathcal{U}(X)$ will be represented by a different set of orthogonal projections $\{\mathcal{U}(\pprj{x})\}$ on the same Hilbert space.  Any such rotation $\mathcal{U}$ can be given a spinor representation (see, e.g., \cite{Hestenes1987,Hestenes1999,Doran2007}) as a two-sided product with a rotor $U$ belonging to the special unitary group, such that $U^\dagger U = U U^\dagger = 1_X$, and $(U^\dagger)^\dagger = U$.  The \emph{involution} ($^\dagger$) is the adjoint with respect to the inner product of the Hilbert space.  While the projections $\{\pprj{x}\}$ correspond to subspaces spanned by vectors $\{\ket{x}\}$ in the Hilbert space, the rotated projections $\{U^\dagger\pprj{x} U\}$ correspond to subspaces spanned by rotated vectors $\{U^\dagger\ket{x}\}$.  In what follows we shall tend to use the shorter algebraic notation $x$ and adopt the equivalent Hilbert space notation $\pprj{x}$ only when it readily simplifies expressions.  

Since the Hilbert space representation of a unitary rotor $U$ generally contains complex numbers in order to satisfy the special unitary group relations, the Hilbert space also becomes \emph{complex}.  However, it is important to note that the complex structure arises solely from the representation of the unitary rotations that specify the relative framework orientations and will not appear directly in any calculable quantity to follow \cite{Note5}.  The representation of the quantum sample space $\mathcal{Q}(X)$ therefore consists of all possible rank-1 projections on the complex Hilbert space in which the classical sample space $X$ is represented.  

\emph{Quantum observables}.---Each classical framework $X$ has an associated Boolean algebra $\Sigma_{X}$ and space of observables $\Sigma_{X}^\mathbb{R}$ exactly as previously discussed.  The space of quantum observables is the collection of all classical observables that are independently constructed in all the classical frameworks in $Q(X)$.  We will denote this space as $\Sigma_{\mathcal{Q}(X)}^\mathbb{R}$.  Quantum observables are therefore constructed entirely with real numbers that have empirical meaning for a laboratory setting; hence, their representations on a complex Hilbert space will be Hermitian operators.  

For observables in the same framework $A,B\in \Sigma_X^\mathbb{R}$, we find that $\mathcal{U}(A)\mathcal{U}(B) = U^\dagger AU U^\dagger B U = U^\dagger A B U = \mathcal{U}(AB)$, meaning that the rotations preserve their algebraic structure.  As a corollary, all observables in $\Sigma_{\mathcal{Q}(X)}^\mathbb{R}$ can be obtained by rotating observables constructed in a single framework $\Sigma_X^\mathbb{R}$; hence, our previous discussion of observables carries over to the quantum theory essentially unaltered.

Furthermore, the independence of the propositions in a framework $X$ remains unaltered by unitary rotation, so every other framework $X'$ has the same number of independent propositions.  Thus, the number of independent propositions is an invariant known as the \emph{quantum dimension}; for a representation it fixes the dimension of the Hilbert space.  Similarly, the identity and zero observables are invariants, so are the same in every framework and unique in the quantum observable algebra.  

Since each different framework forms a separate well-behaved classical sample space, the entire preceding discussion about classical probability theory applies unaltered when restricted to a particular framework in the quantum theory.  All observables constructed in a particular framework will commute with each other.  We expect distinctly quantum features to appear only when comparing elements from different frameworks.

\emph{Noncommutativity}.---The unitary rotations $\mathcal{U}$ are generally \emph{noncommutative} and so introduce noncommutativity into the quantum theory that is not present in the classical theory.  Specifically, given $A,B\in\Sigma_X^\mathbb{R}$, $A' = \mathcal{U}(A)$, and $B' = \mathcal{V}(B)$, then $A'B' = U^\dagger A U V^\dagger B V \neq B'A'$, since $U$ and $V$ do not necessarily commute with each other or with $A$ and $B$.  Such noncommutativity is a manifestation of the fact that the Boolean algebras corresponding to different frameworks are \emph{incompatible} with each other; propositions from one framework cannot form a Boolean logical \emph{and} with propositions from a different framework.  We shall see in the next section, however, that the notion of \emph{disturbance} followed by a logical \emph{and} can be generalized to the noncommutative setting in the form of the projection postulate. 

\emph{Disturbance}.---All nonconditioning disturbances $\mathcal{D}$ in the quantum theory also take the form of unitary rotations $\mathcal{U}$.  Indeed, we shall see that the parallels between the quantum theory and the classical theory with disturbance are quite strong when one interprets all unitary rotations as a form of classical disturbance.

\emph{Time Evolution}.---As an example, the continuous time-evolution of a closed quantum system is specified by a disturbance in the form of a unitary rotation $\mathcal{U}_t$ with corresponding rotor $U_t$, known as a \emph{propagator}.  For nonrelativistic quantum mechanics, the time-dependence of the rotor is specified by the Schr\"{o}dinger equation: $\partial_t U_t = (H /i\hbar) U_t$ and a Hamiltonian observable $H$ that generates the time translation.  We are not concerned with the (well-established) details of continuous time-evolution in this paper, so we will treat any unitary rotations as given in what follows.

\subsubsection{Example: Polarization}
As an example quantum system we shall pick the simplest possible nontrivial system: a qubit.  Specifically, we will consider the polarization degree of freedom of a laser beam.  Suppose we are interested in measuring the linear polarization of the beam with respect to the surface of an optical table.  We denote the polarization direction parallel to the table as ``horizontal'' ($h$) and the direction perpendicular to the table as ``vertical'' ($v$).  Although we casually refer to the polarizations $h$ and $v$ as if they were properties of the light beam, the propositions $h$ and $v$ operationally refer to two independent outcomes of a polarization distinguishing device, such as a polarizing beam splitter, that can be implemented in the laboratory.

The two orthogonal polarizations form a classical sample space $X = \{h,v\}$ and a classical Boolean algebra $\Sigma_X = \{0,h,v,1_X\}$, where $1_X = h + v$, similar to the classical sample space for the marble colors considered in Sec. \ref{sec:cmarble}.  By extending the Boolean algebra over the reals to $\Sigma^\mathbb{R}_X$ as before we can define classical observables $F_X = a\, h + b\, v$ in this sample space, such as the Stokes observable $S_X = h - v$ that distinguishes the polarizations with a sign.  

We can represent the commutative observable algebra $\Sigma^\mathbb{R}_X$ as diagonal $2\times 2$ matrices,
\begin{align}
  h &= \begin{pmatrix}1 & 0 \\ 0 & 0\end{pmatrix}, &
  v &= \begin{pmatrix}0 & 0 \\ 0 & 1\end{pmatrix}, &
  F_X &= \begin{pmatrix}a & 0 \\ 0 & b\end{pmatrix},
\end{align}
which can also be understood as commuting Hermitian operators over a two-dimensional Hilbert space.  The atomic propositions $h = \pprj{h}$ and $v = \pprj{v}$ are projectors that correspond to disjoint subspaces spanned by the orthonormal Jones' polarization basis for the Hilbert space,
\begin{align}
  \ket{h} &= \begin{pmatrix}1 \\ 0\end{pmatrix}, &
  \ket{v} &= \begin{pmatrix}0 \\ 1\end{pmatrix}.
\end{align}

To obtain the full quantum sample space $Q(X)$ from $X$, we introduce the group of possible polarization rotations.  Algebraically, an arbitrary rotation $\mathcal{U}(F_X) = U^\dagger F_X U$ can be readily understood in terms of its rotor $U$, which is an element of the group SU(2) and can be parametrized, for example, in terms of the Cartan decomposition $U_{\alpha,\beta,\gamma} = \exp(i \alpha \sigma_z / 2)\exp(i \beta \sigma_y / 2)\exp(i \gamma \sigma_z / 2)$, which for a qubit happens to correspond to an Euler angle decomposition of a three-dimensional rotation.  Here $i \sigma_z$ and $- i \sigma_y$ are two of the three generators of the Lie algebra for SU(2) in terms of the standard Pauli matrices,  
\begin{align}
  \sigma_y &= \begin{pmatrix}0 & -i \\ i & 0\end{pmatrix}, &
  \sigma_z &= \begin{pmatrix}1 & 0 \\ 0 & -1\end{pmatrix}.
\end{align}
Since the group generators have a complex representation, the unitary rotation $U_{\alpha,\beta,\gamma}$ will also have a complex representation in the Hilbert space,
\begin{subequations}\label{eq:polunitary}
\begin{align}
  e^{i \frac{\alpha}{2} \sigma_z} &= \begin{pmatrix}e^{i \frac{\alpha}{2}} & 0 \\ 0 & e^{-i \frac{\alpha}{2}}\end{pmatrix}, \\
  e^{i \frac{\beta}{2} \sigma_y} &= \begin{pmatrix}\cos\frac{\beta}{2} & \sin\frac{\beta}{2} \\ -\sin\frac{\beta}{2} & \cos\frac{\beta}{2}\end{pmatrix}, \\
  U_{\alpha,\beta,\gamma} &= \begin{pmatrix}e^{i(\alpha + \gamma)/2}\cos\frac{\beta}{2} & e^{i(\alpha - \gamma)/2}\sin\frac{\beta}{2} \\ -e^{-(\alpha-\gamma)/2}\sin\frac{\beta}{2} & e^{-i(\alpha+\gamma)/2}\cos\frac{\beta}{2} \end{pmatrix}.
\end{align}
\end{subequations}
The algebraic involution $U^\dagger_{\alpha,\beta,\gamma}$ is the complex transpose in the matrix representation.  

Physically, the factor $\exp(i \beta \sigma_y / 2)$ corresponds to a rotation of the apparatus around the axis of the light beam by an angle $\beta/2$, while the factors $\exp(i \alpha \sigma_z / 2)$ and $\exp(i \gamma \sigma_z / 2)$ correspond to the action of phase plates that shift the relative phases of $h$ and $v$ by $\alpha/2$ and $\gamma/2$, respectively.  Hence, the ubiquitous quantum phase also appears as a consequence of the unitary rotations.

Using the unitary rotations, we can generate other incompatible frameworks $\mathcal{U}_{\alpha,\beta,\gamma}(X) = \{\mathcal{U}_{\alpha,\beta,\gamma}(h), \mathcal{U}_{\alpha,\beta,\gamma}(v)\}$ in $Q(X)$,
\begin{subequations}
\begin{align}
  \mathcal{U}_{\alpha,\beta,\gamma}(h) &= U^\dagger_{\alpha,\beta,\gamma} h U_{\alpha,\beta,\gamma}, \\
  &= \begin{pmatrix}\cos^2\frac{\beta}{2} & \frac{1}{2}e^{-i\gamma}\sin\beta \\ \frac{1}{2}e^{i\gamma}\sin\beta & \sin^2\frac{\beta}{2}\end{pmatrix}, \nonumber \displaybreak[0]\\
  \mathcal{U}_{\alpha,\beta,\gamma}(v) &= U^\dagger_{\alpha,\beta,\gamma} v U_{\alpha,\beta,\gamma}, \\
  &= \begin{pmatrix}\sin^2\frac{\beta}{2} & -\frac{1}{2}e^{-i\gamma}\sin\beta \\ -\frac{1}{2}e^{i\gamma}\sin\beta & \cos^2\frac{\beta}{2}\end{pmatrix}, \nonumber
\end{align}
\end{subequations}
which depend solely on the two parameters $\beta$ and $\gamma$.  The atomic propositions of such a rotated framework are projectors corresponding to each disjoint subspace spanned by a rotated orthonormal Jones' polarization basis,
\begin{subequations}
\begin{align}
  U^\dagger_{\alpha,\beta,\gamma} \ket{h} &= \begin{pmatrix}e^{-i(\alpha + \gamma)/2}\cos\frac{\beta}{2} \\ e^{-i(\alpha - \gamma)/2}\sin\frac{\beta}{2}\end{pmatrix}, \\
  U^\dagger_{\alpha,\beta,\gamma} \ket{v} &= \begin{pmatrix}-e^{i(\alpha - \gamma)/2}\sin\frac{\beta}{2} \\ e^{i(\alpha + \gamma)/2}\cos\frac{\beta}{2}\end{pmatrix}.
\end{align}
\end{subequations}

Physically, one could in principle construct an apparatus corresponding to such a rotated framework using three laboratory elements: (1) attach a tunable phase plate to the incident port of a polarizing beam splitter with the fast axis aligned to the table, (2) rotate both the beam splitter and attached phase plate with respect to the table, and (3) attach a second tunable phase plate to the incident port of the first phase plate with the fast axis aligned to the table.  Of course this is only one possible parametrization for the unitary rotations; other parametrizations will correspond to other experimental implementations.

It follows that any observable in the full quantum observable space $\Sigma^\mathbb{R}_{Q(X)}$ can be obtained by rotating a classical observable $F_X = a\, h + b\, v$ to the appropriate framework,
\begin{align}\label{eq:qubitobs}
  F_{X'} &= \mathcal{U}_{\alpha,\beta,\gamma}(F_X) = a\, \mathcal{U}_{\alpha,\beta,\gamma}(h) + b\, \mathcal{U}_{\alpha,\beta,\gamma}(v), \\
  &= \begin{pmatrix} \frac{a+b}{2} + \frac{a-b}{2}\cos\beta & \frac{a-b}{2} e^{-i\gamma} \sin\beta \\ \frac{a-b}{2} e^{i\gamma} \sin\beta & \frac{a+b}{2} - \frac{a-b}{2}\cos\beta \end{pmatrix}. \nonumber
\end{align}
We see that a general qubit observable depends on four parameters: the eigenvalues $a$ and $b$, as well as the framework orientation angles $\beta$ and $\gamma$.  The complex representation of an observable stems solely from the unitary rotation of the atomic propositions $h$ and $v$ to a different relative framework.  The observables no longer generally commute since the unitary rotations need not commute.

\subsection{States, densities, and collapse}\label{sec:qcollapse}
\emph{Quantum states}.---A quantum state $P$ is a classical state defined in a particular framework $X$ that is then extended to apply to the entire quantum Boolean algebra $\Sigma_{\mathcal{Q}(X)}$.  The extension of a classical state $P$ that has been defined in a framework $X$ to a proposition $x'=\mathcal{U}(x)\in X'=\mathcal{U}(X)$ in a different framework can be accomplished by heuristically breaking down the state into a composition of the classical state in framework $X$ and \emph{transition probabilities} $D_x(x')$ that connect the framework $X$ to the different framework $X'$,
\begin{align}
  \label{eq:qprob}
  P(x') &= \sum_{x\in X}P(x)D_x(x'). 
\end{align}
The transition probabilities characterize a \emph{disturbance} \eqref{eq:cdisturb} that connects the classical state $P$ to propositions in incompatible frameworks.

To define the transition probabilities, we assume that atomic propositions in the framework $X$ are undisturbed, so $D_x(x) = 1$.  The only classical state with this property is the pure state which has a trace-density \eqref{eq:rho} $\rho = x$.  Hence, we assume that we can consistently write the transition probability $D_x(x')$ in terms of the extension of the \emph{trace} to the full Boolean algebra $\Sigma_{\mathcal{Q}(X)}$,
\begin{align}
  \label{eq:qoverlap}
  D_x(x') &= \Tr{x\, x'}.
\end{align}
Notably, this definition makes the transition between frameworks symmetric.  

\emph{Born rule}.---We pick the trace extension to be the unique measure that satisfies the cyclic property $\Tr{AB} = \Tr{BA}$ for all $A,B\in\Sigma_{\mathcal{Q}(X)}$ and agrees with the classical trace \eqref{eq:ipr} within any specific framework \cite{Takesaki1979}.  On a Hilbert space, \eqref{eq:qoverlap} has the familiar form,
\begin{align}\label{eq:qborn}
  D_{x}(x') &= \Tr{\pprj{x}\pprj{x'}} = |\pipr{x}{x'}|^2, 
\end{align}
which we immediately recognize as the \emph{Born rule} \cite{Born1926}.  Hence, the complex square of the Hilbert space inner product can be seen as a disguised form of the natural extension of the trace to define transition probabilities between propositions in incompatible frameworks.  If we recall that $x' = \mathcal{U}(x) = U^\dagger x U$ we can also write the transition probability \eqref{eq:qborn} in terms of the unitary rotor that connects the two propositions, $D_x(x') = \Tr{\pprj{x}U^\dagger\pprj{x}U} = |\bra{x}U\ket{x}|^2$.

\emph{Density operator}.---We can rewrite \eqref{eq:qprob} in a more familiar form by using the Born rule \eqref{eq:qoverlap} and the full trace-density \eqref{eq:rho} of the original state $\rho = \sum_{x\in X} P(x)\, x$, which is traditionally known as the \emph{density operator},
\begin{align}\label{eq:gleason}
  P(x') &= \sum_{x\in X} P(x)\Tr{x\, x'} = \Tr{\rho\, x'}. 
\end{align}
This form of the probability functional conforms to \emph{Gleason's theorem} \cite{Gleason1957}.  We note, however, that it is the extension of the \emph{trace} that extends the state to the noncommutative quantum setting since the trace-density $\rho$ is identical to a classical trace-density in some particular framework $X$.

\emph{Moments}.---Since the probabilities $P(x')$ are well-defined for a proposition in any framework $x'\in X'$, we can linearly extend $P$ to an expectation functional $\mean{\cdot}$ on the entire quantum observable algebra $\Sigma_{Q(X)}^\mathbb{R}$,  
\begin{align}
  \mean{F_{X'}} = \sum_{x'\in X'} f_{X'}(x') P(x') = \Tr{\rho\, F_{X'}}. 
\end{align}
Similarly, observable moments will be well-defined by the expectation functional,
\begin{align}
  \mean{(F_{X'})^n} = \sum_{x'\in X'} f^n_{X'}(x') P(x') = \Tr{\rho\, (F_{X'})^n}.
\end{align}
Hence, the unitary rotations and resulting extension of the trace completely construct the quantum probability space from a single classical probability space and its associated observables.

\emph{Double-sided \textsc{and}}.---To be consistent with the assumptions made in \eqref{eq:qoverlap}, we must also ensure that conditioning a quantum state on an atomic proposition will collapse the state to a pure state with a trace-density equal to that atomic proposition.  In other words, we must generalize the logical \textsc{and} of the classical case to the noncommutative incompatible frameworks in the quantum case.  The consistent way to do this is through a \emph{double-sided product}: given atomic propositions $x\in X$ and $x'\in X'$ then $x'xx' = \ket{x'}\pipr{x'}{x}\pipr{x}{x'}\bra{x'} = \Tr{x x'}x' = D_x(x')x'$.  

The double-sided product with $x'$ produces a \emph{transition probability} $D_x(x')$ from $x$ to $x'$ as a proportionality factor in addition to collapsing the original proposition $x$ to $x'$.  In this sense, the double-sided product includes a form of \emph{disturbance} in addition to the logical \emph{and} of pure conditioning.  If $X=X'$, so the frameworks coincide, then $x$ and $x'$ will commute; the disturbance will vanish, reducing the transition probability $D_x(x')$ to either $0$ or $1$; and, the classical \textsc{and} will be recovered as a special case.  

\emph{L\"uders' rule}.---Using the double-sided product as a disturbance followed by a logical $\emph{and}$, we find the quantum form of the \emph{invasive conditioning rule} \eqref{eq:cmdisturbpost},
\begin{subequations}\label{eq:qcollapse}
\begin{align}
  \dmean{F_X}_y &= \frac{\mean{y F_X y}}{P(y)} = \Tr{\rho_y\, F_X}, \\
  \rho_y &= \frac{y \rho y}{\Tr{\rho y}},
\end{align}
\end{subequations}
for any Boolean proposition $y$ in a framework algebra $\Sigma_X$ measured prior to the observable $F_X$.  As with the classical case, we use the tilde to indicate the intrinsic quantum invasiveness of the measurement process.  If $\rho$ and $y$ commute, or if $F_X$ and $y$ commute, then the noninvasive classical conditioning rule \eqref{eq:ccollapse} is properly recovered.  This generalization of \eqref{eq:cmdisturbpost} is known as the projection postulate, or \emph{L\"uders' Rule} \cite{Luders1951}.  If $y$ is an atomic proposition in $X$, then $\rho_y = y$ as in the classical case \eqref{eq:ccollapse} and we consistently recover the assumption \eqref{eq:qoverlap}.  

For contrast, \citet{Leifer2011} provide a careful quantum generalization of the \emph{noninvasive} conditioning rule \eqref{eq:ccollapse} using a formalism based around conditional density operators.  They confirm that L\"{u}der's Rule \eqref{eq:qcollapse} cannot be obtained with pure conditioning, so it must imply additional disturbance from the measurement process itself, as indicated here.

\emph{Aharonov-Bergmann-Lebowitz rule}.---Just as with classical invasive conditioning, the order of conditioning will generally matter.  Specifically, substituting a system proposition $z\in\Sigma_X$ into \eqref{eq:qcollapse} yields $\dmean{z}_y = P(yzy)/P(y)$; however, $P(yzy)\neq P(zyz)$, so the ``joint probability'' in the numerator is order-dependent unless $y$ and $z$ commute, just as in \eqref{eq:cgseqcollapsedisturb}.  That is, $\dmean{z}_y$ explicitly describes the case when the conditioning proposition $y$ is measured first as a \emph{preselection}, followed by the proposition $z$.  

To obtain the converse case when the conditioning proposition $z$ is measured second as a \emph{postselection}, we must derive the quantum form of \eqref{eq:cgpostselection}.  As in the classical case, we reinterpret the denominator of \eqref{eq:qcollapse} as a marginalization $P(y) = \sum_z P(yzy)$ of the ordered joint probability that renormalizes the conditioning procedure; the identity $\sum_z z = 1_X$ permits the equality.  With this interpretation, the postselected form of conditioning becomes straightforward,
\begin{align}\label{eq:qpostselection}
  \cmean{z}{y} = \frac{P(yzy)}{\sum_{y'\in Y} P(y'zy')}.
\end{align}
As in the classical case, the different position of the subscript serves to distinguish the two conditioned expectations $\dmean{\,\cdot\,}_z$ and $\cmean{z}{\,\cdot\,}$ corresponding to different measurement orderings.  
  
For a pure state $\rho = x = \pprj{x}$, this postselected conditioning is known as the \emph{Aharonov-Bergmann-Lebowitz (ABL) rule} \cite{Aharonov1964}, and has the form $\ccmean{z}{y}{x} = |\pipr{z}{y}|^2|\pipr{y}{x}|^2 / \sum_{y'\in Y}|\pipr{z}{y'}|^2|\pipr{y'}{x}|^2$.  Unlike L\"{u}ders' rule \eqref{eq:qcollapse}, the generalized ABL rule \eqref{eq:qpostselection} does not perform a simple update to the trace-density $\rho$; moreover, it depends on the entire disturbance of the first measurement via the normalization sum in the denominator.  If $y$ and $z$ commute, then the disturbance vanishes and we again correctly recover the classical case \eqref{eq:ccollapse} that is order-independent.  

\emph{Bayes' rule}.---The two forms of quantum invasive conditioning also lead to a modified form of Bayes' rule that relates the preselected conditioning of a sequence to the postselected conditioning of the same sequence, similarly to the classical case \eqref{eq:cgbayesdisturb},
\begin{align}\label{eq:qbayes}
  \cmean{z}{y} = \dmean{z}_y \frac{P(y)}{\sum_{y'\in Y} P(y'zy')}.
\end{align}
If $y$ and $z$ commute, then the disturbance vanishes and we correctly recover Bayes' rule \eqref{eq:cbayes}.

The unusual form of \eqref{eq:qpostselection} has led to postselected quantum conditioning being largely overlooked.  The lack of symmetry in the density update under such postselected conditioning has even prompted works in multistate-density time-symmetric reformulations of quantum mechanics \cite{Aharonov1988,Aharonov1990,Aharonov2005,Aharonov2008,Aharonov2009,Aharonov2010,Massar2011}, which are outside the scope of this work.  However, we see here that the form of the conditioning is the same as the classically \emph{invasive} postselected conditioning \eqref{eq:cgpostselection}.  Later we shall use a fully generalized form of the ABL rule \eqref{eq:qpostselection} together with CVs to consider the subtle case of postselected averages of observables in some detail, so we delay their consideration for now.

\subsubsection{Example: Polarization state}
A quantum state for a single system is a classical state in some particular framework.  For a two-dimensional framework such as $\{h,v\}$, all probabilities for such a classical state can be completely specified by a mixing angle $\theta$ such that $P(h) = \cos^2(\theta/2)$ and $P(v) = \sin^2(\theta/2)$.  Hence, after rotating the trace-density $\rho = P(h) h + P(v) v$ to an arbitrary framework according to \eqref{eq:qubitobs}, any quantum state trace-density of polarization must have the form,
\begin{align}\label{eq:qubitstate}
  \rho_{\theta,\beta,\gamma} &= \cos^2(\theta/2)\, \mathcal{U}_{\alpha,\beta,\gamma}(h) + \sin^2(\theta/2)\, \mathcal{U}_{\alpha,\beta,\gamma}(v), \\
  &= \frac{1}{2}\begin{pmatrix}1 + \cos\beta \cos\theta & e^{-i\gamma} \sin\beta \cos\theta \\ e^{i\gamma} \sin\beta \cos\theta & 1 - \cos\beta \cos\theta \end{pmatrix}. \nonumber
\end{align}
The $\alpha$ parameter of the rotation disappears in favor of the $\theta$ parameter characterizing the classical state, leaving only three net parameters, in contrast to the four parameters of an arbitrary observable \eqref{eq:qubitobs}.

The expectation functional $\mean{\cdot}_{\theta,\beta,\gamma}$ is then defined from the trace-density $\rho_{\theta,\beta,\gamma}$ and the unique extension of the trace $\text{Tr}$ to the whole observable algebra $\Sigma^\mathbb{R}_{Q(X)}$ according to $\mean{F_{X'}}_{\theta,\beta,\gamma} = \text{Tr}(\rho_{\theta,\beta,\gamma}\, F_{X'})$.  The trace extension is the sum of the diagonal matrix elements in the matrix representation.  Hence for the expectation of an arbitrary observable \eqref{eq:qubitobs} under an arbitrary state \eqref{eq:qubitstate} we find,
\begin{subequations}
\begin{align}
  \mean{\mathcal{U}_{\alpha',\beta',\gamma'}(F_X)}_{\theta,\beta,\gamma} = \frac{a+b}{2} + \frac{a-b}{2}(\cos\theta) \, \Xi, \\
  \Xi = \cos\beta \cos\beta' + \sin\beta \sin\beta' \cos(\gamma - \gamma'),
\end{align}
\end{subequations}
where $\Xi\in[-1,1]$ is an interference factor that depends only on relative orientation between the state framework and the observable framework.  If the frameworks coincide, then $\Xi = 1$ and the classical result is recovered.

\subsection{Detectors and probability observables} \label{sec:qdetector}
\emph{Joint observable space}.---As with the classical case, we can couple a system to a detector by enlarging the sample space to the product space $XY$ of a particular pair of frameworks.  We can then perform \emph{local} unitary rotations on each space independently to form a joint quantum sample space from the classical joint observables $\mathcal{Q}(X)\mathcal{Q}(Y)$.  However, the quantum observable space also admits \emph{global} unitary rotations on the classical joint observables to form a larger joint quantum sample space $\mathcal{Q}(XY)$.  Just as with a single sample space, any two propositions in $\mathcal{Q}(XY)$ can be continuously connected with some global unitary rotation.

The full quantum observable space $\Sigma_{\mathcal{Q}(XY)}^\mathbb{R}$ is constructed from $\mathcal{Q}(XY)$ in the usual way.  Product observables will maintain their product form under local unitary rotations, $\mathcal{U}_X(\mathcal{V}_Y(A_XB_Y)) = \mathcal{U}_X(A_X)\mathcal{V}_Y(B_Y)$.  However, global unitary rotations can create unfactorable correlated joint observables in $\Sigma_{\mathcal{Q}(XY)}^\mathbb{R}$ even from product observables $\mathcal{U}(A_XB_Y)$.

\emph{Joint states}.---Similarly, joint \emph{states} on a classical product framework extend to joint quantum states on the quantum product observable space.  Under local unitary rotations, product states remain product states and classically correlated states between two specific frameworks remain classically correlated.  However, \emph{global} unitary rotations performed on any state can also form \emph{entangled} states that have no analog in the classical theory \cite{Horodecki2009}.  Entangled states have some degree of \emph{local-rotation-independent} correlation between frameworks, so display a stronger degree of correlation than can even be defined with a classically correlated state that is restricted to a single pair of frameworks.  As an extreme example, maximally entangled states are completely local-rotation-independent and perfectly correlated with respect to any pair of frameworks.

\emph{Quantum operations}.---The specifics of entanglement do not concern us here, since any type of correlation is sufficient to represent detector probabilities within the reduced system space.  For the purposes of measurement, we only assume that the correlated state with density $\rho = \mathcal{U}^\dagger(\rho_X\rho_Y) = U\rho_X\rho_Y U^\dagger$ is connected to some initial product state with density $\rho_X\rho_Y$ via a unitary rotation $\mathcal{U}^\dagger$.  Since all quantum states can be continuously connected with some global unitary rotation that acts as a disturbance \eqref{eq:cdisturbcouple}, this is always possible.  Physically, the unitary rotation couples the known detector state $\rho_Y$ to an unknown system state $\rho_X$.  Furthermore, we assume that the initial state of the detector has some (not necessarily unique) pure-state expansion that is meaningful with respect to the preparation procedure $\rho_Y = \sum_{y'\in Y'} P'(y') y'$.  

It then follows that the numerator for the conditioning rules \eqref{eq:qcollapse} and \eqref{eq:qpostselection} becomes,
\begin{align}\label{eq:qmeas}
  \mean{y F_X y} &= \text{Tr}(\rho y F_X y), \\
  &= \text{Tr}_X(\text{Tr}_Y(U \rho_X\rho_YU^\dagger y F_X y)), \nonumber \displaybreak[0]\\
  &= \mean{\mathcal{E}_y(F_X)}_X = \text{Tr}_X(\mathcal{E}^\dagger_y(\rho_X) F_X), \nonumber \displaybreak[0]
\end{align}
with the \emph{operations} $\mathcal{E}_y$ and $\mathcal{E}^\dagger_y$ defined as,
\begin{subequations}\label{eq:qmeasoper}
\begin{align}
  \label{eq:qoper}
  \mathcal{E}_y(F_X) &= \mean{U^\dagger y F_X y U}_Y, \\
  &= \sum_{y'\in Y'} P'(y') \text{Tr}_Y(y' U^\dagger y F_X y U), \nonumber \\
  &= \sum_{y'\in Y'} M^\dagger_{y,y'} F_X M_{y,y'}, \nonumber \displaybreak[0]\\
  \label{eq:qoperad}
  \mathcal{E}^\dagger_y(\rho_X) &= \text{Tr}_Y(yU\rho_X\rho_YU^\dagger y), \\
  &= \sum_{y'\in Y'} P'(y') \text{Tr}_Y(y U \rho_X y' U^\dagger y), \nonumber \\
  &= \sum_{y'\in Y'} M_{y,y'} \rho_X M^\dagger_{y,y'}, \nonumber \displaybreak[0]\\
  \label{eq:qmeasops}
  M_{y,y'} &= e^{i\phi_{y,y'}} \sqrt{P'(y')} \bra{y}U\ket{y'}, \\
  M_{y,y'}^\dagger &= e^{-i\phi_{y,y'}} \sqrt{P'(y')} \bra{y'}U^\dagger\ket{y}.
\end{align}
\end{subequations}
Here, the Hilbert space representations of the \emph{Kraus operators} $\{M_{y,y'}\}$ have the form of partial matrix elements and are only well-defined up to the arbitrary phase factors $e^{i\phi_{y,y'}}$.  We also stress that $\{M_{y,y'}\}$ depend not only on the measured detector outcome $y$, but also on a particular detector \emph{preparation} $y'$.

As a result, we find the quantum versions of the probability observables \eqref{eq:cpovmdisturb},
\begin{align}
  P(y) &= \mean{\mathcal{E}_y(1_X)}_X = \mean{E_y}_X, \\
  \label{eq:qpovm}
  E_y &= \mathcal{E}_y(1_X) = \mean{U^\dagger y U}_Y, \\
  &= \sum_{y'\in Y'} M^\dagger_{y,y'}M_{y,y'}, \nonumber
\end{align}
and the general invasive measurement \eqref{eq:cgcollapsedisturb},
\begin{align}\label{eq:qgcollapse}
  \dmean{F_X}_y &= \frac{\mean{\mathcal{E}_y(F_X)}_X}{\mean{\mathcal{E}_y(1_X)}_X}, \\
  &= \frac{\sum_{y'\in Y'} \text{Tr}_X(\rho_X M^\dagger_{y,y'}F_X M_{y,y'})}{\text{Tr}_X(\rho_X E_y)}. \nonumber 
\end{align}

Similarly to the invasive classical case \eqref{eq:cgoperations}, the measurement of $y$ on the detector must be described by a \emph{quantum operation} $\mathcal{E}_y$ in \eqref{eq:qmeas}, which is a completely positive map \cite{VonNeumann1932,Jauch1968,Davies1970,Kraus1971,Lindblad1976,Kraus1983,Braginski1992,Busch1995,Nielsen2000,Alicki2001,Keyl2002,Breuer2007,Wiseman2010} that performs a \emph{generalized measurement} on the system state corresponding to the detector outcome $y$.  The operation $\mathcal{E}_y$ acting on the identity in \eqref{eq:qpovm} produces a positive operator known as a \emph{quantum effect}, $E_y$.  By construction, the set of operations $\{\mathcal{E}_y\}$ preserves the identity, $\sum_y \mathcal{E}_y(1_X) = 1_X$; hence, the effects form a partition of the identity, $\sum_y E_y = 1_X$, making them probability observables over a particular detector framework exactly as in \eqref{eq:cpovmdisturb}.  

Sequences of measurements emphasize the temporal ordering of operations, just as in the invasive classical case \eqref{eq:cgseqcollapsedisturb}.  Given two sets of quantum operations that define the sequential interaction of two detectors with the system and their subsequent conditioning, $\{\mathcal{E}_y\}$ and $\{\mathcal{E}'_z\}$, the joint probability of the ordered sequence of detector outcomes $(y,z)$ is,
\begin{align}\label{eq:qgseqprob}
  P(y)P(z|y) &= P(yzy) = P(yz1_Xzy),\\
  &= \mean{\mathcal{E}_y(\mathcal{E}'_z(1_X))}_X = \mean{\mathcal{E}_y(E'_z)}_X, \nonumber
\end{align}
where $E'_z = \mathcal{E}'_z(1_X)$.  The proper sequential probability observable $\mathcal{E}_y(E'_z) = \sum_{y'} M^\dagger_{y,y'} E'_z M_{y,y'}$ is not a simple product of the individual probability observables $E_y$ and $E'_z$.  

These sequence probabilities then give us the full generalization of the ABL rule \eqref{eq:qpostselection},
\begin{align}
  \label{eq:qgpostselection}
  \cmean{z}{y} &= \frac{\mean{\mathcal{E}_y(E'_z)}_X}{\mean{\mathcal{E}(E'_z)}_X} = \frac{\mean{\mathcal{E}_y(E'_z)}_X}{\sum_{y''\in Y}\mean{\mathcal{E}_{y''}(E'_z)}_X}, \\
  &= \frac{\sum_{y'\in Y'} \text{Tr}_X(\rho_X M^\dagger_{y,y'}E'_z M_{y,y'})}{\sum_{y''\in Y}\sum_{y'\in Y'} \text{Tr}_X(\rho_X M^\dagger_{y'',y'}E'_z M_{y'',y'})}, \nonumber
\end{align}
and the most general version of the invasive quantum Bayes' rule \eqref{eq:qbayes},
\begin{align}
  \label{eq:qgbayes}
  \cmean{z}{y} &= \dmean{E'_z}_y \frac{\mean{E_y}_X}{\mean{\mathcal{E}(E'_z)}_X}, 
\end{align}
As with \eqref{eq:cgpostselection} and \eqref{eq:qpostselection}, the postselected conditioning \eqref{eq:qgpostselection} depends on the entire disturbance of the first measurement via the \emph{nonselective measurement} $\mathcal{E} = \sum_{y''\in Y}\mathcal{E}_{y''}$ in the denominator.  

The noncommutativity of the detection operations $\mathcal{E}_y$ emphasizes the fact that measurement is an active \emph{process}: an experimenter alters the quantum state by coupling it to a detector and then conditioning on acquired information from the detector.  Without some filtering process that completes the disturbance implied by \eqref{eq:qmeas}, there is no measurement.  The nonselective measurement $\mathcal{E}$ also includes the active disturbance of the measurement process, but does not condition on a particular outcome.  Furthermore, measuring a quantum state in a different order generally disturbs it differently.  The state may also in certain conditions be probabilistically ``uncollapsed'' back to where it started by using the correct conditioning sequence \cite{Korotkov2006,Katz2008,Kim2009}.  In this sense, sequential quantum conditioning is analogous to a stochastic control process that guides the progressive disturbance of a state along some trajectory in the state space \cite{Wiseman2010}.

\emph{Measurement operators}.---Since the quantum operation $\mathcal{E}_y$ performs a measurement, we will refer to its Kraus operators $\{M_{y,y'}\}$ \eqref{eq:qmeasoper} as \emph{measurement operators}.  However, a quantum operation generally has many equivalent double-sided product expansions like \eqref{eq:qoper} in terms of measurement operators.  Each such set of measurement operators $\{M_{y,y'}\}$ corresponds to a specific choice of framework for the preparation of the detector state $\rho_Y = \sum_{y'\in Y'} P(y')\, y'$.  

Given a specific set of measurement operators, the substitution $M_{y,y'}\to U_{y,y'}M_{y,y'}$ with unitary $U_{y,y'}$ will produce the same effect $E_y$ according to \eqref{eq:qpovm} but will correspond to a different operation $\mathcal{E}'_y$.  Hence, we conclude that many measurement operations can produce the same probability observables on the system space \cite{Braunstein1988}.  Therefore, \emph{probability observables are not sufficient to completely specify a quantum measurement}: one needs to specify the full operations as in the classically invasive case \eqref{eq:cgoperations}.  

\emph{Quantum process tomography}.---Just as classical probability observables can be characterized via process tomography, operations can be characterized by \emph{quantum process tomography}.  One performs quantum process tomography by sending known states into a detector, measuring the detector, then measuring the resulting states to see how the state was changed by the detector.  Since quantum operations contain information about disturbance as well as conditioning, quantum process tomography generally requires more characterization measurements than pure classical process tomography.

\emph{Pure operations}.---An initially pure detector state with density $y'$ produces a \emph{pure operation} $\mathcal{E}_y(F_X) = M^\dagger_y F_X M_y$ with a single associated measurement operator $M_y = e^{i\phi_y} \bra{y}U\ket{y'}$ that is unique up to the arbitrary phase factor $e^{i\phi_y}$.  Most laboratory preparation procedures for the detector are designed to produce a pure initial state, so pure operations will be the typical case.  A pure operation has the additional property of partially collapsing a pure state to another pure state.  It is also most directly related to the probability observable $E_y = M^\dagger_y M_y$, since the single measurement operator has a polar decomposition $M_y = U_y E_y^{1/2}$ in terms of the positive root of the probability observable $E_y^{1/2}$.

\emph{Weak measurement}.---If we wish for such a conditioning process to leave the state approximately unchanged, we must make a \emph{weak measurement}, just as in the classical case \eqref{eq:cweak}.  However, a quantum weak measurement requires a strict condition regarding the measurement operations and not just the probability observables due to the additional disturbance in the measurement.  Formally, the measurement operations typically depend on a measurement strength parameter $\epsilon$ such that, 
\begin{align}\label{eq:qweak}
  \forall y\in Y\; \lim_{\epsilon\to 0} \mathcal{E}_y(\epsilon; F_X) = P_Y(y) \mathcal{I}(F_X),
\end{align}
where $\mathcal{I}$ is the identity operation and $P_Y(y)$ is the probability for obtaining the detector outcome $y$ in the absence of interaction.  As with the classical case, the limit as $\epsilon\to 0$ is an idealization known as the \emph{weak measurement limit} and is not strictly achievable in the laboratory.  

The definition \eqref{eq:qweak} implies that subsequent measurements will be unaffected, $\forall y\in Y,\; \lim_{\epsilon\to 0} \dmean{F_X}_y = \mean{F_X}$, and that the probability observables are proportional to the identity in the weak limit, $\forall y\in Y,\; \lim_{\epsilon\to 0} E_y(\epsilon) = P_Y(y) 1_X$, just as in the classical case \eqref{eq:cweak}.  It also follows that any set of measurement operators $\{M_{y,y'}(\epsilon)\}$ that characterize $\mathcal{E}_y(\epsilon)$ must also be proportional to the identity in the weak limit $\forall y\in Y,y'\in Y',\; \lim_{\epsilon\to 0} M_{y,y'}(\epsilon) \propto 1_X$.

Weak measurements are more interesting in the quantum case than in the classical case due to the existence of incompatible frameworks.  Since a weak measurement of an observable does not appreciably affect the quantum state, subsequent measurements on incompatible observables can be made that will probe approximately the same state.  This technique allows (noisy) information about two incompatible frameworks to be gleaned from nearly the same quantum state in a single experiment, which is strictly impossible using strong measurements that collapse the state to a pure state in a particular framework after each measurement.  The penalty for using weak measurements is that many more measurements are needed than in the strong measurement case to overcome the ambiguity of the measurement, as discussed in the classical case.

\subsubsection{Example: Coverslip polarization detector}\label{sec:qpbscover}
\begin{figure*}[th]
  \begin{center}
    \includegraphics{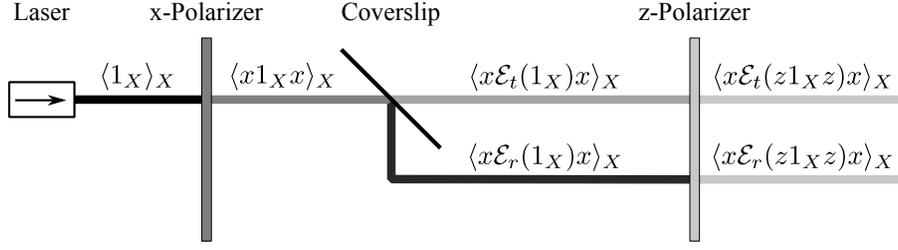}
  \end{center}
  \caption{Coverslip polarization measurement.  A laserbeam passes through a preselection $x$-polarizer, a glass microscope coverslip, and a postselection $z$-polarizer.  The transmission probabilities for each segment of the apparatus are shown.  By assigning appropriate contextual values $f_Y(t)$ and $f_Y(r)$ \eqref{eq:polcv} to the output ports of the coverslip, the polarization observable $F_X = f_X(h) h + f_X(v) v$ can be measured using the equivalent expansion in terms of the appropriate measurement context $F_X = f_Y(t)\mathcal{E}_t(1_X) + f_Y(r)\mathcal{E}_r(1_X)$.  Averaging the same contextual values with pre- and postselected conditional probabilities $\ccmean{z}{t}{x} = \mean{x\mathcal{E}_t(z)x}_X/(\mean{x\mathcal{E}_t(z)x}_X + \mean{x\mathcal{E}_r(z)x}_X)$ and $\ccmean{z}{r}{x} = \mean{x\mathcal{E}_r(z)x}_X/(\mean{x\mathcal{E}_t(z)x}_X + \mean{x\mathcal{E}_r(z)x}_X)$ produces the conditioned average \eqref{eq:polppsca} $\ccmean{z}{F_X}{x} = f_Y(t) \ccmean{z}{t}{x} + f_Y(r) \ccmean{z}{r}{x}$. }
  \label{fig:coverslip}
\end{figure*}

To cement these ideas, we consider the task of indirectly measuring polarization in a particular framework.  For specificity, we will consider the passage of a laser beam with unknown polarization through a glass microscope coverslip, as shown in Fig.~\ref{fig:coverslip}.  Fresnel reflection off the coverslip leads to a disparity between transmission and reflection of the polarizations, so comparing transmitted to reflected light allows a generalized measurement of polarization, as we demonstrated experimentally in \cite{Dressel2011}.

The system sample space we wish to measure is the polarization with respect to the table ($h = \pprj{h}$) and ($v = \pprj{v}$), which could in principle be measured ideally with a polarizing beam splitter.  The detector sample space is the spatial degree of freedom of the transmitted ($t = \pprj{t}$) and reflected ($r = \pprj{r}$) ports of a coverslip rotated to some fixed angle with respect to the incident beam around an axis perpendicular to the table.  The initial state of the detector is the pure state indicating that the beam enters a single incident port ($b = \pprj{b}$) of the coverslip with certainty.  The rotation $\mathcal{U}^\dagger(\rho_X b) = U \rho_X b U^\dagger$ that couples the system to the detector describes the interaction of the beam with the coverslip and has a unitary rotor $U$ corresponding to the polarization-dependent scattering matrix of the coverslip.  Assuming that the scattering preserves beams of pure polarization, so $h$ remains $h$ and $v$ remains $v$, the rotor decouples into a direct sum of rotors that are specific to each polarization,
\begin{align}
  U &= U_h \oplus U_v,
\end{align}
meaning that $U$ has a block-diagonal structure when represented as a matrix.

Selecting each output port of the coverslip produces the two \emph{measurement operators} according to \eqref{eq:qmeasoper},
\begin{subequations}\label{eq:polmeasops}
\begin{align}
  M_t &= \bra{t}U\ket{b} = \begin{pmatrix}\bra{t}U_h\ket{b} & 0 \\ 0 & \bra{t}U_v\ket{b}\end{pmatrix}, \\
  M_r &= \bra{r}U\ket{b} = \begin{pmatrix}\bra{r}U_h\ket{b} & 0 \\ 0 & \bra{r}U_v\ket{b}\end{pmatrix},
\end{align}
\end{subequations}
which characterize the \emph{pure measurement operations} that modify observables according to \eqref{eq:qoper},
\begin{subequations}\label{eq:polpmo}
\begin{align}
  \mathcal{E}_t(F_X) &= M_t^\dagger F_X M_t, \\
  \mathcal{E}_r(F_X) &= M_r^\dagger F_X M_r,
\end{align}
\end{subequations}
and their adjoints that modify the state density according to \eqref{eq:qoperad},
\begin{subequations}\label{eq:polpmoa}
\begin{align}
  \mathcal{E}^\dagger_t(\rho_X) &= M_t \rho_X M_t^\dagger, \\
  \mathcal{E}^\dagger_r(\rho_X) &= M_r \rho_X M_r^\dagger.
\end{align}
\end{subequations}

The pure measurement operations in turn produce \emph{probability observables} according to \eqref{eq:qpovm},
\begin{subequations}\label{eq:polpo}
\begin{align}
  E_t &= \mathcal{E}_t(1_X) = M_t^\dagger M_t, \\
  &= \begin{pmatrix} |\bra{t}U_h\ket{b}|^2 & 0 \\ 0 & |\bra{t}U_v\ket{b}|^2 \end{pmatrix}, \nonumber \\
  E_r &= \mathcal{E}_r(1_X) = M_r^\dagger M_r, \\
  &= \begin{pmatrix} |\bra{r}U_h\ket{b}|^2 & 0 \\ 0 & |\bra{r}U_v\ket{b}|^2 \end{pmatrix}, \nonumber
\end{align}
\end{subequations}
in the same framework as $h$ and $v$.  These probability observables are therefore equivalent to classical probability observables \eqref{eq:cpovmdisturb} specified by the effective characterization probabilities $\tilde{P}(t|h) = |\bra{t}U_h\ket{b}|^2$, $\tilde{P}(r|h) = |\bra{r}U_h\ket{b}|^2$, $\tilde{P}(t|v) = |\bra{t}U_v\ket{b}|^2$, and $\tilde{P}(r|v) = |\bra{r}U_v\ket{b}|^2$.  

The measurement operators \eqref{eq:polmeasops} have a polar decomposition in terms of the roots of the probability observables and an extra unitary phase contribution,
\begin{subequations}\label{eq:polmeasops2}
\begin{align}
  M_t &= \begin{pmatrix}e^{i \phi_{h,t}} \sqrt{\tilde{P}(t|h)} & 0 \\ 0 & e^{i \phi_{v,t}} \sqrt{\tilde{P}(t|v)}\end{pmatrix}, \\
  M_r &= \begin{pmatrix}e^{i \phi_{h,r}} \sqrt{\tilde{P}(r|h)} & 0 \\ 0 & e^{i \phi_{v,r}} \sqrt{\tilde{P}(r|v)}\end{pmatrix}.
\end{align}
\end{subequations}
Any nonzero relative phase, such as $\phi_{h,t} - \phi_{v,t}$, will affect the framework orientation for subsequent measurements; however, it will not contribute to the acquisition of information from the measurement since it does not contribute to the probability observables.  Such relative phase is therefore part of the \emph{disturbance} of the measurement process.

Specifically, the initial state of polarization $P_X$ will be conditioned by a selection of a particular port on the detector according to,
\begin{subequations}
\begin{align}
  \dmean{F_X}_t &= \frac{\mean{\mathcal{E}_t(F_X)}_X}{\mean{\mathcal{E}_t(1_X)}_X} = \frac{\text{Tr}_X(M_t \rho_X M_t^\dagger F_X)}{\text{Tr}_X(\rho_X E_t)}, \\
  \dmean{F_X}_r &= \frac{\mean{\mathcal{E}_r(F_X)}_X}{\mean{\mathcal{E}_r(1_X)}_X} = \frac{\text{Tr}_X(M_r \rho_X M_r^\dagger F_X)}{\text{Tr}_X(\rho_X E_r)}.
\end{align}
\end{subequations}
Although the probabilities in each denominator only depend on the probability observables, the altered states in each numerator depend on the measurement operations and will include effects from the relative phase in the measurement operators \eqref{eq:polmeasops2}.

\subsection{Contextual values} \label{sec:qcv}
\emph{Operation correspondence}.---The introduction of contextual values in the quantum case proceeds identically to the classical case of invasive measurements \eqref{eq:coperexpand}.  Since we must generally represent detector probabilities by \emph{operations} $\{\mathcal{E}_y\}$ within the reduced system space according to \eqref{eq:qpovm} and \eqref{eq:qgseqprob}, we must also generally represent detector observables by \emph{weighted operations} within the reduced system space,
\begin{align}\label{eq:qpovmdetobs}
  \mean{F_Y} &= \sum_{y\in Y} f_Y(y) P(y), \\
  &= \sum_{y\in Y} f_Y(y) \mean{\mathcal{E}_y(1_X)}_X = \mean{\mathcal{F}_X(1_X)}_X, \nonumber \\
  \label{eq:qobsoper}
  \mathcal{F}_X &= \sum_{y\in Y} f_Y(y) \mathcal{E}_y.
\end{align}
If we are concerned with only a single measurement, or are working within a single framework as in the classical formalism, then for all practical purposes the operation $\mathcal{F}_X$ reduces to its associated system observable $F_X = \mathcal{F}_X(1_X)$ as in the classical definition \eqref{eq:ccgrain}.  

\emph{Contextual values}.---We observe a corollary exactly as in the classical case \eqref{eq:cpovmexpand}: if we can expand a \emph{system} observable in terms of the probability observables generated by a particular measurement operation, then that observable can also be expressed as an equivalent \emph{detector} observable,
\begin{align}\label{eq:qpovmexpand}
  F_X &= \sum_{y\in Y} f_Y(y) E_y \implies F_Y = \sum_y f_Y(y) y,
\end{align}
which is the quantum form of our main result originally introduced in \cite{Dressel2010}.  As in the classical case, we dub the required detector labels $f_Y(y)$ the \textbf{contextual values} (CV) of the quantum observable $F_X$ with respect to the \emph{context} of a specific detection scheme as represented in the system space by the measurement operations $\{\mathcal{E}_y\}$.  Since many measurement operations produce the same probability observables $\{\mathcal{E}_y(1_X) = E_y\}$, many detection schemes can use the same CVs to reproduce an observable average.

\emph{Moments}.---As with classically invasive measurements \eqref{eq:coperseq}, higher statistical moments of the observable require more care to measure.  For instance, we require the following equality in order to accurately reproduce the $n$\textsuperscript{th} moment of an observable indirectly using the same CV,
\begin{align}\label{eq:qsecondmom}
  \mean{(F_X)^n}_X &= \sum_{y_1,\ldots,y_n\in Y}f_Y(y_1)\cdots f_Y(y_n) \mean{E_{y_1}\cdots E_{y_n}}_X.
\end{align}
However, as indicated in \eqref{eq:qgseqprob}, performing a sequence of $n$ measurements produces the measurable probability $\mean{\mathcal{E}_{y_1}(\cdots(E_{y_n})\cdots)}_X \neq \mean{E_{y_1}\cdots E_{y_n}}_X$.  Indeed, $\mean{E_{y_1}\cdots E_{y_n}}_X$ will not generally be a well-formed probability.  To obtain the equality \eqref{eq:qsecondmom} with a particular choice of CV, we need the additional constraint that \emph{all the measurement operators must commute with each other}.  As a result, they must be part of the same framework as the system observable and hence commute with that observable as well.  We will call any detector with commuting measurement operators with respect to a particular observable a \emph{fully compatible detector} for that observable.  Evidently, this is a strict requirement for a detector.

Alternatively, as with the classical case, we can change the CVs to define new observables that correspond to powers of the original observable, such as $G_X = (F_X)^n = \sum_{y\in Y} g_Y(y) E_y$.  These new observables can then be measured indirectly using the same experimental setup without the need for measurement sequences.  The CVs $g_Y(y)$ for the $n$\textsuperscript{th} power of $F_X$ will not be a simple power of the CVs $f_Y(y)$ for $F_X$ unless the measurement is unambiguous.

\emph{Correlation functions}.---If a time-evolution unitary rotation $\mathcal{U}_t$ is inserted between different observable measurements, then we obtain a quantum \emph{correlation function} instead,
\begin{align}\label{eq:qopercorrfun}
  \dmean{F_X(0)G_X(t)} &= \mean{\mathcal{F}_X(\mathcal{U}_t(\mathcal{G}_X(1_X)))}_X,
\end{align}
which should be compared to the classical case \eqref{eq:copercorrfun}.  Similarly, $n$-time correlations can be defined with $n-1$ time-evolutions between the observable measurements $\mean{\mathcal{F}_1(\mathcal{U}_{t_1}(\mathcal{F}_2(\cdots \mathcal{U}_{t_{n-1}}(\mathcal{F}_n(1_X))\cdots)))}$.  

\emph{Inversion}.---Since the CVs depend only on the probability observables, which commute with the measured observable for a fully compatible detector, the procedure for determining the CVs will be identical to the classical case.  That is, \emph{the contextual values of a quantum observable exactly correspond to the detector labels for a classically ambiguous detector}.  We shall refer the reader back to the classical inversion \eqref{eq:pseudoinversion} for discussion on how to solve the relation \eqref{eq:qpovmexpand}.  As a reminder, we advocate the pseudoinverse as a principled approach for picking the CVs in the event of redundancy or course-graining.

\emph{Conditioned averages}.---We can construct a general \emph{postselected conditioned average} from the CVs and the fully generalized ABL rule \eqref{eq:qgpostselection} analogously to the classical case \eqref{eq:ccondav},
\begin{align}\label{eq:qcondav}
  \cmean{z}{F_X} &= \sum_y f_Y(y)\, \cmean{z}{y} = \frac{\mean{\mathcal{F}_X(E'_z)}_X}{\mean{\mathcal{E}(E'_z)}_X}, \\
  &= \frac{\sum_{y\in Y}\sum_{y'\in Y'} f_Y(y) \Tr{\rho_X M^\dagger_{y,y'}E'_z M_{y,y'}}}{\sum_{y\in Y}\sum_{y'\in Y'} \Tr{\rho_X M^\dagger_{y,y'}E'_z M_{y,y'}}}. \nonumber
\end{align}
We introduced this type of conditioned average in \cite{Dressel2010} for the typical case of pure operations $\{\mathcal{E}_y\}$ with single associated measurement operators $\{M_y\}$.  

If the postselection is defined in the same framework as the measurement operation, then the nonselective measurement $\mathcal{E}$ in the denominator will reduce to unity, leaving a classical conditioned average,
\begin{align}\label{eq:qcondavcommute}
  \mean{F_X}_z &= \frac{\sum_{y\in Y} f_Y(y) \mean{E_y E'_z}_X}{\mean{E'_z}_X} = \frac{\mean{F_X E'_z}_X}{\mean{E'_z}_X},
\end{align}
of the same form as \eqref{eq:cgcollapse}.  Similarly, the preselected conditioning \eqref{eq:qgcollapse} will also reduce to \eqref{eq:qcondavcommute} for such a case.  This special case cannot exceed the eigenvalue range of the observable: the observable $F_X$ will always reduce to its eigenvalues since either the state or the postselection commute with it.

More generally, however, the combination of amplified CVs and the context-dependent probabilities in the general postselected average \eqref{eq:qcondav} can send it outside the eigenvalue range of the observable.  As we discussed in \cite{Williams2008,Dressel2011}, having such a conditioned average stray outside the eigenvalue range of the observable is equivalent to a violation of a Leggett-Garg inequality that tests the assumptions of macrorealism under noninvasive detection.  As a result, an eigenvalue range violation gives a direct indication of either \emph{nonclassicality} present in a measurement sequence, or intrinsic measurement \emph{disturbance} beyond that of noninvasive classical conditioning as we saw in the example in \S\ref{sec:cmarbledisturb}.  We refer the reader to \cite{Williams2008,Dressel2011} for more detail on this matter.

\emph{Strong-conditioned average}.---There are two other important special cases of the conditioned average \eqref{eq:qcondav} worth mentioning: strong measurement and weak measurement.  The strong measurement case is distinguished by being constrained exclusively to the eigenvalue range of the observable.  Specifically, \eqref{eq:qcondav} reduces to the form,
\begin{align}\label{eq:qcondavstrong}
  \cmean{z}{F_X} &= \frac{\sum_{x\in X} f_X(x) P(x)D_x(z)}{\sum_{x\in X} P(x)D_x(z)}, \\
  &= \frac{\sum_{x\in X} f_X(x) \bra{x}\rho\ket{x}|\pipr{x}{z}|^2}{\sum_{x\in X} \bra{x}\rho\ket{x}|\pipr{x}{z}|^2}, \nonumber
\end{align}
which contains only the eigenvalues $f_X(x)$ of the observable and factored probability products.  However, it cannot be expressed solely in terms of the observable $F_X$ and a conditioned state as in the classical case \eqref{eq:ccondav} due to the disturbances $D_x(z)$.  Only when the state or postselection commutes with the observable does \eqref{eq:qcondavstrong} reduce to a special case of \eqref{eq:qcondavcommute} and become free from disturbance.

\emph{Weak values}.---The weak measurement case is distinguished by being the only case of the quantum postselected conditioned average \eqref{eq:qcondav} that can become \emph{context independent} for any state and postselection (under certain conditions).  The context-independent weak limit of the conditioned average \eqref{eq:qcondav} is the \emph{weak value} \cite{Aharonov1988,Aharonov1990,Aharonov2005,Aharonov2008,Aharonov2009,Aharonov2010,Dressel2010},
\begin{align}\label{eq:qwv}
  \cmean{z}{F_X}^w &= \frac{\mean{E'_z F_X + F_X E'_z}_X}{2 \mean{E'_z}_X},
\end{align}
and is expressed entirely in terms of the system expectation functional $\mean{\cdot}_X$, the postselection probability observable $E'_z$, and the observable $F_X$.  Written in this form it is clear that it is a symmetrized version of the context-independent commuting case \eqref{eq:qcondavcommute}; however, unlike \eqref{eq:qcondavcommute} the weak value \eqref{eq:qwv} is not constrained to the eigenvalue range and can even diverge.  For a pure initial state with trace-density $x$ and pure postselection $z$, the weak value \eqref{eq:qwv} takes the traditional form,
\begin{align}\label{eq:qpwv}
  \wv{z}{F_X}{x} &\to \text{Re}\frac{\bra{z}F_X\ket{x}}{\pipr{z}{x}}.
\end{align}
We will consider under what conditions one can obtain such a weak value in Sec. \ref{sec:wv}.

\subsubsection{Example: Coverslip detector revisited}\label{sec:qpbscover2}
Continuing the example from Sec. \ref{sec:qpbscover} and Fig.~\ref{fig:coverslip}, observables defined in the same framework as the probability observables may be expressed in terms of the probability observables according to \eqref{eq:qpovmexpand} using \emph{contextual values} (CVs), exactly as in the classical example \eqref{eq:marblematrix},
\begin{subequations}
\begin{align}
  F_X &= f_X(h) h + f_X(v) v, \\
  &= f_Y(t) E_t + f_Y(r) E_r, \nonumber \\
  \begin{pmatrix}f_X(h) \\ f_X(v)\end{pmatrix} &= \begin{pmatrix}\tilde{P}(t|h) & \tilde{P}(r|h) \\ \tilde{P}(t|v) & \tilde{P}(r|v)\end{pmatrix}\begin{pmatrix}f_Y(t) \\ f_Y(r)\end{pmatrix}.
\end{align}
\end{subequations}
Inverting this relation according to \eqref{eq:pseudoinversion} produces the unique CVs,
\begin{subequations}\label{eq:polcv}
\begin{align}
  f_Y(t) &= \frac{\tilde{P}(r|v) f_X(h) - \tilde{P}(r|h) f_X(v)}{\tilde{P}(t|h) \tilde{P}(r|v) - \tilde{P}(r|h)\tilde{P}(t|v)}, \\
  f_Y(r) &= -\frac{\tilde{P}(t|v) f_X(h) - \tilde{P}(t|h) f_X(v)}{\tilde{P}(t|h) \tilde{P}(r|v) - \tilde{P}(r|h)\tilde{P}(t|v)}.
\end{align}
\end{subequations}
The denominator is unity when the output ports of the coverslip are perfectly correlated with the polarization.  Otherwise, the denominator is less than one and serves to \emph{amplify} the CVs to compensate for the ambiguity of the detection.  The numerator contains cross-compensation factors that correct bias in the detector; that is, the eigenvalue $f_X(h)$ for $h$ in the contextual value $f_Y(t)$ for $t$ is weighted by the conditional probability $\tilde{P}(r|v)$ corresponding to the complementary quantities of $v$ and $r$, and so forth.

The CVs define the detector observable that is actually being measured in the laboratory,
\begin{align}
  F_Y &= f_Y(t) t + f_Y(r) r.
\end{align}
This detector observable corresponds to a detection \emph{operation} on the system space according to \eqref{eq:qobsoper},
\begin{align}
  \mathcal{F}_X &= f_Y(t) \mathcal{E}_t + f_Y(r) \mathcal{E}_r,
\end{align}
which fully describes the interaction with the detector, subsequent conditioning, and experimental convention for defining the observable.  When no subsequent conditioning is performed on the system, this operation constructs the system observable $F_X = \mathcal{F}_X(1_X) = f_Y(t) E_t + f_Y(r) E_r$, as desired.

Since the pure measurement operations all belong to the same framework and commute with $F_X$, the operation $\mathcal{F}_X$ is also \emph{fully compatible} with the observable $F_X$, meaning it can measure any moment of that observable using the same CVs according to \eqref{eq:qsecondmom},
\begin{align}
  \mean{\mathcal{F}^n_X(1_X)}_X &= \mean{(F_X)^n}_X, \\
  &= \sum_{i_1\dots i_n} f_Y(i_1)\dots f_Y(i_n) \mean{E_{i_1}\dots E_{i_n}}_X. \nonumber
\end{align}
The quantity $\mathcal{F}^n_X(1_X)$ indicates a sequence of $n$ consecutive measurements made by the same coverslip on the beam to construct the observable $(F_X)^n$ for the $n$\textsuperscript{th} moment of $F_X$.  That is, the output from each port of the coverslip is fed back into the coverslip to be measured again.  There are $2^n$ possible outcome sequences $(i_1, \dots, i_n)$ for $n$ traversals through the coverslip, each with probability $\mean{E_{i_1}\dots E_{i_n}}_X$ of occurring.  These probabilities are weighted with appropriate products of corresponding CVs and summed to correctly construct the $n$\textsuperscript{th} moment of $F_X$.

Alternatively, one can change the CVs to directly measure the observable $G_X = (F_X)^n = g_Y(t) E_t + g_Y(r) E_r$ from one traversal of the coverslip.  The required CVs for $G_X$, 
\begin{subequations}\label{eq:polcvnthmom}
\begin{align}
  g_Y(t) &= \frac{\tilde{P}(r|v) (f_X(h))^n - \tilde{P}(r|h) (f_X(v))^n}{\tilde{P}(t|h) \tilde{P}(r|v) - \tilde{P}(r|h)\tilde{P}(t|v)}, \\
  g_Y(r) &= -\frac{\tilde{P}(t|v) (f_X(h))^n - \tilde{P}(t|h) (f_X(v))^n}{\tilde{P}(t|h) \tilde{P}(r|v) - \tilde{P}(r|h)\tilde{P}(t|v)},
\end{align}
\end{subequations}
are not simple powers of the CVs \eqref{eq:polcv} for $F_X$ unless the measurement is unambiguous.

In addition to moments of $F_X$, we can obtain postselected \emph{conditioned averages} of $F_X$ by conditioning on a second measurement outcome characterized by a probability observable $E'_z$ after the measurement by the coverslip according to \eqref{eq:qcondav},
\begin{align}\label{eq:polcondav}
  \cmean{z}{F_X} &= \frac{\mean{\mathcal{F}_X(E'_z)}_X}{\mean{\mathcal{E}(E'_z)}_X},
\end{align}
where $\mathcal{E} = \mathcal{E}_t + \mathcal{E}_r$ is the nonselective measurement by the coverslip.  The second measurement could be a polarizer, another coverslip, or any other method for measuring polarization a second time.

If the initial state is pure with a density $\rho = x = \pprj{x}$ and the final postselection is also pure $z = \pprj{z}$, then \eqref{eq:polcondav} simplifies to a pre- and postselected conditioned average,
\begin{align}\label{eq:polppsca}
  \ccmean{z}{F_X}{x} &= \frac{f_Y(t) |\bra{z}M_t\ket{x}|^2 + f_Y(r) |\bra{z}M_r\ket{x}|^2}{|\bra{z}M_t\ket{x}|^2 + |\bra{z}M_r\ket{x}|^2}.
\end{align}
If we relate both pure states to the reference state $h$ via unitary rotations as defined in \eqref{eq:polunitary}, $x = \mathcal{U}_{\alpha,\beta,\gamma}(h)$ and $z = \mathcal{U}_{\alpha',\beta',\gamma'}(h)$, then the probabilities take the form,
\begin{subequations}\label{eq:polprobs}
\begin{align}
  |\bra{z}M_t\ket{x}|^2 &= \tilde{P}^h(t)\cos^2(\beta/2)\cos^2(\beta'/2) \\
  &\quad + \tilde{P}^v(t)\sin^2(\beta/2)\sin^2(\beta'/2) \nonumber \\
  &\quad + \frac{\sqrt{\tilde{P}^h(t)\tilde{P}^v(t)}}{2}\sin\beta\sin\beta' \times \nonumber\\
  &\quad\quad \cos(\gamma - \gamma' - \phi_{h,t} + \phi_{v,t}), \nonumber \\
  |\bra{z}M_r\ket{x}|^2 &= \tilde{P}^h(r)\cos^2(\beta/2)\cos^2(\beta'/2) \\
  &\quad + \tilde{P}^v(r)\sin^2(\beta/2)\sin^2(\beta'/2) \nonumber \\
  &\quad + \frac{\sqrt{\tilde{P}^h(r)\tilde{P}^v(r)}}{2}\sin\beta\sin\beta' \times \nonumber\\
  &\quad\quad \cos(\gamma - \gamma' - \phi_{h,r} + \phi_{v,r}). \nonumber
\end{align}
\end{subequations}
We see that each probability possesses an interference term that stems from the relative orientations of the incompatible frameworks for the preparation, measurement, and postselection.  In addition, the relative phases in the measurement operators \eqref{eq:polmeasops2} will affect the orientations of the frameworks and further disturb the measurement, as mentioned.  For the classical case, the frameworks coincide, so $\beta,\beta' \in \{0,\pi\}$; the interference term vanishes; and, the probabilities reduce to the conditional probabilities that characterize the probability observables.

The combination of the expanded range of the CVs \eqref{eq:polcv} and the interference term in the probabilities \eqref{eq:polprobs} can make the postselected conditioned averages \eqref{eq:polcondav} counter-intuitively exceed the eigenvalue range of the observable $F_X$.  Such a violation of the eigenvalue range cannot occur from classical conditioning without disturbance as in Sec. \ref{sec:cmarbledisturb}.

\subsubsection{Example: Calcite polarization detector}
\begin{figure*}[th]
  \begin{center}
    \includegraphics{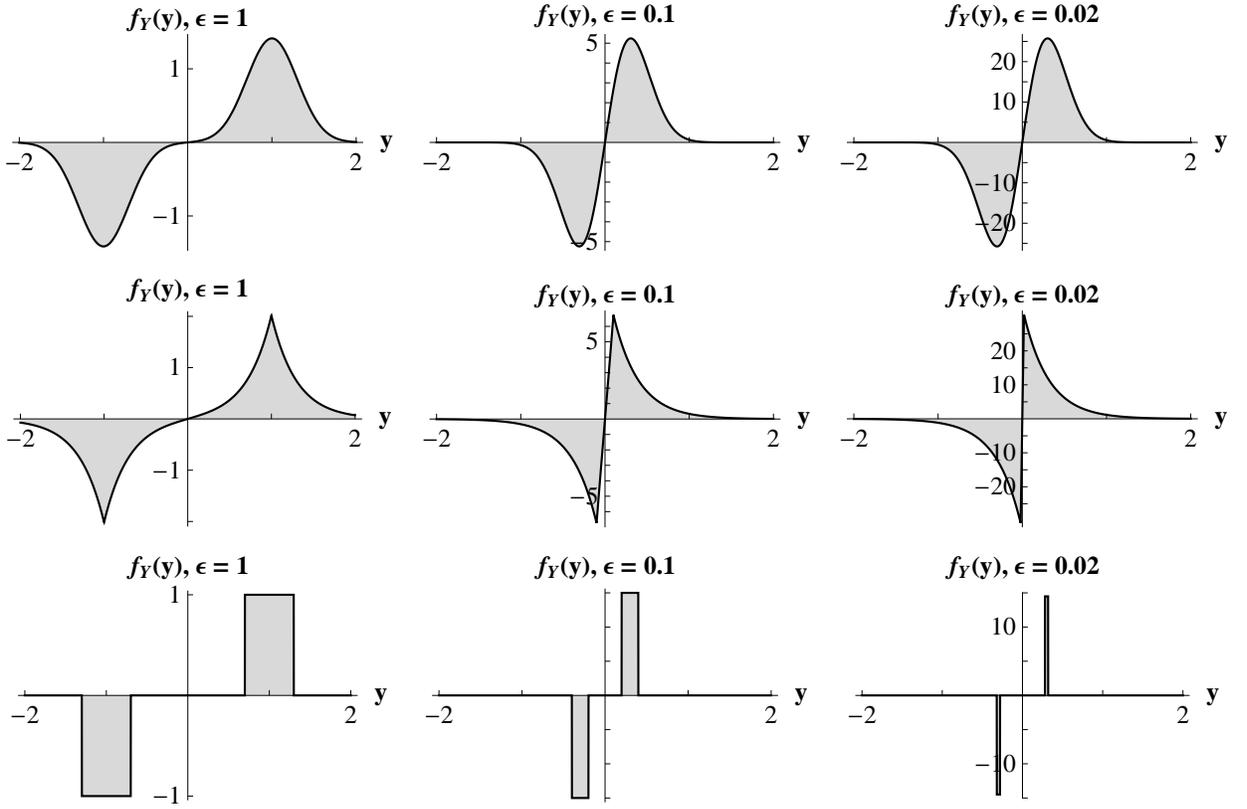}
  \end{center}
  \caption{Preferred CVs $f_Y(y)$ given in \eqref{eq:contpolcvgen} for a calcite position measurement that targets the polarization observable $F_X = h - v$, shown for strong separation ($\epsilon = 1$), wimpy separation ($\epsilon = 0.1$), and weak separation ($\epsilon = 0.02$) of the polarizations.  \emph{Top Row}: Initial Gaussian beam profile.  \emph{Middle Row}: Initial Laplace beam profile.  \emph{Bottom Row}: Initial top-hat beam profile.  Note that the top-hat CVs are the eigenvalues of $\pm 1$ under strong separation, but become amplified as the distributions start to overlap; moreover, the top-hat CVs cancel out in the perfectly ambiguous overlapping region.  The amplification and cancellation behavior of the CVs is more complicated for less definite detector profiles.}
  \label{fig:continuouscv}
\end{figure*}

We can also measure polarization using a von Neumann measurement \cite{VonNeumann1932} that uses a detector with a continuous sample space detector, such as position.  For example, passing a beam of polarized light through a calcite crystal will continuously separate the polarizations $h$ and $v$ along a particular position axis.  Measuring the position profile of the resulting split beam along that axis allows information to be gained about the polarization.

For such a setup, measuring the position with a linear scale corresponds to measuring a detector observable $Q = \int_Y y \, d\pprj{y}$ for a continuous sample space of distinguishable positions.  The observable $Q$ has a conjugate $D_Q$ that satisfies $[Q, D_Q] = i 1_Y$.  The conjugate can thus generate translations in $Q$ with a unitary rotor, $\exp(i q D_Q)Q\exp(-i q D_Q) = Q + [i q D_Q, Q] + [i q D_Q, [i q D_Q, Q]] + \dots = Q + q 1_Y$.  Hence, we can model the calcite crystal as a rotation governed by a unitary rotor of the form 
\begin{align}\label{eq:polrotor}
  U &= \exp(-i (\epsilon_h h - \epsilon_v v) D_Q),
\end{align}
which will translate $h$ polarization by some amount $\epsilon_h$ while simultaneously translating $v$ polarization by some amount $\epsilon_v$ in the opposing direction.  The parameters $\epsilon_h$ and $\epsilon_v$ will depend on the geometry of the crystal with respect to the incident beam.

Suppose the light beam has an initially pure beam profile state described by a density $\rho = \pprj{\psi}$.  The probability for obtaining a particular pure position $y = \pprj{y}$ in the profile would then be $dP_Y(y) = p_Y(y) dy = \text{Tr}(\rho y) dy = |\pipr{y}{\psi}|^2 dy$.  Each complex factor $\pipr{y}{\psi}$ is the ``wave function'' of the transverse beam profile, whose complex square is the probability density with respect to the integral $p_Y(y) = |\pipr{y}{\psi}|^2$.  

If we then pass the beam through the crystal described by the rotor \eqref{eq:polrotor} and measure its position in a pure position state $y = \pprj{y}$, we will have enacted a pure operation on the polarization of the beam that is characterized by a single measurement operator,
\begin{subequations}
\begin{align}
  d\mathcal{E}_y(F_X) &= M(y)^\dagger F_X M(y) dy, \\
  M(y) &= \bra{y}U\ket{\psi}, \\
  &= h \pipr{y - \epsilon_h}{\psi} + v \pipr{y + \epsilon_v}{\psi}, \nonumber
\end{align}
\end{subequations}
with components equal to the initial wave function of the detector profile shifted in position by an appropriate $\epsilon$.  The pure measurement operations define a continuous set of probability observables,
\begin{align}\label{eq:contpolpo}
  dE(y) &= d\mathcal{E}_y(1_X) = M(y)^\dagger M(y) dy, \\
  &= h\, dP_Y(y - \epsilon_h) + v\, dP_Y(y + \epsilon_v), \nonumber
\end{align}
with components equal to the initial transverse beam profile shifted in position by an appropriate $\epsilon$.  Unless the shifts become degenerate with $\epsilon_v = -\epsilon_h$ then these probability observables can be used to indirectly measure any observable in the framework of $h$ and $v$.

Since the observable $\epsilon_h h - \epsilon_v v$ appears as a generator for the rotation $U$, it could be tempting to assert that the detector must specifically measure this observable.  However, only the \emph{framework} in which the generating observable is defined determines which observables can be measured.  The choice of CV, which can be made in postprocessing, will calibrate the detector to measure specific observables in that framework.

\begin{figure*}[th]
  \begin{center}
    \includegraphics{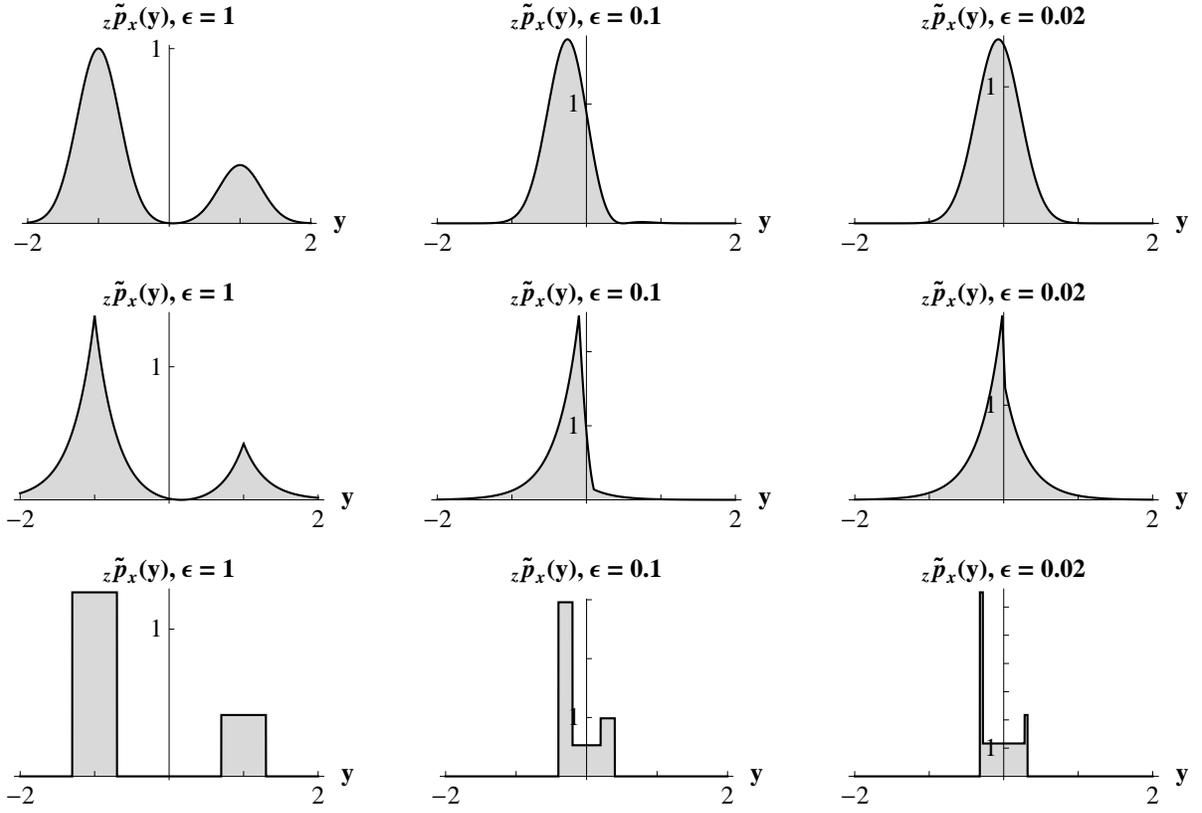}
  \end{center}
  \caption{Pre- and postselected detector probability densities $\tensor[_z]{\tilde{p}}{_x}(y)$ for the calcite position measurement \eqref{eq:contpolpo}, shown for strong separation ($\epsilon = 1$), wimpy separation ($\epsilon = 0.1$), and weak separation ($\epsilon = 0.02$) of the polarizations.  The preselection is $x = \pprj{x}$ with associated vector $\ket{x} = \cos(4\pi/6)\ket{h} + \sin(4\pi/6)\ket{v}$.  The postselection is $z = \pprj{z}$ with associated vector $\ket{z} = (\ket{h} + \ket{v})/\sqrt{2}$.  \emph{Top Row}: Initial Gaussian beam profile.  \emph{Middle Row}: Initial Laplace beam profile.  \emph{Bottom Row}: Initial top-hat beam profile.  Note that the Gaussian profile tilts to approximate a single shifted Gaussian under weak separation, as leveraged in the weak measurement protocol introduced in \cite{Aharonov1988}.}
  \label{fig:continuousprobs}
\end{figure*}

We considered a classical version of similar probability observables in \S\ref{sec:ccontinuousexample}.  Generalizing that derivation only slightly, we can find the preferred contextual values (CVs) $f_Y(y)$ for an arbitrary polarization observable $F_X = f_X(h)\, h + f_X(v)\, v$,
\begin{subequations}
\begin{align}
  f_Y(y) &= f_X(h) \frac{v_+(y) + v_-(y)}{2} \\
          &\quad + f_X(v) \frac{v_+(y) - v_-(y)}{2}, \nonumber \displaybreak[0] \\
  v_+(y) &= \frac{p_Y(y - \epsilon_h) + p_Y(y + \epsilon_v)}{a + b(\epsilon_h,\epsilon_v)},\displaybreak[0]  \\
  \label{eq:contpolcvgen}
  v_-(y) &= \frac{p_Y(y - \epsilon_h) - p_Y(y + \epsilon_v)}{a - b(\epsilon_h,\epsilon_v)}, \displaybreak[0]  \\
  a &= \int_Y p_Y^2(y) \, dy, \displaybreak[0] \\
  b(\epsilon_h,\epsilon_v) &= \int_Y p_Y(y-\epsilon_h)\,p_Y(y+\epsilon_v) \, dy.
\end{align}
\end{subequations}
In particular, one can measure the orthogonal observables $h - v$ and $1_X$ using the expansions,
\begin{align}
  h - v &= \int_Y v_-(y)\,dE(y), \\
  1_X = h + v &= \int_Y v_+(q)\,dE(y).
\end{align}

For the specific case of an initial Gaussian beam centered at zero, we have,
\begin{subequations}\label{eq:contpolgauss}
\begin{align}
  p(y) &= \exp\left(-\frac{y^2}{2 \sigma^2}\right)/\sigma \sqrt{2\pi},\displaybreak[0]  \\
  \epsilon &= (\epsilon_h + \epsilon_v)/2,\displaybreak[0]  \\
  \delta &= (\epsilon_h - \epsilon_v)/2,\displaybreak[0]  \\
  a &= \frac{1}{2\sigma\sqrt{\pi}},\displaybreak[0]  \\
  b(\epsilon) &= a \exp(- (\epsilon/\sigma)^2 ),\displaybreak[0]  \\
  \label{eq:contpolcv}
  v_-(y) &= \sqrt{2}\, \frac{\exp(-\frac{(y-\delta)^2}{2\sigma^2}) \sinh(\frac{\epsilon(y-\delta)}{\sigma^2})}{\sinh( \frac{\epsilon^2}{2 \sigma^2})},\displaybreak[0]  \\
  v_+(y) &= \sqrt{2}\, \frac{\exp(-\frac{(y-\delta)^2}{2\sigma^2}) \cosh(\frac{\epsilon(y-\delta)}{\sigma^2})}{\cosh( \frac{\epsilon^2}{2\sigma^2})},
\end{align}
\end{subequations}
What matters for the measurement is the average translation $\epsilon$ away from the midpoint $(y-\delta)$.  The amplification of the CVs is controlled by the parameter $\epsilon/\sigma$, which serves as an indicator for the ambiguity of the measurement.  When the shift $\epsilon$ is large compared to the width of the Gaussian $\sigma$, then $\epsilon/\sigma \gg 1$; the shifted Gaussians for $h$ and $v$ are distinguishable; the CVs approach the eigenvalues of the measurement; and, the measurement is unambiguous.  When the shift is small compared to the width of the Gaussian, then $\epsilon/\sigma \ll 1$, the Gaussians for $h$ and $v$ largely overlap, the CVs diverge, and the measurement is ambiguous.  Fig.~\ref{fig:continuouscv} shows the CVs \eqref{eq:contpolcv} for the Gaussian initial beam profile, as well as for a Laplace and top-hat profile for comparison.

\begin{figure*}[th]
  \begin{center}
    \includegraphics{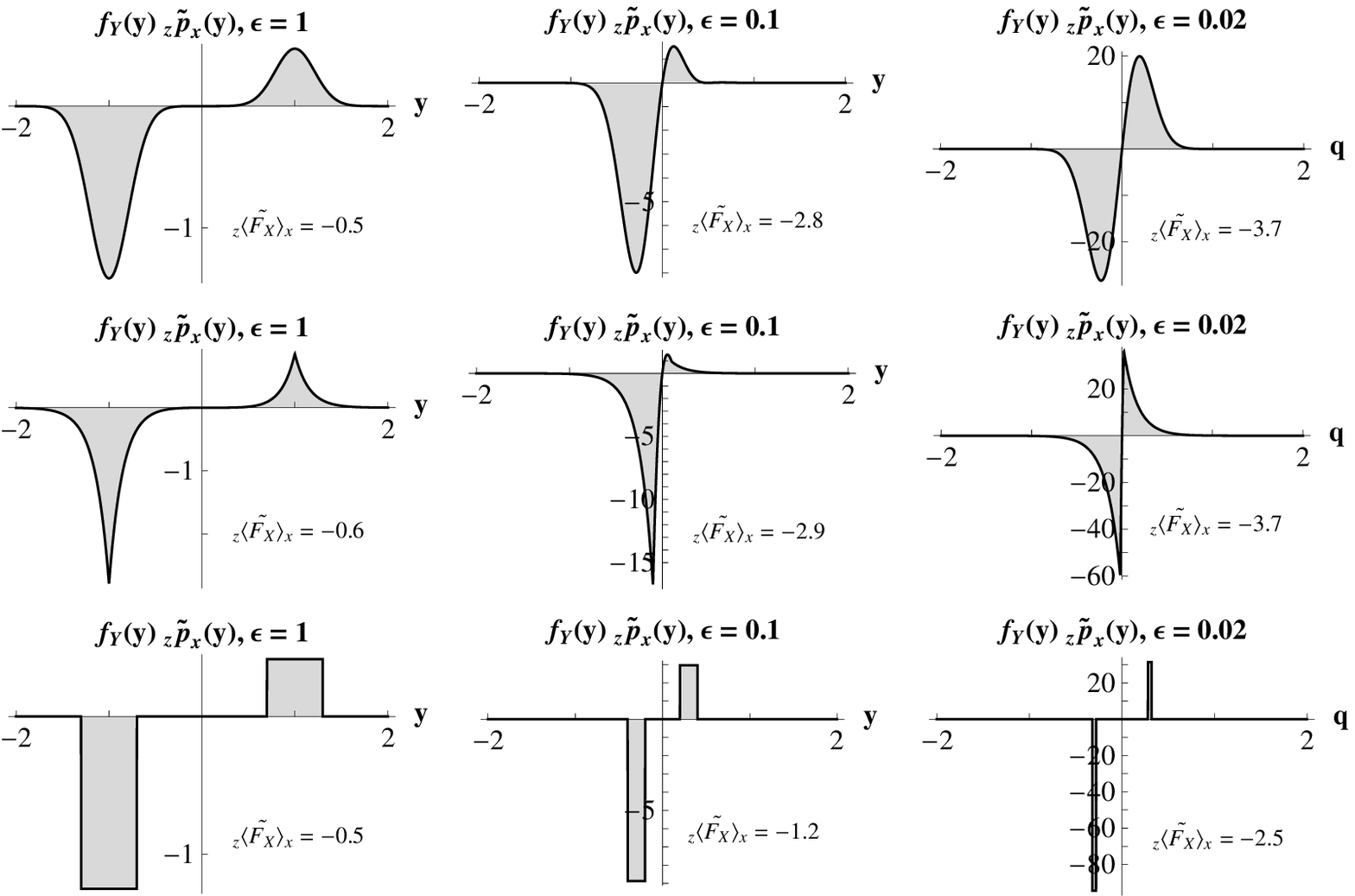}
  \end{center}
  \caption{Pre- and postselected conditioned average densities $f_Y(y) \tensor[_z]{\tilde{p}}{_x}(y)$ for a calcite position measurement targeting the observable $F_X = h - v$ with CVs as in Fig.~\ref{fig:continuouscv}, shown for strong separation ($\epsilon = 1$), wimpy separation ($\epsilon = 0.1$), and weak separation ($\epsilon = 0.02$) of the polarizations.  The conditioned averages $\ccmean{z}{F_X}{x} = \int_Y f_Y(y) \tensor[_z]{\tilde{p}}{_x}(y)\, dy$ are the areas under the curves and are shown inset.  As in Fig.~\ref{fig:continuousprobs}, the preselection is $x = \pprj{x}$, where $\ket{x} = \cos(4\pi/6)\ket{h} + \sin(4\pi/6)\ket{v}$.  The postselection is $z = \pprj{z}$, where $\ket{z} = (\ket{h} + \ket{v})/\sqrt{2}$.  \emph{Top Row}: Initial Gaussian beam profile.  \emph{Middle Row}: Initial Laplace beam profile.  \emph{Bottom Row}: Initial top-hat beam profile.  For sufficiently strong separation all three detector profiles will produce the strong conditioned average $\ccmean{z}{F_X}{x} = -1/2$.  For weak separation all three profiles approximate the weak value $\wv{z}{F_X}{x} = -2 - \sqrt{3} \approx -3.73$.  However, the different detector profiles converge to the weak value at different rates with decreasing $\epsilon$.}
  \label{fig:continuouswv}
\end{figure*}

This sort of detection protocol was used in the original paper on weak values \cite{Aharonov1988} in the form of a Stern-Gerlach apparatus that measures spin analogously to polarization using a continuous momentum displacement generated by a magnetic field.  The initial Gaussian beam profile shifted an amount $\epsilon$ away from the midpoint of the initial beam profile in a direction corresponding to the value of the spin.  Since the beam profile was symmetric about its mean, the generic CVs $f_Y(y) = y/\epsilon$ were implicitly assigned as a linear calibration of the detector, which targets a specific observable analogous to $h - v$.  Motivating this implicit choice was the fact that when $\epsilon$ is sufficiently small, the two overlapping Gaussians produce to a good approximation a single resulting Gaussian with a shifted mean consistent with such a linear scaling, as shown in Fig.~\ref{fig:continuousprobs}.  That such a choice was being made was later pointed out explicitly in \cite{DiLorenzo2008} before we identified the role of the CVs in \cite{Dressel2010} and derived the preferred form \eqref{eq:contpolcv}.  The proposed spin measurement protocol was adapted to a polarization measurement using a calcite crystal, as we have developed in this section, and then verified experimentally \cite{Duck1989,Ritchie1991}.

To produce the weak value from the polarization measurement, we postselect on a second measurement to form a conditioned average.  If the initial polarization state is pure with a density $\rho = x = \pprj{x}$ and the final postselection is also pure $z = \pprj{z}$, then we have the form,
\begin{align}\label{eq:contpolcondav}
  \ccmean{z}{F_X}{x} &= \frac{\int_Y f_Y(y) |\bra{z}M(y)\ket{x}|^2 dy}{\int_Y |\bra{z}M(y)\ket{x}|^2 dy}.
\end{align}
If we choose the symmetric Gaussian case \eqref{eq:contpolgauss} with $\delta = 0$ and take the form of $M(y)$ without additional unitary disturbance, 
\begin{align}
  M(y) &= h \exp\left(-\frac{(y-\epsilon)^2}{4 \sigma^2}\right)/\sqrt{\sigma \sqrt{2\pi}}, \\
  &+ v \exp\left(-\frac{(y+\epsilon)^2}{4 \sigma^2}\right)/\sqrt{\sigma \sqrt{2\pi}}, \nonumber
\end{align}
and relate both pure states to the reference state $h$ via unitary rotations as defined in \eqref{eq:polunitary}, $x = \mathcal{U}_{\alpha,\beta,\gamma}(h)$ and $z = \mathcal{U}_{\alpha',\beta',\gamma'}(h)$, then the postselected probability density $\tensor[_z]{\tilde{p}}{_x}(y)$ takes the form,
\begin{align}
  |\bra{z}M(y)\ket{x}|^2 &= \frac{\exp(- \frac{y^2 + \epsilon^2}{2 \sigma^2})}{2 \sigma \sqrt{2\pi}} \times \\
  &\qquad \big( (1 + \cos\beta\cos\beta')\cosh\frac{y\epsilon}{\sigma^2} \nonumber \\
  &\qquad + (\cos\beta + \cos\beta')\sinh\frac{y\epsilon}{\sigma^2} \nonumber \\
  &\qquad + \sin\beta\sin\beta'\cos(\gamma - \gamma') \big), \nonumber
\end{align}
Choosing the CVs \eqref{eq:contpolcv} to target the observable $h - v$, the conditioned average \eqref{eq:contpolcondav} then takes the form,
\begin{subequations}\label{eq:contcondavgauss}
\begin{align}
  \ccmean{z}{h - v}{x} &= \frac{\cos\beta + \cos\beta'}{1 + \cos\beta\cos\beta' + \Xi(\epsilon,\sigma)}, \\
  \Xi(\epsilon,\sigma) &= \sin\beta\sin\beta'\cos(\gamma-\gamma')\exp\left(-\frac{\epsilon^2}{2 \sigma^2}\right).
\end{align}
\end{subequations}
The interference term $\Xi(\epsilon,\sigma)$ in the denominator is the only part of the conditioned average that depends on the details of the measurement context through the exponential dependence on $\epsilon/\sigma$, which was also noted in \cite{Williams2008,Geszti2010}.  This conditioned average can exceed the eigenvalue range of the observable due to the combination of the amplified CVs and the disturbance linking the incompatible frameworks in the conditional probabilities.  Fig.~\ref{fig:continuouswv} shows the Gaussian measurement of the conditioned average \eqref{eq:contcondavgauss}, as well as top-hat and triangular measurements for comparison.

The conditioned average \eqref{eq:contcondavgauss} has two limiting cases that eliminate the explicit context-dependence: (1) In the strong-measurement limit, $\epsilon/\sigma \to \infty$, the interference term vanishes, leaving a conditioned average of projective measurements that always stays in the eigenvalue range of the observable. (2) In the weak-measurement limit, $\epsilon/\sigma \to 0$, the conditioned average reduces to the \emph{weak value},
\begin{align}\label{eq:calcitewv}
  \wv{z}{h-v}{x} &= \text{Re}\frac{\bra{z}(h - v)\ket{x}}{\pipr{z}{x}}, \\
  &= \frac{\cos\beta + \cos\beta'}{1 + \cos\beta\cos\beta' + \sin\beta\sin\beta'\cos(\gamma-\gamma')}. \nonumber
\end{align}
The weak value is distinguished by being the only case that can be written entirely in terms of the observable, the post-selection, and the pre-selected state without reference to the intermediate measurement.  In this sense, it is the only context-independent form of the conditioned average.  However, we shall see in \S\ref{sec:wv} that the weak value is not guaranteed as a limit point of the conditioned average in the weak-measurement limit.

\subsubsection{Example: Three-box paradox}
We can also use contextual values and the general conditioned average to analyze an often repeated paradox related to the logic of weak values: the three-box paradox \cite{Aharonov1991,Resch2004,Ravon2007,Aharon2008}.  Suppose one has three boxes, only one of which may be occupied by some quantum particle.  The boxes form a classical sample space, $X = \{a,b,c\}$, with Boolean algebra $\Sigma_X = \{0, a, b, c, a+b, b+c, c+a, 1_X\}$, with $1_X = a + b + c$.  Suppose that the boxes are preselected in the pure state with density $x = \pprj{x}$ and associated Hilbert space vector $\ket{x} = (\ket{a} + \ket{b} + \ket{c}) / \sqrt{3}$ and then later postselected with the pure projector $z = \pprj{z}$ and associated vector $\ket{z} = (\ket{a} + \ket{b} - \ket{c})/\sqrt{3}$.  The postselected state has a transition probability from the preselected state of $D_x(z) = |\pipr{z}{x}|^2 = 1/9$.

According to the weak value definition \eqref{eq:qpwv}, the weak values of the box-occupation observables for this pre- and postselected situation are,
\begin{subequations}\label{eq:threeboxwv}
\begin{align}
  \wv{z}{a}{x} &= 1, \\
  \wv{z}{b}{x} &= 1, \\
  \wv{z}{c}{x} &= -1.
\end{align}
\end{subequations}
These values have occasionally been interpreted as the counterfactual conditional probabilities of box-occupation given the double boundary conditions; that is, the box-occupation was not checked in between the pre- and postselection, but if it \emph{had} been checked without disturbing the system, then these probabilities would have been observed.  Part of the paradox is that the weak value for $c$ is negative, despite the fact that the eigenvalues for the occupation projector $c$ are $1$ and $0$ and cannot produce such a negative conditioned average unless \emph{negative conditional probabilities} average the eigenvalues.  Moreover, if the weak values do represent counterfactual probabilities, then the weak values for $a$ and $b$ both indicate a counterfactual certainty of occupation, and hence require a negative counterfactual probability for $c$ to correctly maintain the probability normalization condition.

Operationally, the weak value is an idealized limit point of a pre- and postselected conditioned average.  Since measuring it is not strictly achievable in the laboratory, we prefer to analyze this situation by considering a specific measurement context containing experimentally observable quantities.  In particular, we shall consider a detector for the three-box occupation that has the three outcomes $1$, $2$, and $3$.  The measurement operations are fully characterized by the single measurement operators,
\begin{subequations}
\begin{align}
  M_1 &= a\, \sqrt{(1+\epsilon)/3} + b\, \sqrt{(1-\epsilon)/3} + c\, \sqrt{1/3}, \\
  M_2 &= a\, \sqrt{(1-\epsilon)/3} + b\, \sqrt{1/3} + c\, \sqrt{(1+\epsilon)/3}, \\
  M_3 &= a\, \sqrt{1/3} + b\, \sqrt{(1+\epsilon)/3} + c\, \sqrt{(1-\epsilon)/3},
\end{align}
\end{subequations}
corresponding to the probability observables $E_1 = M_1^2$, $E_2 = M_2^2$, and $E_3 = M_3^2$.  For the particular pre- and postselection under consideration, these measurement operators produce the generalized ABL conditional probabilities,
\begin{subequations}
 \begin{align}
   \ccmean{z}{1}{x} &= \frac{\text{Tr}(z M_1 x M_1)}{\sum_{i=1}^3 \text{Tr}(z M_i x M_i)}, \\
   &= \frac{3 - 2\sqrt{1 + \epsilon} - 2\sqrt{1 - \epsilon} + 2\sqrt{1 - \epsilon^2}}{9 - 2\sqrt{1 + \epsilon} - 2\sqrt{1 - \epsilon} - 2\sqrt{1 - \epsilon^2}}, \nonumber \\
   &= \frac{1}{3} - \frac{\epsilon^2}{3} + O(\epsilon^3), \nonumber \displaybreak[0]\\
   \ccmean{z}{2}{x} &= \frac{\text{Tr}(z M_2 x M_2)}{\sum_{i=1}^3 \text{Tr}(z M_i x M_i)}, \\
   &= \frac{3 - 2\sqrt{1 + \epsilon} + 2\sqrt{1 - \epsilon} - 2\sqrt{1 - \epsilon^2}}{9 - 2\sqrt{1 + \epsilon} - 2\sqrt{1 - \epsilon} - 2\sqrt{1 - \epsilon^2}}, \nonumber \\
   &= \frac{1}{3} - \frac{2\epsilon}{3} + \frac{\epsilon^2}{6} + O(\epsilon^3), \displaybreak[0]\nonumber \\
   \ccmean{z}{3}{x} &= \frac{\text{Tr}(z M_3 x M_3)}{\sum_{i=1}^3 \text{Tr}(z M_i x M_i)}, \\
   &= \frac{3 + 2\sqrt{1 + \epsilon} - 2\sqrt{1 - \epsilon} - 2\sqrt{1 - \epsilon^2}}{9 - 2\sqrt{1 + \epsilon} - 2\sqrt{1 - \epsilon} - 2\sqrt{1 - \epsilon^2}}, \nonumber \\
   &= \frac{1}{3} + \frac{2\epsilon}{3} + \frac{\epsilon^2}{6} + O(\epsilon^3), \nonumber
 \end{align}
\end{subequations}
These detection probabilities are all positive and well-formed, since they are operationally accessible quantities.

If we target a particular observable $O_X = o_X(a) a + o_X(b) b + o_X(c) c$ for the three boxes, we can solve for the appropriate CVs by inverting the matrix equation,
\begin{align}
  \begin{pmatrix}o_X(a) \\ o_X(b) \\ o_X(c) \end{pmatrix} &= \frac{1}{3}\begin{pmatrix}1+\epsilon & 1-\epsilon & 1 \\ 1-\epsilon & 1 & 1+\epsilon \\ 1 & 1+\epsilon & 1-\epsilon \end{pmatrix}\begin{pmatrix}o_Y(1) \\ o_Y(2) \\ o_Y(3)\end{pmatrix},
\end{align}
producing,
\begin{align}
  \begin{pmatrix}o_Y(1) \\ o_Y(2) \\ o_Y(3)\end{pmatrix} &= \frac{o_X(a) + o_X(b) + o_X(c)}{3} \\
      &\qquad + \frac{1}{\epsilon}\begin{pmatrix}o_X(a) - o_X(b) \\ o_X(c) - o_X(a) \\ o_X(b) - o_X(c)\end{pmatrix}. \nonumber
\end{align}
In particular, we can use these CVs to expand the box-occupation observables in terms of the probability observables,
\begin{subequations}
\begin{align}
  a &= \left(\frac{1}{3} + \frac{1}{\epsilon}\right)E_1 + \left(\frac{1}{3} - \frac{1}{\epsilon}\right)E_2 + \frac{1}{3} E_3, \\
    &= \frac{1}{3}1_X + \frac{1}{\epsilon}(E_1 - E_2), \nonumber \\
  b &= \left(\frac{1}{3} - \frac{1}{\epsilon}\right)E_1 + \frac{1}{3} E_2 + \left(\frac{1}{3} + \frac{1}{\epsilon}\right)E_3, \\
    &= \frac{1}{3}1_X + \frac{1}{\epsilon}(E_3 - E_1), \nonumber \\
  c &= \frac{1}{3}E_1 + \left(\frac{1}{3} + \frac{1}{\epsilon}\right)E_2 + \left(\frac{1}{3} - \frac{1}{\epsilon}\right) E_3, \\
    &= \frac{1}{3}1_X + \frac{1}{\epsilon}(E_2 - E_3). \nonumber
\end{align}
\end{subequations}
Hence, all three box-occupation observables can be measured simultaneously from the same set of probabilities for the three detector outcomes.  Notably, the CVs assigned to each outcome can be \emph{negative} for sufficiently small $\epsilon$, even though all eigenvalues are positive or zero.  Hence the values being averaged can be negative and thus can lead to negative averages in principle.

Computing the appropriate conditioned averages we find to $O(\epsilon^3)$,
\begin{subequations}
\begin{align}
  \ccmean{z}{a}{x} &= 1 - \frac{\epsilon}{2} - \frac{\epsilon^2}{4} + O(\epsilon^3), \\
  \ccmean{z}{b}{x} &=  1 + \frac{\epsilon}{2} - \frac{\epsilon^2}{4} + O(\epsilon^3), \\
  \label{eq:threeboxca}
  \ccmean{z}{c}{x} &=  -1 + \frac{\epsilon^2}{2} + O(\epsilon^3).
\end{align}
\end{subequations}
which shows that the weak values \eqref{eq:threeboxwv} are the $\epsilon\to 0$ limit of the conditioned averages with this specific measurement context.  

The paradox of the negative weak value \eqref{eq:threeboxwv} can therefore be largely resolved in the following sense: the combination of the amplified negative CVs and the disturbance in the detector probabilities linking pre- and postselection frameworks leads to the negative result for $\ccmean{z}{c}{x}$ given sufficiently small $\epsilon$.  No negative probabilities are required to obtain the negative limit point since negative CVs are being averaged in the weak limit and not eigenvalues.  All operationally accessible probabilities are positive and well-behaved: the negative CVs are assigned by the experimenter and highlighted by the disturbance in the well-behaved probabilities.  

We leave the reader to ponder how to interpret the operationally accessible negative conditioned average \eqref{eq:threeboxca}.  However, we note that with at least this measurement context the conditioned averages do obey the equality,
\begin{align}
  \ccmean{z}{a}{x} + \ccmean{z}{b}{x} + \ccmean{z}{c}{x} = 1,
\end{align}
for all values of $\epsilon$.  The three sets of CVs sum to unity for each detector outcome, leaving only the normalized sum of detector probabilities $\ccmean{z}{1}{x} + \ccmean{z}{2}{x} + \ccmean{z}{3}{x} = 1$.  For more discussion of this paradox, see, for example, \cite{Aharonov1991,Resch2004,Ravon2007,Aharon2008}.

\subsection{Deriving the weak value}\label{sec:wv}
\emph{Weak value controversy}.---As we have seen for the case of the calcite detector \eqref{eq:calcitewv} and the three box paradox \eqref{eq:threeboxwv}, the weak value \eqref{eq:qwv} seems to arise naturally as the weak limit of postselected conditioned averages.  Indeed, much of the existing literature on weak values (e.g. \cite{Aharonov1988,Aharonov1990,Aharonov2005,Aharonov2008,Aharonov2009,Aharonov2010,Tollaksen2007}) operates under the assumption that it is the only weak limit of a conditioned average, or that it is a well-defined property of a pre- and postselected ensemble prior to the ensemble being measured.  However, a conditioned average does not necessarily converge to the weak value in the weak measurement limit, as has been noted independently by several groups \cite{Jozsa2007,DiLorenzo2008,Parrott2009,Dressel2010,Geszti2010,Wu2011a}, making its interpretation as a well-defined property worthy of more careful consideration.  To obtain correct laboratory predictions for a conditioned average, the formula \eqref{eq:qcondav} must be used, which generally requires the specification of the detection strategy and the protocol for assigning CVs to target a specific observable.  

Despite the interpretational controversy, the weak value \eqref{eq:qwv} is distinguished by being a \emph{context independent} weak limit of the conditioned average that is easy to compute theoretically and appears quite commonly in typical laboratory situations.  The formal expression of the weak value can also appear in other measurement scenarios, such as in ``modular values'' \cite{Kedem2010a}, or even perturbative corrections to energy spectra \cite{Note6}, which makes it an independently interesting quantity to study.

We will now demonstrate how the weak value \eqref{eq:qwv} can be uniquely defined from the general conditioned average \eqref{eq:qcondav} by imposing a set of sufficient conditions that the measurement should satisfy.  

\emph{Preliminaries}.---First we note from \eqref{eq:qmeasops} that each measurement operator has a polar decomposition, $M_{y,y'} = U_{y,y'} |M|_{y,y'}$, in terms of a unitary operator $U_{y,y'}$ and a positive operator $|M|_{y,y'}$.  It then follows that,
\begin{align}\label{eq:maction}
  M^\dagger_{y,y'} E'_z M_{y,y'} &= |M|_{y,y'} U^\dagger_{y,y'} E'_z U_{y,y'} |M|_{y,y'}, \\
  &= \acomm{|M|^2_{y,y'}}{\mathcal{U}_{y,y'}(E'_z)}/2  \nonumber \\
  &\quad - \comm{|M|_{y,y'}}{\comm{|M|_{y,y'}}{\mathcal{U}_{y,y'}(E'_z)}}/2, \nonumber
\end{align}
where $\acomm{A}{B} = AB + BA$ is the anticommutator, $\comm{A}{B} = AB - BA$ is the commutator, and $\mathcal{U}_{y,y'}(E'_z) = U^\dagger_{y,y'}E'_z U_{y,y'}$ is a unitary rotation of the postselection.

\emph{Sufficient conditions}.---Next we make the following sufficient assumptions regarding the dependence of the relevant quantities on the measurement strength parameter $\epsilon$:  
\begin{enumerate}
  \item The measurement operators $M_{y,y'}$ are analytic functions of $\epsilon$, and thus have well defined Taylor expansions around $\epsilon=0$ such that they are proportional to the identity in the weak limit, $\forall y,y', \, \lim_{\epsilon\to 0}M_{y,y'} \propto 1_X$.  
  \item The unitary parts of the measurement operators $U_{y,y'} = \exp(i G_{y,y'}(\epsilon))$ are generated by Hermitian operators of order $\epsilon^k$, $G_{y,y'}(\epsilon) = \epsilon^k G^{(k)}_{y,y'} + O(\epsilon^{k+1})$, for some integer $k\geq 1$.  Furthermore, each $U_{y,y'}$ must commute with either the system state or the postselection, $\forall {y,y'},\, [U_{y,y'},\rho_X] = 0$, or $\forall {y,y'},\, [U_{y,y'},E'_z] = 0$.  
  \item The equality $F_X = \sum_y f_Y(\epsilon; y) E_y(\epsilon)$ must be satisfied, where the CVs $f_Y(\epsilon; y)$ are selected according to the pseudo-inverse prescription.
  \item The minimum nonzero order in $\epsilon$ for all $|M|_{y,y'}(\epsilon)$ is $\epsilon^n$ such that assumption $(3)$ can also be satisfied for some CVs by the truncation to order $\epsilon^n$.  That is, for all $y$,$y'$, $|M|_{y,y'} = c_{y,y'} 1_X + |M|^{(n)}_{y,y'} \epsilon^n + O(\epsilon^{n+1})$, where $\sum_{y'} c_{y,y'}^2 = P_Y(y)$ is the detector probability in absence of interaction, and some of the $|M|^{(n)}_{y,y'}$ may vanish.  
  \item The probability observables $E_y(\epsilon) = \sum_{y'} M^\dagger_{y,y'}(\epsilon)M_{y,y'}(\epsilon)$ commute with the observable $F_X$.  
\end{enumerate}

\emph{Theorem}.---Given the above sufficient conditions, we have the following theorem: \emph{in the weak limit $\epsilon\to0$ the context dependence of the conditioned average \eqref{eq:qcondav} vanishes and the weak value \eqref{eq:qwv} is uniquely defined}.

\emph{Proof}.---To prove the theorem, we expand \eqref{eq:qcondav} to the minimum necessary order of $\epsilon^n$ and then take the weak limit as $\epsilon\to 0$.  First, we expand \eqref{eq:maction} to order $\epsilon^n$ using assumptions (1), and (4),
\begin{align}\label{eq:mactionepsilon}
  M^\dagger_{y,y'} E'_z M_{y,y'} &= c_{y,y'}^2 \mathcal{U}_{y,y'}(E'_z) \\
  & + c_{y,y'} \acomm{|M|^{(n)}_{y,y'}}{\mathcal{U}_{y,y'}(E'_z)} \epsilon^n + O(\epsilon^{n+1}). \nonumber
\end{align}
Generally, the remaining unitary rotation of the postselection will disturb the weak limit.  However, if $\comm{U_{y,y'}}{E'_z} = 0$ as in assumption (2), then $\mathcal{U}_{y,y'}(E'_z) = E'_z$ and the unitary disturbance disappears.  If instead $\comm{U_{y,y'}}{\rho_X} = 0$, then we can apply the state to \eqref{eq:mactionepsilon} and find,
\begin{align}\label{eq:mactionepsilon2}
  \mean{M^\dagger_{y,y'} E'_z M_{y,y'}}_X &= c_{y,y'}^2 \mean{\mathcal{U}_{y,y'}(E'_z)}_X \\
  & + c_{y,y'} \mean{\acomm{|M|^{(n)}_{y,y'}}{\mathcal{U}_{y,y'}(E'_z)}}_X \epsilon^n  \nonumber \\
  & + O(\epsilon^{n+1}). \nonumber
\end{align}
Since $\mean{\mathcal{U}_{y,y'}(E'_z)}_X = \text{Tr}_X(\mathcal{U}^\dagger(\rho_X) E'_z) = \mean{E'_z}_X$, the first term simplifies.  The unitary rotation in the second term expands to $\mathcal{U}_{y,y'}(E'_z) = E'_z + O(\epsilon^k)$, and the $O(\epsilon^k)$ correction can be absorbed into the overall $O(\epsilon^{n+1})$ correction.  

Therefore, after summing over $y'$ we find up to corrections of order $\epsilon^{n+1}$,
\begin{align}\label{eq:mactionepsilon3}
  \sum_{y'} \mean{M^\dagger_{y,y'} E'_z M_{y,y'}}_X &= \mean{\acomm{E_y(\epsilon)}{E'_z}}_X/2, 
\end{align}
where the probability observable has the expansion to order $\epsilon^n$,
\begin{align}
  E_y(\epsilon) &= \sum_{y'} |M|_{y,y'}^2(\epsilon), \\
  &= \sum_{y'} (c_{y,y'}^2 1_X + 2 c_{y,y'} |M|^{(n)}_{y,y'}\epsilon^n + O(\epsilon^{n+1})). \nonumber
\end{align}

Inserting \eqref{eq:mactionepsilon3} into \eqref{eq:qcondav}, we find,
\begin{align}
  \cmean{z}{F_X} &= \frac{\mean{\acomm{F_X}{E'_z}/2}_X + \sum_y f_Y(\epsilon; y) O(\epsilon^{n+1})}{\mean{\acomm{1_X}{E'_z}/2}_X + O(\epsilon^{n+1})}, 
\end{align}
where we have simplified $\sum_y f_Y(\epsilon; y) E_y(\epsilon) = F_X$ in the numerator, and $\sum_y E_y(\epsilon) = 1_X$ in the denominator.  Hence, unless the CVs in the numerator have poles larger than $1/\epsilon^n$ the correction terms of order $\epsilon^{n+1}$ will vanish, producing \eqref{eq:qwv} in the weak limit $\epsilon\to 0$, as claimed.  The last step in obtaining \eqref{eq:qwv}, therefore, is to show that the pseudoinverse solution for $f_Y$ that was indicated by assumption (3) cannot have poles larger than $1/\epsilon^n$.  The following lemmas will show this, which will prove the main theorem.

\emph{Lemma preliminaries}.---First, we note that $F_X$ commutes with $\{E_y(\epsilon)\}$ by assumption (5).  As such, we will replace the CV definition $F_X = \sum_y f_Y(\epsilon; y) E_y(\epsilon)$ with an equivalent matrix equation,
\begin{subequations}
\begin{align}
  \label{eq:proofcv}
  \vec{f}_X &= \mathcal{S} \vec{f}_Y, \\
  \label{eq:proofS}
  \mathcal{S} &= \begin{pmatrix}\mean{E_y(\epsilon)}_x & \cdots \\ \vdots & \ddots \end{pmatrix}.
\end{align}
\end{subequations}
The pseudoinverse is constructed from the singular value decomposition $\mathcal{S} = \mathcal{U} \Sigma \mathcal{V}^T$ as $\mathcal{S}^+ = \mathcal{V}\Sigma^+ \mathcal{U}^T$, where $\mathcal{U}$ and $\mathcal{V}$ are orthogonal matrices such that $\mathcal{U}^T \mathcal{U} = \mathcal{V} \mathcal{V}^T = 1$, $\Sigma$ is the singular value matrix composed of the square roots of the eigenvalues of $\mathcal{S}\mathcal{S}^T$, and $\Sigma^+ $ is composed of the inverse nonzero elements in $\Sigma^T$.  

Next, we note that the truncation of the matrix $\mathcal{S}$ to order $\epsilon^n$ has the form,
\begin{equation}
  \mathcal{S}' = \mathcal{P} + \epsilon^n \mathcal{S}_n,
\end{equation}
where $\mathcal{P} = (P_Y(y) \vec{1}, \cdots)$ is a matrix whose rows are identical and whose columns contain the interaction-free detector probabilities $P_Y(y)$, and $\mathcal{S}_n = (\vec{E}^{(n)}_1, \cdots)$ is a matrix whose rows all sum to zero.  Furthermore, since the solution to the equation $\vec{f}_X = \mathcal{S}' \vec{f}'_Y$ is assumed to exist by assumption (4), then the dimension of the detector, $N$, must be greater than or equal to the dimension of the system, $M$.  We then have the following two lemmas.

\emph{Lemma 1}.---\emph{The singular values of the truncated matrix} $\mathcal{S}'$ \emph{have maximum leading order} $\epsilon^n$.

\emph{Proof}.---The singular values of $\mathcal{S}'$ are $\sigma_k = \sqrt{\lambda_k}$, where $\lambda_k$ are $M$ eigenvalues of $\mathcal{H} = \mathcal{S}^T \mathcal{S}$, with its other $N-M$ eigenvalues being zero.  Since $\mathcal{P}^T \mathcal{S}_n = 0$, this matrix has the simple form $\mathcal{H} = \mathcal{P}^T\mathcal{P} + \epsilon^{2n} \mathcal{S}_n^T \mathcal{S}_n$, where $(\mathcal{P}^T \mathcal{P})_{ij} = M P_Y(i) P_Y(j)$ is $M||\vec{p}||^2$ times the projection operator onto the probability vector $\vec{p} = (P_Y(y),\cdots)$, and $(\mathcal{S}_n^T \mathcal{S}_n)_{ij} = \vec{E}^{(n)}_i \cdot \vec{E}^{(n)}_j$.  We will use $\mathcal{H}$ to determine the singular values of $\mathcal{S}'$.
  
Differentiating the eigenvalue relation $\mathcal{H}(\epsilon^{2n}) \vec{u}_k(\epsilon^{2n}) = \lambda_k(\epsilon^{2n}) \vec{u}_k(\epsilon^{2n})$ and the eigenvector normalization condition $\vec{u}_k(\epsilon^{2n}) \cdot \vec{u}_k(\epsilon^{2n}) = 1$ with respect to $\epsilon^{2n}$ produces the following deformation equation that describes how the eigenvalues of $\mathcal{H}$ continuously change with increasing $\epsilon^{2n}$,
\begin{eqnarray}\label{eq:deform}
  \dot{\lambda}_k(\epsilon^{2n}) = ||\mathcal{S}_n \vec{u}_k(\epsilon^{2n})||^2.
\end{eqnarray}
Integrating this equation produces the following perturbative expansion of the eigenvalues for small $\epsilon$,
\begin{eqnarray}\label{eq:perturb}
  \lambda_k(\epsilon^{2n}) &=& \lambda_k(0) + \epsilon^{2n} ||\mathcal{S}_n \vec{u}_k(0)||^2 + O(\epsilon^{4n}). 
\end{eqnarray} 
Hence, to prove the lemma it is sufficient to show that $\lambda_k(0)$ and $\mathcal{S}_n \vec{u}_k(0)$ cannot both vanish unless $\lambda_k(\epsilon^{2n}) = 0$ for all $\epsilon$.

Since $\mathcal{H}(0) = \mathcal{P}^T \mathcal{P}$ is a projection operator, $\lambda_1(0) = M ||\vec{p}||^2$ is its only nonzero eigenvalue with associated eigenvector $\vec{u}_1(0) = \vec{p}/||\vec{p}||$.  Hence, $\sigma_1(\epsilon^{2n}) \approx \sqrt{M} ||\vec{p}|| > 0$ to leading order.  For $k\neq 1$, $\lambda_k(0) = 0$ and $\vec{u}_k(0)$ can be chosen arbitrarily to span the degenerate $(N-1)$-dimensional subspace orthogonal to $\vec{u}_1(0)$.  Suppose $\mathcal{S}_n \vec{u}_k(0) = 0$ for some $k$.  It follows that $\mathcal{H}(\epsilon^{2n}) \vec{u}_k(0) = \mathcal{P}^T \mathcal{P}\vec{u}_k(0) + \epsilon^{2n} \mathcal{S}^T_n \mathcal{S}_n \vec{u}_k(0) = 0$ since $\vec{u}_k(0)$ is orthogonal to $\vec{u}_1(0) \propto \vec{p}$.  Therefore, $\vec{u}_k(0)$ is an eigenvector of $\mathcal{H}(\epsilon^{2n})$ with eigenvalue $0$ for any $\epsilon$.  Since $\mathcal{H}$ is symmetric, its eigenvectors form an orthogonal set for any $\epsilon$, so we must have the identification $\vec{u}_k(\epsilon^{2n}) = \vec{u}_k(0)$.  As a result, the associated eigenvalue vanishes for any $\epsilon$, $\lambda_k(\epsilon^{2n}) = \lambda_k(0) = 0$, which proves the lemma.  

\emph{Lemma 2}.---\emph{The pseudoinverse solution} $\vec{f}_Y$ \emph{to} \eqref{eq:proofcv} \emph{cannot have poles larger than} $1/\epsilon^n$.

\emph{Proof}.---In order to satisfy \eqref{eq:proofcv}, we have the equivalent condition for each component of $\mathcal{U}^T\vec{f}_X = \Sigma \mathcal{V}^T \vec{f}_Y$, 
\begin{align}\label{eq:proofsvcond}
  (\mathcal{U}^T \vec{f}_X)_k = \Sigma_{kk} (\mathcal{V}^T \vec{f}_Y)_k.
\end{align}
Therefore, all singular values $\Sigma_{kk}$ corresponding to nonzero components of $\mathcal{U}^T \vec{f}_X$ must also be nonzero; we shall call these the \emph{relevant} singular values.  Singular values which are not relevant will not contribute to the solution $\vec{f}_Y = \mathcal{V} \Sigma^+ \mathcal{U}^T \vec{f}_X$.  We can see this since $(\vec{f}_Y)_j = (\mathcal{V} \Sigma^+ \mathcal{U}^T \vec{f}_X)_j = \sum_k \mathcal{V}_{jk} \Sigma^+_{kk} (\mathcal{U}^T \vec{f}_X)_k$, so any zero element of $\mathcal{U}^T\vec{f}_X$ will eliminate the inverse irrelevant singular value $\Sigma^+_{kk}$ from the solution for $(\vec{f}_Y)_j$.  

Since the orthogonal matrices $\mathcal{U}$ and $\mathcal{V}$ do not contain any poles, and since $\vec{f}_X$ is $\epsilon$-independent, then the only poles in the solution $\vec{f}_Y = \mathcal{S}^+ \vec{f}_X = \mathcal{V} \Sigma^+  \mathcal{U}^T \vec{f}_Y$ must come from the inverses of the relevant singular values in $\Sigma^+$.  If a singular value $\Sigma_{kk}$ has leading order $\epsilon^m$, then its inverse $\Sigma^+_{kk} = 1/\Sigma_{kk}$ has leading order $1/\epsilon^m$; therefore, to have a pole of order higher than $1/\epsilon^n$ then there must be at least one relevant singular value with a leading order greater than $\epsilon^n$.  However, if that were the case then the truncation $\mathcal{S}'$ of $\mathcal{S}$ to order $\epsilon^n$ could not satisfy \eqref{eq:proofsvcond} since to that order it would have a relevant singular value of zero according to the previous lemma, contradicting assumption (4) about needing to satisfy the CV definition with the minimum nonzero order in $\epsilon$.  Therefore, the pseudoinverse solution $\vec{f}_Y = \mathcal{S}^+\vec{f}_X$ can have no pole with order higher than $1/\epsilon^n$ and the lemma is proven.  

\emph{Exceptions}.---As the main theorem indicates, the weak value will arise as the weak limit of a conditioned average in many common laboratory situations, which explains its seeming stability in the literature.  However, if the sufficiency conditions of the theorem are not met, then a different weak limit may be found.  For example, if there is $\epsilon$-dependent unitary disturbance in the measurement, then the postselection can be effectively rotated to a different framework for each measurement outcome, which creates additional terms in the weak limit.  Similarly, if the CVs are $\epsilon$-dependent and diverge more rapidly than $1/\epsilon^n$ then additional terms will become relevant in the weak limit.  (See, for example, Ref.~\cite{Wiseman2002}.)  This latter case can happen either from a pathological choice of CVs by the experimenter in the case of redundancy, or from a set of probability observables that cannot satisfy the constraint equation $F_X = \sum_y f_Y(\epsilon; y) E_y(\epsilon)$ with their lowest nonzero order in $\epsilon$.  Such probability observables that do not satisfy the constraint equation to lowest order are poorly correlated with the observable in the weak limit.  We refer the reader to \cite{Dressel2012} for more discussion on the uniqueness issue of weak values.  The theorem presented here is a slight generalization of the one presented therein.

\section{Conclusion} \label{sec:conclusion}
In this work, we have detailed the contextual-value approach to the generalized measurement of observables that we originally introduced in the letter \cite{Dressel2010} and further developed in \cite{Dressel2011,Dressel2011b,Dressel2012}.  This approach completes the well-established operational theory of state measurements by directly relating the state-transformations to traditional observables.  Each such operation typically corresponds to a distinguishable outcome of a correlated detection apparatus.  An experimenter can construct an observable from such an apparatus by assigning values to its outcomes.  The assigned values can be generally amplified from the eigenvalues of the constructed observable due to ambiguity in the measurement, and thus form a generalized spectrum that depends on the specific measurement context.  Hence, we call these values \emph{contextual values} for the constructed observable that allow its indirect measurement using such a correlated detector. 

Constructing an observable using contextual values requires only classical probability theory, according to \eqref{eq:cpovmexpand}.  Hence, the technique may be used wherever Bayesian probability theory applies.  We have outlined an algebraic approach to operational measurements from within Bayesian probability theory to encourage applications along these lines.

We have also shown how to construct a quantum probability space as the orbit of a classical probability space under the special unitary group.  This point of view illustrates that quantum observables can be constructed from contextual values in precisely the same way \eqref{eq:qpovmexpand} as their classical counterparts.  The approach also highlights the similarity between L\"{u}der's rule \eqref{eq:qcollapse} for updating a quantum state and \emph{invasive} classical conditioning \eqref{eq:cmdisturbpost}, which leads to a similarity between quantum operations \eqref{eq:qmeasoper} and classically invasive measurement operations \eqref{eq:cgoperations}.  Numerous physical examples have been given.

By putting all observable measurements on the same footing, the contextual values formalism subsumes not only projective measurements but also weak measurements as special cases.  To emphasize this point, we have analyzed the quantum weak measurement protocol introduced by Aharonov et al. \cite{Aharonov1988} in detail as an example using a calcite crystal and a polarized laser beam.  We have also derived the quantum weak value \eqref{eq:qwv} as a limit point of a general pre- and postselected conditioned average \eqref{eq:qcondav} as the measurement strength goes to zero and have given sufficient conditions for the convergence to hold.  Like the classically invasive conditioned average \eqref{eq:ccondav}, the quantum conditioned average, with the quantum weak value as a special case, can exceed the eigenvalue bounds of the observable.

The use of contextual values considerably clarifies and formalizes the process of measuring observables, particularly within a laboratory setting.  The elements of the formalism directly describe operationally accessible quantities that can be tomographically calibrated.  As such, the technique should be of considerable interest to experimentalists working on measurement and control of both quantum and classical systems.  Furthermore, the formalism prompts interesting theoretical questions about the foundations of quantum mechanics by highlighting its myriad similarities to classical probability theory.

\begin{acknowledgments}
  We acknowledge helpful discussions with Shantanu Agarwal and support from the National Science Foundation under Grant No. DMR-0844899, and the US Army Research Office under grant Grant No. W911NF-09-0-01417.
\end{acknowledgments}

\bibliographystyle{apsrev}


\end{document}